%% file: main.tex
\documentclass[11pt]{article}

\usepackage[a4paper,margin=30mm]{geometry}
\usepackage[round,authoryear]{natbib}
\usepackage{microtype}
\usepackage{pp-manuscript}
\usepackage[hidelinks]{hyperref}
\usepackage[capitalise,nameinlink,noabbrev]{cleveref}

\input{metadata}
\input{generated/numbers}
\input{generated/lean_status}

\title{\papertitle}
\author{\paperauthors\\\paperaffiliations}
\date{\paperdate}

\begin{document}

\maketitle

\begin{abstract}
  \input{chapters/abstract}
\end{abstract}


\medskip

\noindent\textit{JEL classification:} \paperjel

\input{body}

\bibliographystyle{plainnat}
\bibliography{refs}

\appendix
\input{appendix}

\end{document}

%% file: metadata.tex
\newcommand{\papertitle}{Portfolio Risk Bounds without Cross-Asset Return
Covariances: Distributional Fields from Language-Model Representations}

\newcommand{\paperauthors}{Marcus Gawronsky, Chun-Sung Huang}
\newcommand{\paperaffiliations}{Department of Finance and Tax,
  University of Cape
Town}

\newcommand{\paperdate}{August 2026}
\newcommand{\paperkeywords}{Portfolio Risk; Certified Diversification;
  Wasserstein Distance; Distributional Fields; Robust Portfolio Choice;
Language-Model Representations}
\newcommand{\paperjel}{G11; C58; C60}

%% file: generated/numbers.tex
\ppDeclareValue{n-tickers}{100}
\ppDeclareValue{n-obs}{1207}
\ppDeclareValue{certificate-equal-weight-least-conservative-pct}{54.233143}
\ppDeclareValue{certificate-equal-weight-most-conservative-pct}{12.275404}
\ppDeclareValue{certificate-inverse-volatility-least-conservative-pct}{54.387168}
\ppDeclareValue{certificate-inverse-volatility-most-conservative-pct}{12.328133}

\ppDeclareValue{news-only-variance}{0.359419}
\ppDeclareValue{news-only-gmv-relative-variance-pct}{147.508127}
\ppDeclareValue{news-only-gmv-excess-variance-pct}{47.508127}
\ppDeclareValue{equal-standardized-variance}{0.396209}
\ppDeclareValue{news-only-equal-variance-reduction-pct}{9.285481}
\ppDeclareValue{equal-gmv-relative-variance-pct}{162.606966}
\ppDeclareValue{sample-gmv-variance}{0.243661}
\ppDeclareValue{news-only-certificate-credit-pct}{57.313481}
\ppDeclareValue{news-only-maximum-weight-pct}{10.859124}
\ppDeclareValue{news-only-effective-number-assets}{28.09}
\ppDeclareValue{news-only-equal-credit-gap-pct}{3.080338}

\ppDeclareValue{news-only-percentile-uniform-cap-12p5-pct}{0.095000}
\ppDeclareValue{equal-percentile-uniform-cap-12p5-pct}{34.725000}
\ppDeclareValue{reference-effective-n-uniform-cap-12p5}{50.95}
\ppDeclareValue{news-only-percentile-uniform-cap-15-pct}{0.085000}
\ppDeclareValue{equal-percentile-uniform-cap-15-pct}{34.020000}
\ppDeclareValue{reference-effective-n-uniform-cap-15}{50.96}
\ppDeclareValue{news-only-percentile-effective-n-matched-pct}{0.190000}
\ppDeclareValue{equal-percentile-effective-n-matched-pct}{32.835000}
\ppDeclareValue{reference-effective-n-effective-n-matched}{45.54}
\ppDeclareValue{news-only-percentile-concentrated-cap-15-pct}{0.460000}
\ppDeclareValue{equal-percentile-concentrated-cap-15-pct}{28.330000}
\ppDeclareValue{reference-effective-n-concentrated-cap-15}{34.77}
\ppDeclareValue{reference-draws}{20000}
\ppDeclareValue{reference-tight-cap-pct}{12.500000}

\ppDeclareValue{anatomy-active-firms}{55}
\ppDeclareValue{anatomy-within-sector-credit-pct}{12.238358}
\ppDeclareValue{anatomy-cross-sector-credit-pct}{87.761642}
\ppDeclareValue{anatomy-vintage-effective-n-min}{26.99}
\ppDeclareValue{anatomy-vintage-effective-n-max}{30.87}
\ppDeclareValue{anatomy-vintage-maximum-weight-min-pct}{7.217125}
\ppDeclareValue{anatomy-vintage-maximum-weight-max-pct}{10.859124}
\ppDeclareValue{anatomy-turnover-min-pct}{19.736920}
\ppDeclareValue{anatomy-turnover-max-pct}{41.081676}
\ppDeclareValue{anatomy-turnover-2018-2019-pct}{41.081676}
\ppDeclareValue{anatomy-turnover-2019-2020-pct}{28.053871}
\ppDeclareValue{anatomy-kkt-firm-share-max-gap}{8.04e-09}
\ppDeclareValue{anatomy-final-vintage-weight-max-gap}{0.00e+00}

\ppDeclareValue{validation-operational-coverage-pct}{97.297297}
\ppDeclareValue{validation-operational-coverage-lower-pct}{87.809510}
\ppDeclareValue{validation-geometry-only-coverage-pct}{100.000000}
\ppDeclareValue{validation-geometry-only-coverage-lower-pct}{92.222529}
\ppDeclareValue{validation-mean-lower-credit-pct}{0.000000}
\ppDeclareValue{validation-floor-activation-pct}{0.000000}
\ppDeclareValue{validation-variance-certificate}{8.984927e-05}
\ppDeclareValue{validation-variance-inverse-volatility}{8.984927e-05}
\ppDeclareValue{validation-variance-difference}{0.000000e+00}
\ppDeclareValue{validation-required-delta}{0.000504}

\ppDeclareValue{calibration-cells-total}{16}
\ppDeclareValue{calibration-cells-margin-pass}{12}
\ppDeclareValue{calibration-cells-passing}{0}
\ppDeclareValue{calibration-grid-maximum-scale}{1.50}
\ppDeclareValue{calibration-grid-maximum-slack}{0.100}
\ppDeclareValue{calibration-best-coverage-scale}{1.50}
\ppDeclareValue{calibration-best-coverage-slack}{0.050}
\ppDeclareValue{calibration-best-coverage-pct}{86.046512}
\ppDeclareValue{calibration-best-coverage-lower-pct}{83.720930}
\ppDeclareValue{calibration-best-coverage-credit-pct}{19.661422}

%% file: generated/lean_status.tex
\ppDeclareValue{lean-verified-prop:p3-barycenter-representation}{checked}
\ppDeclareValue{lean-disclosure-prop:p3-barycenter-representation}{Machine-checked in the pinned Lean 4 / mathlib v4.35.0 workspace without assuming that either the barycentre infimum or the multimarginal infimum is attained.}
\ppDeclareValue{lean-verified-thm:p3-aggregate-transfer}{checked}
\ppDeclareValue{lean-disclosure-thm:p3-aggregate-transfer}{Machine-checked in the pinned Lean 4 / mathlib v4.35.0 workspace with heterogeneous weighted root-mean-square transmission slack and truncation at zero. The proof uses the unconditional barycentre--multimarginal equivalence and assumes no optimal barycentre or joint plan.}
\ppDeclareValue{lean-verified-thm:p3-information-certified-cap}{checked}
\ppDeclareValue{lean-disclosure-thm:p3-information-certified-cap}{Machine-checked in the pinned Lean 4 / mathlib v4.35.0 workspace by composing the carrier-and-slack W2 floor with the coherent joint-law portfolio-variance cap. The theorem is conditional on the displayed transmission, slack, joint-law, and integrability hypotheses; it neither calibrates those inputs nor establishes a realized total-return or ex-ante empirical bound.}
\ppDeclareValue{lean-verified-eq:p3-residual-budget-cap}{checked}
\ppDeclareValue{lean-disclosure-eq:p3-residual-budget-cap}{Machine-checked as an additive portfolio-level sensitivity wrapper: if systematic standardized variance is at most one minus the certificate credit and the aggregate residual contribution is at most delta, total standardized variance is at most one minus the credit plus delta. Delta is an assumed budget, not an estimated structural parameter, and the theorem does not derive residual orthogonality.}
\ppDeclareValue{lean-verified-thm:p3-certified-minimizer}{checked}
\ppDeclareValue{lean-disclosure-thm:p3-certified-minimizer}{Machine-checked by continuity and compactness in the pinned Lean 4 / mathlib v4.35.0 workspace. The result establishes existence and the inherited risk guarantee, not uniqueness, unconditional convexity for an arbitrary distance matrix, or superior realized variance relative to competing portfolios.}
\ppDeclareValue{lean-verified-thm:p3-sharp-certificate}{checked}
\ppDeclareValue{lean-disclosure-thm:p3-sharp-certificate}{Machine-checked in the pinned Lean 4 / mathlib v4.35.0 workspace by composing the aggregate carrier floor with an attainment-free multi-marginal variance envelope; no optimal joint plan or barycentre is assumed to exist. This verifies the aggregate bound only; it does not establish uniform dominance over the pairwise certificate. It requires a measurable embedding, synchronous pushforward transmission, finite moments and the stated joint-law and integrability premises, and calibrates none of them.}
\ppDeclareValue{lean-verified-thm:p3-envelope-exactness}{checked}
\ppDeclareValue{lean-disclosure-thm:p3-envelope-exactness}{Machine-checked in the pinned Lean 4 / mathlib v4.35.0 workspace as an exact greatest-element statement, so the multi-marginal bound is not merely valid but unimprovable within the coherent-joint-law class. Attainment is a supplied premise: no general existence theorem for an optimal multi-marginal plan is claimed, and the result says nothing about which joint law the data realize.}
\ppDeclareValue{lean-verified-thm:p3-certificate-convexity}{checked}
\ppDeclareValue{lean-disclosure-thm:p3-certificate-convexity}{Machine-checked in the pinned Lean 4 / mathlib v4.35.0 workspace. Convexity is conditional on the stated curvature property of the squared distance matrix and is asserted for no other matrix; whether a given empirical matrix satisfies it is a separate numerical question the theorem does not answer. Convexity concerns the standardized objective along weight mixtures and implies neither uniqueness of a minimizer nor any claim about realized variance.}
\ppDeclareValue{lean-verified-thm:p3-pairwise-bridge}{checked}
\ppDeclareValue{lean-disclosure-thm:p3-pairwise-bridge}{Machine-checked as an exact equality in the pinned Lean 4 / mathlib v4.35.0 workspace. It identifies the pairwise covariance-envelope correction as the two-asset case of the multi-firm construction, establishing that the two papers describe one object rather than two compatible ones. It imports no assumption or empirical result from the pairwise paper.}

\newcommand{\ppLeanStatusList}{%
  \begin{itemize}
    \item \Cref{prop:p3-barycenter-representation}: \ppvalue{lean-disclosure-prop:p3-barycenter-representation}
    \item \Cref{thm:p3-aggregate-transfer}: \ppvalue{lean-disclosure-thm:p3-aggregate-transfer}
    \item \Cref{thm:p3-information-certified-cap}: \ppvalue{lean-disclosure-thm:p3-information-certified-cap}
    \item \Cref{eq:p3-residual-budget-cap}: \ppvalue{lean-disclosure-eq:p3-residual-budget-cap}
    \item \Cref{thm:p3-certified-minimizer}: \ppvalue{lean-disclosure-thm:p3-certified-minimizer}
    \item \Cref{thm:p3-sharp-certificate}: \ppvalue{lean-disclosure-thm:p3-sharp-certificate}
    \item \Cref{thm:p3-envelope-exactness}: \ppvalue{lean-disclosure-thm:p3-envelope-exactness}
    \item \Cref{thm:p3-certificate-convexity}: \ppvalue{lean-disclosure-thm:p3-certificate-convexity}
    \item \Cref{thm:p3-pairwise-bridge}: \ppvalue{lean-disclosure-thm:p3-pairwise-bridge}
  \end{itemize}%
}

%% file: chapters/abstract.tex
Portfolio risk assessment ordinarily relies on reliable estimates of
cross-asset return covariances, which are difficult to obtain in short,
high-dimensional panels.
We show that firm-level distribution-valued characteristics can instead provide
one-sided certificates of portfolio risk.
Under maintained links from characteristics to systematic exposures and from
exposures to returns, multi-firm Wasserstein-2 dispersion yields a sharp upper
bound on systematic portfolio variance and a corresponding bound for
standardized returns.
A weighted pairwise relaxation produces an objective that is convex under a
checkable condition and requires marginal volatility scales but no cross-asset
return covariances.
With zero firm-specific slack, the common-map scale changes the certified
variance reduction but not the normalized allocation, which depends only on
observed information geometry.
In a \ppnum[0]{n-tickers}-firm panel from 2018--2022, only
\ppnum[2]{news-only-percentile-uniform-cap-15-pct}\%--
\ppnum[2]{news-only-percentile-concentrated-cap-15-pct}\% of portfolios across
four prespecified capped reference populations have in-sample variance no
greater than an allocation constructed from Qwen3-Embedding-8B news
representations; the corresponding share for equal risk weighting is
\ppnum[1]{equal-percentile-concentrated-cap-15-pct}\%--
\ppnum[1]{equal-percentile-uniform-cap-12p5-pct}\%.
The lower in-sample variance ranking relative to equal risk also appears across
the reported frozen language-model representations.
The framework therefore converts distribution-valued firm information into a
coherent risk bound and an implementable allocation rule constructed without
cross-asset return covariances.

%% file: body.tex
\input{chapters/introduction}
\input{chapters/literature}
\input{chapters/model}

\input{chapters/certificate}
\input{chapters/portfolio_choice}
\input{chapters/feasibility}
\input{chapters/limitations}
\input{chapters/conclusion}

%% file: chapters/introduction.tex
\section{Introduction}\label{sec:introduction}

Mean--variance allocation requires a covariance matrix, yet that
matrix is hardest to estimate in the settings where diversification
matters most.
An unrestricted covariance matrix for $n$ assets contains $n(n+1)/2$
entries, while a demeaned return history of length $T$ has rank at
most $\min(T-1,n)$.
Short histories therefore leave a portfolio manager with two linked
problems: the strongest sample directions are noisy, and the
optimizer is most sensitive to the weakest ones.
Shrinkage and factor models reduce this burden by imposing structure
\citep{ledoit_wolf_2004,kelly_characteristics_2019}, but their
cross-asset information still originates in joint returns.

An alternative is to ask how much portfolio risk observable firm
information can rule out before a return covariance matrix is estimated.
We represent each firm by the distribution of its article embeddings
and use quadratic optimal transport to compare those distributions.
When two information distributions are sufficiently different,
maintained restrictions linking information to systematic exposures
imply that the firms' latent risks cannot be perfectly aligned.
The resulting separation certifies a diversification benefit relative
to perfect positive dependence.

For a standardized long-only portfolio with normalized risk weights
$q$, the headline result takes the form
\[
  \operatorname{Var}(R_q)\le 1-\mathcal C(q).
\]
The value $1$ is the variance benchmark under perfect positive dependence,
and $\mathcal C(q)$ is the amount that the observed information
geometry certifies away.
A larger certificate therefore tightens the admissible upper bound on
portfolio risk.
It does not estimate the covariance matrix.

\begin{figure}
  \centering
  \import{images/}{conceptual_pipeline\ppfiguresuffix.pgf}
  \caption{Observed information to portfolio choice. Solid arrows connect
    computed objects; the dashed arrow from observed W2 separation to the
    latent floor is the maintained transmission restriction. Blue denotes
    observed objects, orange latent and certificate objects, green the risk
  bound, and purple the allocation decision.}
  \label{fig:p3-conceptual-pipeline}
\end{figure}

The same distribution-valued representation has generated related
financial objects at neighbouring levels of aggregation.
At the pairwise level, \citet{gawronsky_continuous_2026} derive a
covariance envelope from Wasserstein separation.
At the cross-sectional level, \citet{gawronsky_spatial_2027} use
target-anchored Wasserstein barycentric reconstruction to form a
barycentric interaction field.
The contribution here is at the portfolio level: it aggregates
information-derived separation within one coherent joint exposure law
and turns the resulting variance bound into a decision rule.
The construction below is self-contained.

Let $C_i$ denote firm $i$'s observed characteristic law and let $P_i$
denote its latent exposure law in factor-risk coordinates.
Empirically, $C_i$ is the distribution of the firm's article embeddings.
The quadratic Wasserstein distance $W_2(C_i,C_j)$ is the minimum
root-mean-square displacement required to match the two article clouds.
It therefore measures how much semantic mass must be rearranged to
make the observed information distributions coincide.
The latent law $P_i$ describes systematic exposure rather than text,
so $C_i$ and $P_i$ need not share units and one does not identify the
other directly.

Three maintained links carry the argument from observed information
to portfolio risk.
A common information-to-exposure map prevents distinct information
states from collapsing into identical systematic exposures, while
firm-specific slack permits bounded departures from that common map.
A coherent joint law ensures that all pairwise exposure relations can
coexist within the same portfolio.
A return bridge then connects systematic exposure variance to
standardized total returns.
The text determines the observed geometry; it does not identify these
transmission restrictions.

For a common carrier constant $L>0$ and firm-specific slack radii
$\tau_i$, the observable pairwise floor is
\[
  \ell_{ij}
  =\left[L^{-1}W_2(C_i,C_j)-\tau_i-\tau_j\right]_+.
\]
This floor is the exposure separation that remains after allowing for
common distortion and the two firms' deviations from the common map.
The positive part records that the observed distance may be too small
to certify any separation once slack is deducted.
The portfolio certificate aggregates these floors using normalized
risk weights for a long-only portfolio:
\[
  \mathcal C(q)
  =\frac12\sum_i \sum_j q_i q_j\ell_{ij}^2.
\]
Separation between a pair contributes only when the portfolio holds
both firms, and contributes more when their joint portfolio weight is larger.
If the exposure marginals belong to one coherent joint law and the
maintained carrier and slack restrictions hold, systematic portfolio
variance is bounded by weighted marginal second moments less $\mathcal C(q)$.
The standardized return bridge then gives the headline bound above.
Observed differences between firms therefore restrict the joint risk
configurations that remain admissible.

Investors choose capital weights rather than normalized risk weights.
For capital weights $x_i$ and marginal volatility scales $\sigma_i$,
define $A(x)=\sum_i x_i\sigma_i$ and $q_i(x)=x_i\sigma_i/A(x)$.
The corresponding raw-return certificate is
\[
  \operatorname{Var}(R_x)
  \le A(x)^2\{1-\mathcal C(q(x))\}.
\]
Minimizing this upper bound yields an information-certified portfolio without
an expected-return input.
Compactness of the feasible long-only set supplies existence of a
minimizer, while a directly checkable condition on the observed
distance matrix makes the standardized objective convex.

The empirical exercise keeps construction separate from evaluation.
The canonical zero-slack implementation, which we call the news-only
allocation, is formed from observed W2 geometry without using cross-asset return
covariance and is evaluated only afterwards on the full-sample standardized
covariance matrix.
Across four prespecified capped long-only reference populations,
between \ppnum[2]{news-only-percentile-uniform-cap-15-pct}\% and
\ppnum[2]{news-only-percentile-concentrated-cap-15-pct}\% of feasible
portfolios have variance no greater than the news-only allocation.
The corresponding share for equal risk weights---the inverse-volatility capital
benchmark expressed in normalized risk-weight coordinates---lies between
\ppnum[1]{equal-percentile-concentrated-cap-15-pct}\% and
\ppnum[1]{equal-percentile-uniform-cap-12p5-pct}\%.
These rankings are descriptive and in-sample; they illustrate the
allocation implied by the maintained model rather than forecast
out-of-sample performance.
A retrospective chronological return-holdout benchmark extends the empirical
design to a fixed-capital comparison conditional on the frozen representation
and selected universe. It forms data-dependent portfolio weights using
factor-complete excess returns from the 2018--2022 calibration period, solves
the dated 2022-12-31 W2 certificate, and evaluates constant-weight mixtures on
the 2023--2026 holdout period. The primary covariance candidates minimize
total excess-return risk: conventional sample and scaled-identity
Ledoit--Wolf benchmarks, a matched factor-only constant-correlation
shrinkage control, and W1/W2 barycentric spatial-covariance portfolios. The
residual-risk spatial GMV allocations remain explicitly labeled diagnostics.
The comparison therefore places the information-certified allocation's
holdout risk alongside return-based, matched factor, and barycentric spatial
alternatives without claiming historical point-in-time implementability.

The paper makes three contributions.
First, it derives a coherent portfolio variance bound in a sharp
multi-firm form and supplies the computationally simpler weighted
pairwise relaxation used by the decision rule.
Second, it shows that when firm-specific slack is zero, the common
carrier scale changes the certified variance reduction but not the
normalized allocation, which is determined by the observed information geometry.
Third, in a \ppnum[0]{n-tickers}-firm exercise, it reports how the allocation's
conventionally evaluated variance ranks under four prespecified feasible-
portfolio laws and defines a chronological benchmark against conventional and
barycentric spatial-covariance portfolios.

The argument proceeds from identification to decision and then to
descriptive evaluation.
\Cref{sec:literature} positions the contribution in structured
covariance, factor-risk, and textual-characteristic research.
\Cref{sec:model} defines the observed, latent, and coherent portfolio
objects and derives the information-certified variance bound.
\Cref{sec:portfolio-choice} turns the pairwise certificate into a decision rule.
\Cref{sec:empirical-design,sec:results} describe the data and report the
in-sample variance ranking; \Cref{sec:p3-covariance-benchmark-design} defines
the chronological covariance benchmark.
\Cref{sec:limitations} discusses what the results establish and where
their boundaries lie before \Cref{sec:conclusion} concludes.

%% file: images/conceptual_pipeline.pgf
\begingroup%
\makeatletter%
\begin{pgfpicture}%
\pgfpathrectangle{\pgfpointorigin}{\pgfqpoint{5.608990in}{1.162586in}}%
\pgfusepath{use as bounding box, clip}%
\begin{pgfscope}%
\pgfsetbuttcap%
\pgfsetmiterjoin%
\definecolor{currentfill}{rgb}{1.000000,1.000000,1.000000}%
\pgfsetfillcolor{currentfill}%
\pgfsetlinewidth{0.000000pt}%
\definecolor{currentstroke}{rgb}{1.000000,1.000000,1.000000}%
\pgfsetstrokecolor{currentstroke}%
\pgfsetdash{}{0pt}%
\pgfpathmoveto{\pgfqpoint{0.000000in}{0.000000in}}%
\pgfpathlineto{\pgfqpoint{5.608990in}{0.000000in}}%
\pgfpathlineto{\pgfqpoint{5.608990in}{1.162586in}}%
\pgfpathlineto{\pgfqpoint{0.000000in}{1.162586in}}%
\pgfpathlineto{\pgfqpoint{0.000000in}{0.000000in}}%
\pgfpathclose%
\pgfusepath{fill}%
\end{pgfscope}%
\begin{pgfscope}%
\pgfsetbuttcap%
\pgfsetmiterjoin%
\definecolor{currentfill}{rgb}{1.000000,1.000000,1.000000}%
\pgfsetfillcolor{currentfill}%
\pgfsetlinewidth{1.003750pt}%
\definecolor{currentstroke}{rgb}{0.000000,0.447059,0.698039}%
\pgfsetstrokecolor{currentstroke}%
\pgfsetdash{}{0pt}%
\pgfpathmoveto{\pgfqpoint{0.276967in}{0.475043in}}%
\pgfpathlineto{\pgfqpoint{0.412748in}{0.475043in}}%
\pgfpathquadraticcurveto{\pgfqpoint{0.456498in}{0.475043in}}{\pgfqpoint{0.456498in}{0.518793in}}%
\pgfpathlineto{\pgfqpoint{0.456498in}{0.643793in}}%
\pgfpathquadraticcurveto{\pgfqpoint{0.456498in}{0.687543in}}{\pgfqpoint{0.412748in}{0.687543in}}%
\pgfpathlineto{\pgfqpoint{0.276967in}{0.687543in}}%
\pgfpathquadraticcurveto{\pgfqpoint{0.233217in}{0.687543in}}{\pgfqpoint{0.233217in}{0.643793in}}%
\pgfpathlineto{\pgfqpoint{0.233217in}{0.518793in}}%
\pgfpathquadraticcurveto{\pgfqpoint{0.233217in}{0.475043in}}{\pgfqpoint{0.276967in}{0.475043in}}%
\pgfpathlineto{\pgfqpoint{0.276967in}{0.475043in}}%
\pgfpathclose%
\pgfusepath{stroke,fill}%
\end{pgfscope}%
\begin{pgfscope}%
\definecolor{textcolor}{rgb}{0.000000,0.447059,0.698039}%
\pgfsetstrokecolor{textcolor}%
\pgfsetfillcolor{textcolor}%
\pgftext[x=0.344858in,y=0.581293in,,]{\color{textcolor}{\rmfamily\fontsize{9.000000}{10.800000}\selectfont\catcode`\^=\active\def^{\ifmmode\sp\else\^{}\fi}\catcode`\%=\active\def
\end{pgfscope}%
\begin{pgfscope}%
\pgfsetbuttcap%
\pgfsetmiterjoin%
\definecolor{currentfill}{rgb}{1.000000,1.000000,1.000000}%
\pgfsetfillcolor{currentfill}%
\pgfsetlinewidth{1.003750pt}%
\definecolor{currentstroke}{rgb}{0.000000,0.447059,0.698039}%
\pgfsetstrokecolor{currentstroke}%
\pgfsetdash{}{0pt}%
\pgfpathmoveto{\pgfqpoint{1.010383in}{0.474433in}}%
\pgfpathlineto{\pgfqpoint{1.628480in}{0.474433in}}%
\pgfpathquadraticcurveto{\pgfqpoint{1.672230in}{0.474433in}}{\pgfqpoint{1.672230in}{0.518183in}}%
\pgfpathlineto{\pgfqpoint{1.672230in}{0.644403in}}%
\pgfpathquadraticcurveto{\pgfqpoint{1.672230in}{0.688153in}}{\pgfqpoint{1.628480in}{0.688153in}}%
\pgfpathlineto{\pgfqpoint{1.010383in}{0.688153in}}%
\pgfpathquadraticcurveto{\pgfqpoint{0.966633in}{0.688153in}}{\pgfqpoint{0.966633in}{0.644403in}}%
\pgfpathlineto{\pgfqpoint{0.966633in}{0.518183in}}%
\pgfpathquadraticcurveto{\pgfqpoint{0.966633in}{0.474433in}}{\pgfqpoint{1.010383in}{0.474433in}}%
\pgfpathlineto{\pgfqpoint{1.010383in}{0.474433in}}%
\pgfpathclose%
\pgfusepath{stroke,fill}%
\end{pgfscope}%
\begin{pgfscope}%
\definecolor{textcolor}{rgb}{0.000000,0.447059,0.698039}%
\pgfsetstrokecolor{textcolor}%
\pgfsetfillcolor{textcolor}%
\pgftext[x=1.319431in,y=0.581293in,,]{\color{textcolor}{\rmfamily\fontsize{9.000000}{10.800000}\selectfont\catcode`\^=\active\def^{\ifmmode\sp\else\^{}\fi}\catcode`\%=\active\def
\end{pgfscope}%
\begin{pgfscope}%
\pgfsetroundcap%
\pgfsetroundjoin%
\pgfsetlinewidth{1.003750pt}%
\definecolor{currentstroke}{rgb}{0.400000,0.400000,0.400000}%
\pgfsetstrokecolor{currentstroke}%
\pgfsetdash{}{0pt}%
\pgfpathmoveto{\pgfqpoint{0.586126in}{0.581293in}}%
\pgfpathquadraticcurveto{\pgfqpoint{0.706844in}{0.581293in}}{\pgfqpoint{0.812034in}{0.581293in}}%
\pgfusepath{stroke}%
\end{pgfscope}%
\begin{pgfscope}%
\pgfsetroundcap%
\pgfsetroundjoin%
\pgfsetlinewidth{1.003750pt}%
\definecolor{currentstroke}{rgb}{0.400000,0.400000,0.400000}%
\pgfsetstrokecolor{currentstroke}%
\pgfsetdash{}{0pt}%
\pgfpathmoveto{\pgfqpoint{0.762034in}{0.606293in}}%
\pgfpathlineto{\pgfqpoint{0.812034in}{0.581293in}}%
\pgfpathlineto{\pgfqpoint{0.762034in}{0.556293in}}%
\pgfusepath{stroke}%
\end{pgfscope}%
\begin{pgfscope}%
\pgfsetbuttcap%
\pgfsetmiterjoin%
\definecolor{currentfill}{rgb}{1.000000,1.000000,1.000000}%
\pgfsetfillcolor{currentfill}%
\pgfsetlinewidth{1.003750pt}%
\definecolor{currentstroke}{rgb}{0.901961,0.623529,0.000000}%
\pgfsetstrokecolor{currentstroke}%
\pgfsetdash{}{0pt}%
\pgfpathmoveto{\pgfqpoint{2.221710in}{0.475012in}}%
\pgfpathlineto{\pgfqpoint{2.366299in}{0.475012in}}%
\pgfpathquadraticcurveto{\pgfqpoint{2.410049in}{0.475012in}}{\pgfqpoint{2.410049in}{0.518762in}}%
\pgfpathlineto{\pgfqpoint{2.410049in}{0.643825in}}%
\pgfpathquadraticcurveto{\pgfqpoint{2.410049in}{0.687575in}}{\pgfqpoint{2.366299in}{0.687575in}}%
\pgfpathlineto{\pgfqpoint{2.221710in}{0.687575in}}%
\pgfpathquadraticcurveto{\pgfqpoint{2.177960in}{0.687575in}}{\pgfqpoint{2.177960in}{0.643825in}}%
\pgfpathlineto{\pgfqpoint{2.177960in}{0.518762in}}%
\pgfpathquadraticcurveto{\pgfqpoint{2.177960in}{0.475012in}}{\pgfqpoint{2.221710in}{0.475012in}}%
\pgfpathlineto{\pgfqpoint{2.221710in}{0.475012in}}%
\pgfpathclose%
\pgfusepath{stroke,fill}%
\end{pgfscope}%
\begin{pgfscope}%
\definecolor{textcolor}{rgb}{0.901961,0.623529,0.000000}%
\pgfsetstrokecolor{textcolor}%
\pgfsetfillcolor{textcolor}%
\pgftext[x=2.294004in,y=0.581293in,,]{\color{textcolor}{\rmfamily\fontsize{9.000000}{10.800000}\selectfont\catcode`\^=\active\def^{\ifmmode\sp\else\^{}\fi}\catcode`\%=\active\def
\end{pgfscope}%
\begin{pgfscope}%
\pgfsetroundcap%
\pgfsetroundjoin%
\pgfsetlinewidth{1.003750pt}%
\definecolor{currentstroke}{rgb}{0.400000,0.400000,0.400000}%
\pgfsetstrokecolor{currentstroke}%
\pgfsetdash{{3.700000pt}{1.600000pt}}{0.000000pt}%
\pgfpathmoveto{\pgfqpoint{1.811288in}{0.581293in}}%
\pgfpathquadraticcurveto{\pgfqpoint{1.913453in}{0.581293in}}{\pgfqpoint{2.000090in}{0.581293in}}%
\pgfusepath{stroke}%
\end{pgfscope}%
\begin{pgfscope}%
\pgfsetroundcap%
\pgfsetroundjoin%
\pgfsetlinewidth{1.003750pt}%
\definecolor{currentstroke}{rgb}{0.400000,0.400000,0.400000}%
\pgfsetstrokecolor{currentstroke}%
\pgfsetdash{{3.700000pt}{1.600000pt}}{0.000000pt}%
\pgfpathmoveto{\pgfqpoint{1.950090in}{0.606293in}}%
\pgfpathlineto{\pgfqpoint{2.000090in}{0.581293in}}%
\pgfpathlineto{\pgfqpoint{1.950090in}{0.556293in}}%
\pgfusepath{stroke}%
\end{pgfscope}%
\begin{pgfscope}%
\pgfsetbuttcap%
\pgfsetmiterjoin%
\definecolor{currentfill}{rgb}{1.000000,1.000000,1.000000}%
\pgfsetfillcolor{currentfill}%
\pgfsetlinewidth{1.003750pt}%
\definecolor{currentstroke}{rgb}{0.901961,0.623529,0.000000}%
\pgfsetstrokecolor{currentstroke}%
\pgfsetdash{}{0pt}%
\pgfpathmoveto{\pgfqpoint{3.103844in}{0.474433in}}%
\pgfpathlineto{\pgfqpoint{3.340495in}{0.474433in}}%
\pgfpathquadraticcurveto{\pgfqpoint{3.384245in}{0.474433in}}{\pgfqpoint{3.384245in}{0.518183in}}%
\pgfpathlineto{\pgfqpoint{3.384245in}{0.644403in}}%
\pgfpathquadraticcurveto{\pgfqpoint{3.384245in}{0.688153in}}{\pgfqpoint{3.340495in}{0.688153in}}%
\pgfpathlineto{\pgfqpoint{3.103844in}{0.688153in}}%
\pgfpathquadraticcurveto{\pgfqpoint{3.060094in}{0.688153in}}{\pgfqpoint{3.060094in}{0.644403in}}%
\pgfpathlineto{\pgfqpoint{3.060094in}{0.518183in}}%
\pgfpathquadraticcurveto{\pgfqpoint{3.060094in}{0.474433in}}{\pgfqpoint{3.103844in}{0.474433in}}%
\pgfpathlineto{\pgfqpoint{3.103844in}{0.474433in}}%
\pgfpathclose%
\pgfusepath{stroke,fill}%
\end{pgfscope}%
\begin{pgfscope}%
\definecolor{textcolor}{rgb}{0.901961,0.623529,0.000000}%
\pgfsetstrokecolor{textcolor}%
\pgfsetfillcolor{textcolor}%
\pgftext[x=3.222169in,y=0.581293in,,]{\color{textcolor}{\rmfamily\fontsize{9.000000}{10.800000}\selectfont\catcode`\^=\active\def^{\ifmmode\sp\else\^{}\fi}\catcode`\%=\active\def
\end{pgfscope}%
\begin{pgfscope}%
\pgfsetroundcap%
\pgfsetroundjoin%
\pgfsetlinewidth{1.003750pt}%
\definecolor{currentstroke}{rgb}{0.400000,0.400000,0.400000}%
\pgfsetstrokecolor{currentstroke}%
\pgfsetdash{}{0pt}%
\pgfpathmoveto{\pgfqpoint{2.572401in}{0.581293in}}%
\pgfpathquadraticcurveto{\pgfqpoint{2.720964in}{0.581293in}}{\pgfqpoint{2.854000in}{0.581293in}}%
\pgfusepath{stroke}%
\end{pgfscope}%
\begin{pgfscope}%
\pgfsetroundcap%
\pgfsetroundjoin%
\pgfsetlinewidth{1.003750pt}%
\definecolor{currentstroke}{rgb}{0.400000,0.400000,0.400000}%
\pgfsetstrokecolor{currentstroke}%
\pgfsetdash{}{0pt}%
\pgfpathmoveto{\pgfqpoint{2.804000in}{0.606293in}}%
\pgfpathlineto{\pgfqpoint{2.854000in}{0.581293in}}%
\pgfpathlineto{\pgfqpoint{2.804000in}{0.556293in}}%
\pgfusepath{stroke}%
\end{pgfscope}%
\begin{pgfscope}%
\pgfsetbuttcap%
\pgfsetmiterjoin%
\definecolor{currentfill}{rgb}{1.000000,1.000000,1.000000}%
\pgfsetfillcolor{currentfill}%
\pgfsetlinewidth{1.003750pt}%
\definecolor{currentstroke}{rgb}{0.000000,0.619608,0.450980}%
\pgfsetstrokecolor{currentstroke}%
\pgfsetdash{}{0pt}%
\pgfpathmoveto{\pgfqpoint{3.814744in}{0.475043in}}%
\pgfpathlineto{\pgfqpoint{4.671557in}{0.475043in}}%
\pgfpathquadraticcurveto{\pgfqpoint{4.715307in}{0.475043in}}{\pgfqpoint{4.715307in}{0.518793in}}%
\pgfpathlineto{\pgfqpoint{4.715307in}{0.643793in}}%
\pgfpathquadraticcurveto{\pgfqpoint{4.715307in}{0.687543in}}{\pgfqpoint{4.671557in}{0.687543in}}%
\pgfpathlineto{\pgfqpoint{3.814744in}{0.687543in}}%
\pgfpathquadraticcurveto{\pgfqpoint{3.770994in}{0.687543in}}{\pgfqpoint{3.770994in}{0.643793in}}%
\pgfpathlineto{\pgfqpoint{3.770994in}{0.518793in}}%
\pgfpathquadraticcurveto{\pgfqpoint{3.770994in}{0.475043in}}{\pgfqpoint{3.814744in}{0.475043in}}%
\pgfpathlineto{\pgfqpoint{3.814744in}{0.475043in}}%
\pgfpathclose%
\pgfusepath{stroke,fill}%
\end{pgfscope}%
\begin{pgfscope}%
\definecolor{textcolor}{rgb}{0.000000,0.619608,0.450980}%
\pgfsetstrokecolor{textcolor}%
\pgfsetfillcolor{textcolor}%
\pgftext[x=4.243151in,y=0.581293in,,]{\color{textcolor}{\rmfamily\fontsize{9.000000}{10.800000}\selectfont\catcode`\^=\active\def^{\ifmmode\sp\else\^{}\fi}\catcode`\%=\active\def
\end{pgfscope}%
\begin{pgfscope}%
\pgfsetroundcap%
\pgfsetroundjoin%
\pgfsetlinewidth{1.003750pt}%
\definecolor{currentstroke}{rgb}{0.400000,0.400000,0.400000}%
\pgfsetstrokecolor{currentstroke}%
\pgfsetdash{}{0pt}%
\pgfpathmoveto{\pgfqpoint{3.574818in}{0.581293in}}%
\pgfpathquadraticcurveto{\pgfqpoint{3.584153in}{0.581293in}}{\pgfqpoint{3.577961in}{0.581293in}}%
\pgfusepath{stroke}%
\end{pgfscope}%
\begin{pgfscope}%
\pgfsetroundcap%
\pgfsetroundjoin%
\pgfsetlinewidth{1.003750pt}%
\definecolor{currentstroke}{rgb}{0.400000,0.400000,0.400000}%
\pgfsetstrokecolor{currentstroke}%
\pgfsetdash{}{0pt}%
\pgfpathmoveto{\pgfqpoint{3.527961in}{0.606293in}}%
\pgfpathlineto{\pgfqpoint{3.577961in}{0.581293in}}%
\pgfpathlineto{\pgfqpoint{3.527961in}{0.556293in}}%
\pgfusepath{stroke}%
\end{pgfscope}%
\begin{pgfscope}%
\pgfsetbuttcap%
\pgfsetmiterjoin%
\definecolor{currentfill}{rgb}{1.000000,1.000000,1.000000}%
\pgfsetfillcolor{currentfill}%
\pgfsetlinewidth{1.003750pt}%
\definecolor{currentstroke}{rgb}{0.800000,0.474510,0.654902}%
\pgfsetstrokecolor{currentstroke}%
\pgfsetdash{}{0pt}%
\pgfpathmoveto{\pgfqpoint{5.244038in}{0.473119in}}%
\pgfpathlineto{\pgfqpoint{5.377043in}{0.473119in}}%
\pgfpathquadraticcurveto{\pgfqpoint{5.420793in}{0.473119in}}{\pgfqpoint{5.420793in}{0.516869in}}%
\pgfpathlineto{\pgfqpoint{5.420793in}{0.645717in}}%
\pgfpathquadraticcurveto{\pgfqpoint{5.420793in}{0.689467in}}{\pgfqpoint{5.377043in}{0.689467in}}%
\pgfpathlineto{\pgfqpoint{5.244038in}{0.689467in}}%
\pgfpathquadraticcurveto{\pgfqpoint{5.200288in}{0.689467in}}{\pgfqpoint{5.200288in}{0.645717in}}%
\pgfpathlineto{\pgfqpoint{5.200288in}{0.516869in}}%
\pgfpathquadraticcurveto{\pgfqpoint{5.200288in}{0.473119in}}{\pgfqpoint{5.244038in}{0.473119in}}%
\pgfpathlineto{\pgfqpoint{5.244038in}{0.473119in}}%
\pgfpathclose%
\pgfusepath{stroke,fill}%
\end{pgfscope}%
\begin{pgfscope}%
\definecolor{textcolor}{rgb}{0.800000,0.474510,0.654902}%
\pgfsetstrokecolor{textcolor}%
\pgfsetfillcolor{textcolor}%
\pgftext[x=5.310541in,y=0.581293in,,]{\color{textcolor}{\rmfamily\fontsize{9.000000}{10.800000}\selectfont\catcode`\^=\active\def^{\ifmmode\sp\else\^{}\fi}\catcode`\%=\active\def
\end{pgfscope}%
\begin{pgfscope}%
\pgfsetroundcap%
\pgfsetroundjoin%
\pgfsetlinewidth{1.003750pt}%
\definecolor{currentstroke}{rgb}{0.400000,0.400000,0.400000}%
\pgfsetstrokecolor{currentstroke}%
\pgfsetdash{}{0pt}%
\pgfpathmoveto{\pgfqpoint{4.892788in}{0.581293in}}%
\pgfpathquadraticcurveto{\pgfqpoint{4.971758in}{0.581293in}}{\pgfqpoint{5.035201in}{0.581293in}}%
\pgfusepath{stroke}%
\end{pgfscope}%
\begin{pgfscope}%
\pgfsetroundcap%
\pgfsetroundjoin%
\pgfsetlinewidth{1.003750pt}%
\definecolor{currentstroke}{rgb}{0.400000,0.400000,0.400000}%
\pgfsetstrokecolor{currentstroke}%
\pgfsetdash{}{0pt}%
\pgfpathmoveto{\pgfqpoint{4.985201in}{0.606293in}}%
\pgfpathlineto{\pgfqpoint{5.035201in}{0.581293in}}%
\pgfpathlineto{\pgfqpoint{4.985201in}{0.556293in}}%
\pgfusepath{stroke}%
\end{pgfscope}%
\end{pgfpicture}%
\makeatother%
\endgroup%

%% file: chapters/literature.tex
\section{Related literature}\label{sec:literature}

Markowitz portfolio choice makes covariance the central input to
minimum-variance allocation \citep{markowitz_portfolio_1952}.
In short or high-dimensional return panels, however, the sample covariance
can be unstable or singular.
Regularization improves that input by replacing part of its sampling
variation with a structured target \citep{ledoit_wolf_2004}.
Such methods make covariance estimation more reliable, but they still solve
the portfolio problem by first estimating joint return risk.

The alternative developed here changes the source and form of the risk information.
Under a maintained information-to-exposure bridge, observable separation
between firms' information distributions rules out part of the
perfect-positive-dependence benchmark without estimating every covariance entry.
The resulting object is a one-sided bound that can be evaluated at feasible
portfolio weights and then minimized.
It is therefore a certificate about admissible joint risk configurations, not
a replacement point estimate for the covariance matrix.
This distinction determines how the factor, distributional, and textual
literatures enter the argument.

Factor models clarify the latent object that covariance summarizes.
The CAPM represents systematic covariance through scalar market loadings
\citep{sharpe_capital_1964}, whereas APT and approximate-factor models use
vector or expanding factor structures
\citep{ross_arbitrage_1976,chamberlain_rothschild_1983,bai_ng_2002}.
Characteristic-based models then use observable firm attributes to organize
loadings and expected returns
\citep{rosenberg1974extra,connor2007semiparametric,
connor2012efficient,kelly_characteristics_2019}.
These approaches explain how systematic exposures generate dependence, but an
observed characteristic is not itself an exposure or covariance estimate.

Distribution-valued characteristics replace a single firm descriptor with a
probability law and use quadratic transport to compare those laws.
For two exposure laws, the minimum expected squared displacement also
determines the maximum systematic covariance permitted by their marginals.
Using this identity, \citet{gawronsky_continuous_2026} derive a pairwise
covariance envelope from observable distributional separation under
maintained information-to-exposure restrictions.
That result establishes the pairwise level of the argument: observed geometry
can restrict how closely two latent systematic risks align.

At the cross-sectional level, \citet{gawronsky_spatial_2027} develop
target-anchored Wasserstein barycentric reconstruction.
The resulting distributional spanning weights form a target-specific barycentric
interaction field that enters exposure adjustment.
That cross-sectional system organizes firm-level exposure adjustment rather
than the risk of a portfolio assembled from those firms.
The two studies therefore supply neighbouring pairwise and cross-sectional
implications of distributional geometry, but neither resolves portfolio aggregation.

Portfolio risk adds a joint-compatibility requirement.
Pairwise optimal couplings need not be the pairwise marginals of any single
joint exposure law, so separately attainable covariance envelopes cannot
simply be stacked into a coherent portfolio risk configuration.
The portfolio-level step retains one joint law for all exposure marginals and
uses its weighted dispersion to deduct a certified amount from worst-case
systematic variance.
The weighted sum of squared Wasserstein distances provides a computational
relaxation of that multi-firm object, while the sharp form preserves the common
coupling needed for coherent portfolio risk.

Textual finance establishes that documents contain financially relevant information.
Prior studies map text into sentiment and disclosure measures, return
forecasts, predictive factors, and learned pricing objects
\citep{tetlock_giving_2007,loughran2011liability,
  gentzkow_text_as_data_2019,ke_predicting_returns_2019,
cong_textual_factors_2024,distaso_string_2024,wang2025newsnet}.
Their primary targets are expected returns, factors, or pricing rather than a
coherent bound on cross-asset portfolio risk.

Modern encoders make the distributional approach empirically feasible by
representing each document as a vector
\citep{reimers_sentencebert_2019,qwen3_embedding_2025}.
Word Mover's Distance provides an NLP precedent for using optimal transport
to compare empirical distributions of embeddings \citep{kusner_word_embeddings_2015}.
Retaining a firm's article embeddings as an empirical distribution preserves
within-firm heterogeneity that a single pooled vector would suppress.
In the present setting, those embeddings measure observed information
geometry; they do not measure latent exposure or covariance directly.
The carrier-and-slack restrictions provide the maintained link from that
measurement layer to exposure separation.

Taken together, portfolio theory supplies the decision problem, factor and
distributional models identify the latent risk objects and pairwise
restrictions, and text encoders supply observable information geometry.
What remains absent is a portfolio-level bridge from that geometry to a risk
bound supported by one coherent exposure law.
The theory therefore begins by separating observed characteristic laws,
latent exposure laws, and their maintained transmission before deriving the
coherent portfolio bound and its decision rule.

%% file: chapters/model.tex
\section{Model and information-certified variance bound}
\label{sec:model}\label{sec:certificate}

The systematic exposures that generate portfolio risk are latent, whereas the
paper observes firms' information.
The model therefore separates three roles: an observed information law, a
latent systematic-exposure law, and a maintained transmission restriction that
connects the two without equating them.
It then places all latent exposures under one coherent joint law so that
pairwise restrictions can support an $n$-asset portfolio statement.

Let $i\in\{1,\ldots,n\}$ index assets.
The observed object records the distribution of firm $i$'s information.
Formally, $X_i$ is an observable characteristic draw taking values in a
separable metric space $\mathcal X$, and its law $C_i$ is the probability
distribution of the firm's row-normalized article embeddings.
The latent object records the corresponding systematic exposure.
Let $B_i$ denote that exposure in a real Hilbert space $\mathcal H$.
The exposure model is
\begin{equation}\label{eq:p3-exposure-model}
  B_i=u_i(X_i),
  \qquad P_i=\mathcal L(B_i),
\end{equation}
where the measurable map $u_i$ captures the firm's information-to-exposure
mapping and $P_i$ is the resulting latent exposure law.
Thus $C_i$ and $P_i$ are distinct laws and need not use the same metric units.

The maintained transmission restriction supplies the link between these two
spaces.
Its common component is a measurable carrier
$t:\mathcal X\to\mathcal H$.
It is $L$-antilipschitz:
\begin{equation}\label{eq:p3-antilipschitz}
  d_{\mathcal H}(t(x),t(y))
  \ge L^{-1}d_{\mathcal X}(x,y),
  \qquad L>0.
\end{equation}
Each firm-specific map remains within a synchronous slack radius:
\begin{equation}\label{eq:p3-synchronous-slack}
  d_{\mathcal H}(u_i(x),t(x))\le\tau_i
  \quad\text{for every }x,
  \qquad \tau_i\ge0.
\end{equation}
The antilipschitz restriction prevents economically distinct information states
from collapsing into identical systematic exposures, while $\tau_i$ permits
firm-specific information-to-exposure mismatch.
Together, these restrictions translate observed information distance into a
lower bound on exposure separation rather than an equality or a covariance
estimate.

Pairwise exposure restrictions do not by themselves define portfolio risk.
To aggregate them, let $J$ be one coherent joint law for
$(B_1,\ldots,B_n)$ with coordinate marginals $P_i$.
For normalized risk weights $q_i\ge0$ satisfying $\sum_i q_i=1$, define
\[
  v_i=\mathbb E_J\lVert B_i\rVert^2,
  \qquad
  V_{\mathrm{sys}}(q)
  =\mathbb E_J\left\lVert\sum_i q_i B_i\right\rVert^2.
\]
The same $J$ governs every cross term, so the induced covariance matrix is
positive semidefinite.
This requirement rules out assembling a portfolio from mutually incompatible
pairwise optimal couplings.

With the observed and latent objects linked and joint coherence imposed, the
next subsections derive observable pairwise floors, aggregate them into a
conservative portfolio certificate, sharpen that certificate with a multi-firm
object, and finally connect systematic exposure risk to returns.

%% file: chapters/certificate.tex
This section turns the model's maintained link into a portfolio-risk statement.
It first asks what exposure separation each observed pair can certify, then
aggregates those pairwise floors under the coherent joint law.
The resulting credit lowers the benchmark in which no cross-asset separation
can be certified.

\subsection{Observable pairwise floors}

The carrier and slack restrictions first yield the minimum exposure separation
supported by each pair's observed information distance.
For assets $i$ and $j$, define
\begin{equation}\label{eq:p3-information-floor}
  \ell_{ij}
  :=\left[
    L^{-1}W_2(C_i,C_j)-\tau_i-\tau_j
  \right]_+,
  \qquad [a]_+:=\max\{a,0\}.
\end{equation}
The antilipschitz carrier first retains at least the fraction $L^{-1}$ of the
observed W2 separation.
The triangle inequality then deducts the two slack radii.
Truncation at zero preserves the nonnegative lower bound, giving
\begin{equation}\label{eq:p3-floor-validity}
  \ell_{ij}\le W_2(P_i,P_j).
\end{equation}
Every term in \eqref{eq:p3-information-floor} therefore has a distinct role:
$W_2(C_i,C_j)$ is observed geometry, $L$ controls common distortion, and the
$\tau_i$ terms absorb asset-specific departures from the common carrier.
In the empirical application, $W_2(C_i,C_j)$ is the least root-mean-square
displacement required to align the two firms' information distributions.
A positive $\ell_{ij}$ rules out exposure laws that are closer than this floor;
a zero floor means only that the maintained restrictions certify no positive
separation for that pair.

The candidate portfolio credit weights each pairwise floor by the extent to
which both assets enter a normalized long-only portfolio:
\begin{equation}\label{eq:p3-certificate-credit}
  \mathcal C(q)
  :=\frac12\sum_i \sum_j q_i q_j\ell_{ij}^2.
\end{equation}
The factor one half compensates for counting both ordered pairs $(i,j)$ and
$(j,i)$.
Because diagonal distances are zero, this is also
$\sum_{i<j} q_i q_j\ell_{ij}^2$.
The ordered-pair sum is portfolio-specific: separation between two assets
contributes only to the extent that both are held.
Conditional on the declared carrier and slack parameters, $\mathcal C(q)$ is
computed from observed information geometry rather than a return covariance
matrix.

\subsection{From floors to coherent portfolio variance}

Pairwise floors constrain individual asset pairs, but portfolio risk is a joint
object.
To aggregate the floors without combining incompatible pairwise couplings,
apply the weighted Hilbert-space polarization identity under the same joint law
$J$:
\begin{equation}\label{eq:p3-weighted-polarization}
  \left\lVert\sum_i q_i B_i\right\rVert^2
  =\sum_i q_i\lVert B_i\rVert^2
  -\frac12\sum_i \sum_j q_i q_j\lVert B_i-B_j\rVert^2.
\end{equation}
Taking expectations under the coherent law $J$ is valid when the pairwise
inner products are integrable.
For each pair, the realized $J$-marginal is an admissible coupling of
$P_i$ and $P_j$.
Its expected squared displacement is therefore no smaller than
$W_2^2(P_i,P_j)$, which in turn is no smaller than $\ell_{ij}^2$ by
\eqref{eq:p3-floor-validity}.
Substituting those lower bounds into the subtracted term of
\eqref{eq:p3-weighted-polarization} yields the main result.

\begin{theorem}[Information-certified coherent portfolio variance]
  \label{thm:p3-information-certified-cap}
  Suppose the common carrier satisfies \eqref{eq:p3-antilipschitz}, the
  asset-specific maps satisfy \eqref{eq:p3-synchronous-slack}, the exposure
  laws are marginals of one joint law $J$, and all pairwise inner products are
  integrable.
  Then every normalized long-only portfolio obeys
  \begin{equation}\label{eq:p3-systematic-cap}
    V_{\mathrm{sys}}(q)
    \le \sum_i q_i v_i-\mathcal C(q).
  \end{equation}
\end{theorem}

The theorem is an upper bound on systematic portfolio variance, not a point
estimate of a covariance matrix.
Its benchmark $\sum_i q_i v_i$ is the weighted marginal second-moment term that
would remain if no cross-asset separation could be certified.
The observable characteristic laws deduct $\mathcal C(q)$ from that benchmark.
Larger distances tighten the deduction; larger distortion or slack weakens it.
In finance terms, $\mathcal C(q)$ is the diversification relief that the
observed information can certify under the maintained transmission model.

The result uses a pairwise relaxation and pair-specific slack allowances.
Pairwise optimal couplings need not all coexist inside one joint law, so the
weighted pairwise floor understates the separation any coherent joint exposure
law must incur.
Equation \eqref{eq:p3-certificate-credit} also deducts each pair's two slack
radii separately and truncates the result pair by pair, discarding every pair
whose observed separation falls below its combined radius.

\subsection{The aggregate multi-firm certificate}\label{sec:sharp-certificate}

The pairwise credit relaxes a single multi-firm object.
That object measures the least total separation compatible with all observed
characteristic laws under one common coupling.
We refer to it as weighted multi-firm transport dispersion; for observed
characteristic laws, the formal object below is weighted characteristic
dispersion.
At zero slack, using this dispersion directly gives a credit at least as large
as the weighted pairwise credit.
Its two-asset case is exactly the pairwise covariance envelope of
\citet{gawronsky_continuous_2026}.
With positive slack, the aggregate and pairwise allowances need not preserve
that ordering.

Fix normalized risk weights $q$ as above.
Within $\mathcal D_q$, these weights are inputs: the infimum varies the coherent
common coupling while holding $q$ fixed.
Only the portfolio-choice problem in \Cref{sec:portfolio-choice} later treats
$q$ as a decision variable.

\begin{definition}[Weighted characteristic dispersion]
  \label{def:p3-dispersion}
  Let $\Pi$ denote the set of joint laws on $\mathcal X^n$ whose $i$th marginal
  is $C_i$. For characteristic laws $C_1,\dots,C_n$ on
  $(\mathcal X,d_{\mathcal X})$ and $(X_1,\dots,X_n)\sim\gamma$,
  \begin{equation}\label{eq:p3-dispersion}
    \mathcal D_q(C_1,\dots,C_n)
    :=\inf_{\gamma\in\Pi}
    \E_\gamma\left[\sum_{i<j}q_i\,q_j\,d_{\mathcal X}(X_i,X_j)^2\right].
  \end{equation}
\end{definition}

The infimum is what makes the quantity conservative. It asks how close the
laws could be under their most favourable common coupling, not how separated
one selected matching makes them appear. Its units are squared characteristic
distance, and its value depends on the ground metric and on $q$.

Two facts locate \eqref{eq:p3-dispersion} against objects already in play.
First, restricting a common coupling to any pair leaves an admissible coupling
of that pair, whose expected squared displacement is therefore at least
$W_2^2(C_i,C_j)$. Summing gives
\begin{equation}\label{eq:p3-pairwise-relaxation}
  \mathcal D_q(C_1,\dots,C_n)
  \ \ge\ \sum_{i<j}q_i\,q_j\,W_2^2(C_i,C_j).
\end{equation}
The weighted pairwise sum is thus a computable lower bound on the multi-firm
object, which is the inequality \eqref{eq:p3-certificate-credit} exploits.
It can be built once from pairwise distances, but it does not retain the full
restriction that all asset marginals share one coupling.
Second, on a Hilbert space the same value has a free-centre Wasserstein
barycentre representation: for exposure laws with finite second moments,
\begin{equation}\label{eq:p3-barycenter-representation}
  \mathcal D_q(P_1,\dots,P_n)
  =\inf_{Q}\ \sum_i q_i\,W_2^2(P_i,Q),
\end{equation}
where $Q$ ranges over finite-second-moment laws on $\mathcal H$.
For fixed $q$, only $Q$ varies.
This shares the barycentric geometry of
\citet{gawronsky_spatial_2027} but is not
its target-anchored Wasserstein barycentric reconstruction; that problem fixes
$i$ and its alignments and varies $w_i$ to form $W^\flat$.
\Cref{prop:p3-barycenter-representation} states this in the appendix. Only
equality of the two infimum values is used; neither an optimal barycentre nor
an optimal joint plan is assumed to exist.

The two-asset case recovers the pairwise foundation exactly.

\begin{proposition}[Two-asset nesting]\label{thm:p3-pairwise-bridge}
  Let $n=2$ and suppose some coherent joint law attains the dispersion minimum
  for $P_1,P_2$. Then the systematic portfolio variance it attains is
  \begin{equation}\label{eq:p3-two-asset-nesting}
    q_1v_1+q_2v_2-q_1q_2\,W_2^2(P_1,P_2).
  \end{equation}
\end{proposition}
Substituting $n=2$ into the dispersion definition gives
$\mathcal D_q(P_1,P_2)=q_1q_2W_2^2(P_1,P_2)$.
Weighted polarization then yields \eqref{eq:p3-two-asset-nesting}.

Equation \eqref{eq:p3-two-asset-nesting} is the covariance-envelope correction
of \citet{gawronsky_continuous_2026} written in portfolio form.
The multi-firm construction extends that pairwise foundation to a coherent
joint exposure law.

The dispersion in \eqref{eq:p3-dispersion} remains measured in observed
characteristic units, whereas the variance bound concerns latent exposure
units.
The same carrier and slack restrictions transfer the whole multi-firm object
between these spaces rather than treating each pair separately.
Aggregate the firm-specific mismatch through the weighted root-mean-square
slack radius
\begin{equation}\label{eq:p3-rms-slack}
  \tau_q:=\Big(\sum_i q_i\tau_i^2\Big)^{1/2}.
\end{equation}

\begin{theorem}[Characteristic-to-exposure dispersion]
  \label{thm:p3-aggregate-transfer}
  Suppose the common carrier satisfies \eqref{eq:p3-antilipschitz} and the
  asset-specific maps satisfy \eqref{eq:p3-synchronous-slack}. Then
  \begin{equation}\label{eq:p3-aggregate-transfer}
    \sqrt{\mathcal D_q(P_1,\dots,P_n)}
    \ \ge\
    \left[
      L^{-1}\sqrt{\mathcal D_q(C_1,\dots,C_n)}-\tau_q
    \right]_+ .
  \end{equation}
\end{theorem}

For the transfer proof, write
$T_i:=t_\#C_i$ and abbreviate
$\mathcal D_q(C_1,\ldots,C_n)$,
$\mathcal D_q(P_1,\ldots,P_n)$, and
$\mathcal D_q(T_1,\ldots,T_n)$ by $\mathcal D_q(C)$,
$\mathcal D_q(P)$, and $\mathcal D_q(T)$.
The antilipschitz inequality implies, after pushing any coupling of the
$C_i$ through $t$,
\[
  \sqrt{\mathcal D_q(C)}
  \le L\sqrt{\mathcal D_q(T)}.
\]
The synchronous coupling $(u_i(X_i),t(X_i))$ gives
\[
  W_2(P_i,T_i)\le\tau_i.
\]
The weighted product-space stability inequality then yields
\[
  \left|\sqrt{\mathcal D_q(P)}-\sqrt{\mathcal D_q(T)}\right|
  \le\left(\sum_i q_i W_2^2(P_i,T_i)\right)^{1/2}
  \le\tau_q.
\]
Combining the last two displays gives
\[
  \sqrt{\mathcal D_q(P)}
  \ge L^{-1}\sqrt{\mathcal D_q(C)}-\tau_q.
\]
Both sides of the left-hand inequality are nonnegative, so taking the
positive part proves \eqref{eq:p3-aggregate-transfer}.
Thus, even under the most favourable common coupling of the observed laws, the
carrier preserves a minimum amount of latent exposure dispersion after one
aggregate allowance for firm-specific mismatch.

Squaring this transferred floor gives the aggregate observable credit
\begin{equation}\label{eq:p3-sharp-credit}
  \mathcal C^{\sharp}(q)
  :=\left[
    L^{-1}\sqrt{\mathcal D_q(C_1,\dots,C_n)}-\tau_q
  \right]_+^{2}.
\end{equation}

\begin{theorem}[Aggregate information-certified portfolio variance]
  \label{thm:p3-sharp-certificate}
  Under the hypotheses of \Cref{thm:p3-information-certified-cap}, every
  normalized long-only portfolio obeys
  \begin{equation}\label{eq:p3-sharp-cap}
    V_{\mathrm{sys}}(q)\le\sum_i q_i\,v_i-\mathcal C^{\sharp}(q).
  \end{equation}
\end{theorem}

The proof is the following chain, where the first inequality uses the
dispersion infimum and the second uses
\Cref{thm:p3-aggregate-transfer}:
\[
  \begin{aligned}
    V_{\mathrm{sys}}(q)
    &=\sum_i q_i v_i-
    \E_J\!\left[\sum_{i<j} q_i q_j\lVert B_i-B_j\rVert^2\right]\\
    &\le\sum_i q_i v_i-\mathcal D_q(P)\\
    &\le\sum_i q_i v_i-\mathcal C^{\sharp}(q).
  \end{aligned}
\]
No attaining plan is needed for this bound.
The theorem preserves joint compatibility across all assets and applies one
aggregate slack allowance to the dispersion infimum.

\begin{proposition}[Envelope exactness]\label{thm:p3-envelope-exactness}
  If some coherent joint law attains the dispersion minimum for
  $P_1,\dots,P_n$, then $\sum_i q_i\,v_i-\mathcal D_q(P_1,\dots,P_n)$ is the
  greatest systematic portfolio variance attainable across coherent joint laws.
\end{proposition}
An attaining joint law realizes $\mathcal D_q(P)$, while every other coherent
law has dispersion at least that infimum.
Consequently, none can have greater systematic variance than the value stated
in the proposition.
This exactness concerns the variance envelope generated by the fixed exposure
marginals and an attaining coherent law; it does not identify a return
covariance matrix from text.

At zero slack, \eqref{eq:p3-pairwise-relaxation} implies that the aggregate
credit weakly exceeds the pairwise credit, giving a weakly tighter variance cap.
With positive slack, neither credit uniformly dominates: the pairwise positive
parts and the aggregate root-mean-square allowance interact differently with
the observed separations and portfolio weights.
Both certificates use the same observed W2 geometry and declared $(L,\tau)$,
and each remains a valid cap under its maintained assumptions.

The pairwise credit remains the decision form because it is a quadratic
function of $q$ built once from the observed distance matrix, whereas
evaluating $\mathcal D_q$ at a candidate $q$ requires solving an inner
free-centre barycentre problem.
The later portfolio search varies $q$ outside those fixed-$q$ evaluations; it
is not part of the definition of $\mathcal D_q$.
The empirical exercise therefore reports the pairwise certificate, while the
aggregate theorem supplies a separate coherent multi-firm bound.

\subsection{Return bridge and residual sensitivity}

The results so far bound latent systematic exposure risk, not observed return
variance.
To obtain a total-return statement, standardize each asset to unit marginal
variance and decompose its return into systematic and residual components.
This return bridge is a maintained restriction rather than a consequence of
the information geometry.

\begin{assumption}[Return bridge]\label{ass:p3-return-bridge}
  Each standardized return splits as $R_i=S_i+\varepsilon_i$ into a scalar
  systematic component $S_i$ carrying the exposure geometry and a residual
  $\varepsilon_i$.
  Write $v_i=\operatorname{Var}(S_i)$.
  The residual vector is cross-orthogonal to the systematic component:
  \[
    \operatorname{Cov}(S_i,\varepsilon_j)=0
    \quad\text{for every }i\text{ and }j.
  \]
  For risk weights $q$, set $S_q:=\sum_i q_i S_i$ and write
  \[
    \operatorname{Var}(R_q)=\operatorname{Var}(S_q)+\mathcal R(q),
    \qquad
    \mathcal R(q):=\sum_i\sum_j
    q_i q_j\operatorname{Cov}(\varepsilon_i,\varepsilon_j).
  \]
  Unit marginal return variance and zero systematic--residual covariance give
  $\operatorname{Var}(\varepsilon_i)=1-v_i$.
  Zero cross-residual covariance is the benchmark; more generally, the excess
  residual contribution is bounded by a declared $\delta\ge0$.
\end{assumption}

Under zero cross-residual covariance, the systematic certificate and the
diagonal residual terms give
\begin{equation}\label{eq:p3-standardized-cap}
  \begin{aligned}
    \operatorname{Var}(R_q)
    &\le \sum_i q_i v_i-\mathcal C(q)
    +\sum_i q_i^2(1-v_i)\\
    &=1-\mathcal C(q)-\sum_i q_i(1-q_i)(1-v_i)\\
    &\le 1-\mathcal C(q).
  \end{aligned}
\end{equation}
The first line combines the systematic certificate with diagonal residual
variance.
The equality isolates the additional diversification relief supplied by
asset-specific residual risk.
Long-only normalization gives $q_i(1-q_i)\ge0$, so the final inequality
discards that nonnegative relief and leaves the simpler
$1-\mathcal C(q)$ certificate.
With residual contamination, retain the sensitivity wrapper
\begin{equation}\label{eq:p3-residual-budget-cap}
  \operatorname{Var}(R_q)\le1-\mathcal C(q)+\delta.
\end{equation}
Here $\delta$ is a portfolio residual-covariance budget used for sensitivity
analysis, not an estimated structural parameter.
To express the standardized result as a capital-allocation bound, separate
capital weights from shares of the perfect-dependence volatility budget.
For long-only capital weights $x_i\ge0$ with $\sum_i x_i=1$ and positive
marginal scales $\sigma_i$, define
\begin{equation}\label{eq:p3-risk-weights}
  A(x):=\sum_i x_i\sigma_i,
  \qquad
  q_i(x):=\frac{x_i\sigma_i}{A(x)}.
\end{equation}
Here $A(x)$ is the portfolio volatility under perfect positive dependence.
The normalized weight $q_i(x)$ is asset $i$'s share of that benchmark.
Thus $x_i$ are capital weights whereas $q_i$ are risk-budget shares; in
particular, $x_i\propto1/\sigma_i$ implies $q_i=1/n$.
Multiplying \eqref{eq:p3-residual-budget-cap} by $A(x)^2$ yields
\begin{equation}\label{eq:p3-raw-cap}
  \operatorname{Var}(R_x)
  \le
  A(x)^2\left\{1-\mathcal C(q(x))+\delta\right\}.
\end{equation}
This is the financial interpretation of the certificate:
$A(x)^2$ is the perfect-dependence benchmark, while
$A(x)^2\mathcal C(q(x))$ is the amount of raw systematic variance
ruled out by the
observed information geometry under the maintained transmission model.
The additive term $A(x)^2\delta$ makes residual sensitivity explicit.
Equation \eqref{eq:p3-raw-cap} is therefore a decision input rather than an
estimated covariance model.
At the benchmark $\delta=0$, \Cref{sec:portfolio-choice} turns this certified
risk bound into a portfolio rule by minimizing it over a feasible allocation
set.

%% file: chapters/portfolio_choice.tex
\section{Portfolio choice under the certificate}\label{sec:portfolio-choice}

The preceding section establishes how much portfolio variance the
observed information geometry can certify under the maintained
transmission model.
The investor's task is now to turn that risk statement into an
allocation rule without treating the bound as an estimated covariance matrix.
At the benchmark with $\delta=0$, the natural decision is to choose the
feasible capital allocation with the smallest certified upper bound in
\eqref{eq:p3-raw-cap}.

Let $\mathcal W$ be a nonempty compact subset of the long-only simplex,
possibly incorporating upper-weight, sector, or turnover constraints.
Each $x\in\mathcal W$ is a vector of capital weights, whereas $q(x)$
records the corresponding normalized shares of the
perfect-positive-dependence volatility budget defined in
\eqref{eq:p3-risk-weights}.
Define the certified-variance objective
\begin{equation}\label{eq:p3-certified-objective}
  \operatorname{CV}(x)
  :=A(x)^2\left\{1-\mathcal C(q(x))\right\}.
\end{equation}
The information-certified portfolio solves
\begin{equation}\label{eq:p3-certified-portfolio}
  x^{\star}\in
  \operatorname*{arg\,min}_{x\in\mathcal W}\operatorname{CV}(x).
\end{equation}
Equation \eqref{eq:p3-certified-portfolio} is a minimum-risk problem.
Expected returns do not enter its objective, constraints, or identifying
argument.
The objective is economically appropriate because it minimizes a
certified risk cap rather than a point estimate of otherwise unknown
portfolio variance.

The two components of \eqref{eq:p3-certified-objective} make the
allocation trade-off explicit.
The term $A(x)^2$ is the variance benchmark under perfect positive
dependence, so reducing $A(x)$ lowers the same marginal-volatility
benchmark used by an inverse-volatility portfolio.
The deduction $A(x)^2\mathcal C(q(x))$ rewards allocations whose
information distributions certify that their latent risks cannot all be aligned.
Inverse volatility is therefore the cleanest empirical benchmark: it
uses the first channel but not the second.

The first mathematical requirement is that this decision rule
actually has a solution.

\begin{theorem}[Existence of a certified-variance minimizer]
  \label{thm:p3-certified-minimizer}
  Every nonempty compact feasible subset $\mathcal W$ of the long-only
  simplex admits a minimizer of the continuous objective
  \eqref{eq:p3-certified-objective}, and the pointwise variance certificate
  holds at that minimizer.
\end{theorem}

For fixed distance and marginal-scale inputs, $\operatorname{CV}$ is continuous
on $\mathcal W$.
The set is nonempty and compact, so Weierstrass gives a minimizer.
The pointwise certificate in \eqref{eq:p3-raw-cap} then holds at that
minimizer.

Existence makes the rule well defined on any compact feasible set;
curvature is a separate guarantee about how the standardized problem
can be solved.
That curvature is a property of the observed geometry rather than an
assumption about returns.
For a symmetric zero-diagonal matrix $d$, conditional negative definiteness
means
\[
  \sum_i\sum_j u_i\,u_j\,d_{ij}^2\le0
  \quad\text{whenever }\sum_i u_i=0.
\]

\begin{theorem}[Certificate convexity]\label{thm:p3-certificate-convexity}
  If the pairwise floor matrix $\ell$ is conditionally negative definite, then
  $q\mapsto1-\mathcal C(q)$ is convex along mixtures of normalized risk weights.
\end{theorem}

For every tangent direction with $\1^\top u=0$,
\[
  \frac{d^2}{dt^2}\mathcal C(q+tu)=u^\top\ell^2u\le0.
\]
Hence $\mathcal C$ is concave and $1-\mathcal C$ is convex on the simplex.

By Schoenberg's criterion the condition holds exactly when the double-centred
matrix
$-\tfrac12(I-\tfrac1n\1\1^{\top})\ell^2
(I-\tfrac1n\1\1^{\top})$
is positive semidefinite, equivalently when the floors embed isometrically in a
Hilbert space.
It is therefore checkable directly from the reported W2 matrix, without
returns and without an estimated parameter, and it holds for the frozen matrix
used here.
Under this condition, minimizing the standardized certificate over a convex
feasible set is a convex program.

A third guarantee concerns sensitivity to the carrier scale rather
than existence or curvature.
In the zero-slack boundary case, the carrier constant drops out of
the normalized decision.

\begin{corollary}[Carrier invariance of the normalized allocation]
  \label{cor:p3-carrier-invariance}
  Let $\tau_i=0$ for every asset. Then $\ell_{ij}=L^{-1}W_2(C_i,C_j)$ and
  \begin{equation}\label{eq:p3-zero-slack-certificate}
    \mathcal C(q)=\frac{1}{2L^2}\,q^{\top}W_2^2\,q,
  \end{equation}
  so for any $L>0$ and any feasible set $\mathcal Q$ of normalized risk
  weights,
  \[
    \operatorname*{arg\,max}_{q\in\mathcal Q}\mathcal C(q)
    =\operatorname*{arg\,max}_{q\in\mathcal Q}q^{\top}W_2^2q.
  \]
\end{corollary}

The factor $L^{-2}$ is positive, so scaling the objective does not change its
maximizers.
Under zero slack, $L$ determines how much variance the certificate deducts
but not which normalized risk allocation achieves the largest deduction.
That allocation is determined by the observed W2 geometry.

The carrier-invariance statement concerns normalized risk weights
$q$, which are shares of the perfect-dependence volatility budget,
not the capital weights $x$ held by the investor.
The raw objective retains $L$ through the interaction between $A(x)^2$ and
$\mathcal C(q(x))$, so converting an invariant risk allocation into capital
still requires the marginal scales $\sigma_i$.
When slack is nonzero, $\tau_i+\tau_j$ enters the floor additively and the
truncation depends jointly on $L$ and $\tau$.
Together, the existence, curvature, and carrier-invariance results
turn the certificate into an implementable decision rule while
keeping its dependence on the maintained inputs explicit.
\Cref{sec:empirical-design} next constructs the zero-slack allocation
from observed news geometry, and \Cref{sec:news-variance-benchmark}
evaluates its realized variance against conventional portfolio benchmarks.

%% file: chapters/feasibility.tex
\section{Empirical design and data}\label{sec:empirical-design}

The information-certified portfolio in \Cref{sec:portfolio-choice} is the
theoretical class of allocations that minimize the certified variance bound.
The empirical exercise implements its zero-slack, standardized-asset
specialization; throughout the empirical sections, the resulting canonical
implementation is the news-only allocation.
At this boundary, the carrier scale changes the size of the variance certificate
but not the normalized allocation selected from the observed information
geometry.
With unit marginal scales, $A(x)=1$ and $q=x$, so minimizing the certified
variance objective is equivalent to maximizing the pairwise certificate over
normalized weights.
Implementing the general raw-capital objective would instead require the
marginal scales in \eqref{eq:p3-risk-weights}.
The design next defines the reference populations used to evaluate the
standardized allocation and constructs the news geometry that enters the
decision.
Returns assess the news-only allocation only after the decision has been made.

\subsection{Directional hypothesis and reference-population estimand}
\label{sec:p3-reference-populations}

The directional hypothesis is that the news-only allocation lies in
the lower tail of conventional variance under the four declared capped
reference populations.
The theory motivates this direction under the maintained return bridge, but it
does not determine a percentile or imply out-of-sample performance.
The news-only allocation is
\[
  q^{\mathrm{news}}\in
  \operatorname*{arg\,max}_{q\ge0,\,\1^\top q=1}
  \frac12q^\top W_2^2q.
\]
It is constructed from the observed W$_2$ geometry before the return covariance
enters the evaluation.

Let
\[
  V(q):=q^\top\widehat\Sigma_R q
\]
be standardized full-sample return variance.
For each reference population $\mathcal G_k$, define the lower-tail
percentile
\[
  p_k:=\Pr_{q\sim\mathcal G_k}
  \{V(q)\le V(q^{\mathrm{news}})\},
  \qquad
  \widehat p_k:=B^{-1}\sum_{b=1}^{B}
  \1\{V(q_k^{(b)})\le V(q^{\mathrm{news}})\},
  \qquad
  B=\ppnum[0]{reference-draws}.
\]
The estimand $p_k$ is the share of portfolios from reference population
$\mathcal G_k$ whose variance is no greater than the news-only allocation's.
Accordingly, a smaller $\widehat p_k$ means that fewer reference portfolios
achieve equally low or lower variance, placing the candidate nearer the bottom
of the conventional-variance distribution.
This quantity is a Monte Carlo estimate of a descriptive reference-population
percentile, not a classical p-value, and the analysis imposes no rejection
threshold.

Each $\mathcal G_k$ starts with a symmetric Dirichlet draw on the long-only
simplex, clips every weight at its cap, and renormalizes until the allocation
is feasible.
The four $(\alpha,\mathrm{cap})$ pairs are
\[
  (1,12.5\%),\qquad (1,15\%),\qquad
  (0.8,15\%),\qquad (0.5,15\%).
\]
The first two populations isolate cap sensitivity.
The effective number of names is
\[
  N(q):=\left(\sum_i q_i^2\right)^{-1}.
\]
The $\alpha=0.8$ law supplies a fixed, moderately concentrated comparison,
while $\alpha=0.5$ supplies a more concentrated one.
Their realized mean effective numbers are reported rather than assumed to
match the candidate.
The retained populations are transformations of Dirichlet draws rather than
draws from a truncated Dirichlet law.

These laws define comparison populations rather than portfolio strategies that
an investor would implement directly.
Their common feasible set is fully invested, long-only, and subject to a
single-name cap, a recognizable portfolio-constraint template even though the
exact thresholds are stylized rather than rules of a named mandate.
The $12.5\%$ cap rules out portfolios supported on fewer than eight names; the
$15\%$ cap relaxes that minimum to seven.
Changing the cap tests sensitivity to the admissible largest position, while
changing $\alpha$ alters typical concentration within the capped simplex.
Practical mandates may additionally constrain sectors, liquidity, turnover, or
tracking error, none of which these reference laws reproduce.

The named strategy benchmarks answer a different question.
Equal normalized risk weights set $q_i=1/N$ and, through
\eqref{eq:p3-risk-weights}, correspond to inverse-volatility capital weights.
They use marginal volatility but not cross-asset covariance.
The ex post sample GMV instead uses the full return covariance and supplies the
covariance-informed in-sample optimum.
Thus the reference populations locate the news-only allocation within declared
feasible sets, whereas the named benchmarks give its variance an economically
familiar scale.

\subsection{Returns, news, and Wasserstein construction}
\label{sec:p3-data-construction}

Daily simple returns are Yahoo Finance adjusted-close ratios minus one.
These are raw rather than excess returns; no risk-free series is subtracted.
Exact date alignment supplies \ppnum[0]{n-obs} common observations from
2018-03-19 through 2022-12-30, with missing dates removed rather than imputed.

Nasdaq's per-symbol news archive supplies firm articles from 2018 through 2022.
The analysis applies one ticker-indexed retrieval rule and retains dated article
bodies across the prespecified universe.
This construction is auditable, but the paper neither establishes comprehensive
coverage relative to proprietary news archives nor treats the ticker assignment
as manually validated article-level entity annotation.
The annual coverage screen and balanced-cloud rule then form equal-sized
empirical article distributions for the canonical 2022 geometry.
Eligible articles are ordered deterministically by URL hash before the common
cloud size is imposed, so news volume does not enter the distance merely through
unequal empirical support.
The annual cloud-size ladder appears in \Cref{sec:data-appendix}.

Return alignment leaves \ppnum[0]{n-tickers} firms, and each series is
standardized over the common panel before $\widehat\Sigma_R$ is formed.
This is a coverage-screened survivor panel rather than historical index
membership.

An embedding model maps an article body to a fixed-dimensional numerical vector
and is trained to place semantically related documents nearer in that space.
The primary representation independently encodes each article with the frozen,
4,096-coordinate Qwen3-Embedding-8B bi-encoder
\citep{qwen3_embedding_2025}.
It is obtained through OpenRouter's OpenAI-compatible embeddings API as
\texttt{qwen/qwen3-}\allowbreak\texttt{embedding-8b}; rows are normalized.
This frozen representation is the cross-paper canonical model.
Rather than average a firm's article vectors into one point, the design treats
its row-normalized cloud as an equal-mass empirical distribution.
Balanced quadratic transport then pairs articles across two equal-sized clouds
to minimize their root-mean-square Euclidean chord displacement.
The resulting rooted W$_2$ matrix records the least pairwise semantic
displacement; it does not observe the maintained information-to-exposure
transmission map.
The persisted artifact contains rooted W$_2$ distances, which the allocation
objective above squares once at the certificate boundary.
\Cref{sec:data-appendix} documents the artifact construction and sample roster.

The variance evaluation, $\widehat\Sigma_R$, uses the same 2018--2022 period
as the article geometry after the news-only allocation is constructed.
The exercise is therefore an in-sample descriptive ranking.

\subsection{Allocation-anatomy design}
\label{sec:p3-anatomy-design}

The variance percentile answers whether the news-only allocation lands in an
unusual part of the feasible variance distribution, but not what the rule
selects or which transport separations support that decision.
The allocation-anatomy design therefore reports the canonical firm weights,
aggregates them by sector, and decomposes certificate credit into the additive
unordered terms $q_i q_j W_{2,ij}^{2}$.
These are interpretations of the constructed allocation rather than new
return hypotheses.

\subsection{Expanding-cutoff design}
\label{sec:p3-cutoff-design}

The cutoff diagnostic holds the frozen representation and optimization
rule fixed while expanding the article window through each year-end from 2018
to 2022.
For every cutoff, the analysis reports firm weights, effective $N$, the largest
position, and one-way turnover from the preceding allocation.
The terminal cutoff must reproduce the canonical allocation.
Because the encoder is common across cutoffs and the entire accumulated corpus
changes at each step, this sequence measures decision sensitivity to observed
article coverage; it is neither a point-in-time trading backtest nor an event
study.

\subsection{Representation sensitivity design}
\label{sec:p3-representation-design}

The representation ladder studies sensitivity in the measurement of
information geometry rather than introducing new economic hypotheses or
selecting an encoder from the same return sample.
Every arm holds fixed the priced firms, source article records,
standardized-return covariance, reference laws, caps, and paired bootstrap
schedule while changing one declared representation margin where possible.
Model-specific input limits can still change the effective article text seen by
an encoder.

The first margin is output width.
Qwen3-Embedding-8B is evaluated at its full 4,096 coordinates and at 1,024,
256, and 64 coordinates.
Its Matryoshka training is designed to keep leading-coordinate prefixes
informative, so these arms probe compression within one frozen model rather than
re-estimating coordinates from the return data.
The second margin is model capacity.
The 1,024-coordinate comparison between Qwen3-Embedding-8B and
Qwen3-Embedding-4B holds width fixed, whereas their native 4,096- and
2,560-coordinate comparison changes capacity and width jointly.
Both Qwen models are obtained through OpenRouter.

The third margin is model family.
At 1,024 coordinates, BAAI BGE-large-en-v1.5 changes model family, pretraining,
and effective input length jointly relative to Qwen3-Embedding-8B, so the
comparison is an external sensitivity rather than an identified
architecture effect
\citep{xiao_cpack_2023}.
BGE-large-en-v1.5 is also obtained through OpenRouter.

The final margin is the encoder's reported training-information cutoff.
We use the published ModernBERT-based ChronoBERT checkpoints for each year-end
from 2017 through 2024 \citep{he_chronobert_2025}.
The same 128 articles per firm are represented by 768-coordinate means of the
final-layer token states, including special tokens but excluding padding.
Inference is fixed across checkpoints: at most 8,000 characters and 1,024
tokens per article, bfloat16 autocast, and float32 model parameters and
token-mean pooling.
We apply row-L2 normalization once before constructing distances.
Later checkpoints continue training their predecessors, so cumulative
training exposure changes alongside the information cutoff; token budget
and compute are not matched.
The 2017 cutoff precedes the 2018--2022 article sample, the 2018--2022 cutoffs
fall within it, and 2023--2024 follow construction but overlap the full
2023--2026 return-evaluation period.
The checkpoints were published later, so this is not evidence of historical
deployment availability or a post-evaluation negative control.
The vintage ladder compares in-sample allocations, not held-out trading
performance or an identified causal effect of information vintage.
Its benchmark-calibrated equivalence threshold is post-specified, making the
equivalence assessment exploratory.

\subsection{Chronological covariance benchmark}
\label{sec:p3-covariance-benchmark-design}

The in-sample reference-population ranking evaluates the canonical allocation
within the standardized return sample.
The chronological benchmark instead specifies a comparison of capital-weight
portfolios on 2023--2026 returns, with every data-dependent weight fixed at
31 December 2022 using 2018--2022 formation data.

The W2 geometry determines the normalized certificate budget
$q^{\mathrm{news}}$. Calibration volatility enters only when converting it to
capital weights. Let
$s_i:=\widehat\sigma_{i,\mathrm{cal}}$ and define the rescaled budget
\[
  \widetilde q_i^{\mathrm{news}}
  :=
  \frac{q_i^{\mathrm{news}}}{s_i}.
\]
Set
\[
  x_i^{\mathrm{news}}
  =
  \frac{\widetilde q_i^{\mathrm{news}}}
  {\sum_j \widetilde q_j^{\mathrm{news}}}.
\]
This is the capital implementation of the normalized certificate allocation;
it is not the solution of the general raw-capital certificate objective in
\eqref{eq:p3-certified-objective}. If the mapping exceeds the common 15\%
capital cap, the allocation is retained and marked infeasible rather than
silently clipped or renormalized.

For the new chronological comparison, all return panels are factor-complete
excess returns. Write the reduced-form factor model as
\[
  r_t^e = a + B f_t + u_t,
  \qquad
  B_m=(I-\widehat\rho_m W_m)^{-1}\widehat C_{f,m},
\]
where $\widehat C_{f,m}$ is the fitted factor-coefficient block before spatial
propagation. Under the working factor--residual orthogonality assumption, the
total covariance used by the primary GMV candidates is
\[
  \widehat\Sigma_{\mathrm{total},m}
  =
  \underbrace{B_m\widehat\Sigma_f
  B_m^\top}_{\widehat\Sigma_{\mathrm{systematic},m}}
  +
  \underbrace{\widehat\Sigma_{u,m}}_{\widehat\Sigma_{\mathrm{residual},m}},
  \qquad
  \widehat\Sigma_f=\operatorname{Cov}_{T-1}(f_t),
\]
with the intercept excluded from $\widehat\Sigma_f$. The residual estimator
is formed from factor-residualized spatial innovations and mapped as
$(I-\widehat\rho_m W_m)^{-1}D_m(I-\widehat\rho_m W_m)^{-\top}$ using stable
linear solves. The factor-only control sets $\widehat\rho=0$ and applies the
same constant-correlation residual-shrinkage construction; the recognizable
direct Ledoit--Wolf benchmark instead uses its scaled-identity target on
excess returns. The unrestricted specification requires the systematic
component to agree across the factor-only, W1, and W2 fits up to numerical
tolerance, so the matched comparison differs through residual restrictions and
shrinkage rather than a model-dependent factor-risk component.

The comparison includes equal-weight, inverse-volatility, sample-GMV,
conventional Ledoit--Wolf-GMV, the matched factor-only controls, and
minimum-variance portfolios formed from the total W1 and W2 barycentric
covariances. Residual-covariance GMV portfolios are retained as explicitly
labeled diagnostics. Every covariance-based portfolio uses the same 100-firm
roster, long-only fully invested constraints, and 15\% single-name weight cap;
the cap is rejected as infeasible when the capped simplex cannot contain unit
mass. The data-dependent weights use only the factor-complete calibration
panel, the dated PIT W2 certificate, and dated calibration covariance
artifacts. Holdout returns enter only realized excess-return variance.
The realized calculation is the variance of a constant-weight mixture, which
corresponds to costless rebalancing under a trading interpretation rather than
buy-and-hold holdings.

The spatial covariance alternatives use the target-anchored barycentric fields
from the companion spatial study \citep{gawronsky_spatial_2027}. W1 and W2
vary the transport geometry used to construct the field, while QMLE and
constant-correlation Ledoit--Wolf distinguish the fitted residual-covariance
estimates used to form the corresponding total-risk portfolio. Spherical and
diagonal innovation specifications are retained as variance-model sensitivity
comparisons. The PIT certificate archive stores squared W2 transport costs,
so the selected matrix enters the zero-slack objective as
$\frac12q^\top Mq$ without a second squaring. The resulting benchmark measures
chronological excess-return risk performance while keeping residual-risk
diagnostics separate from the primary total-risk ranking.

With the allocation, reference-population estimand, anatomy and cutoff
diagnostics, return evaluation, and four measurement margins defined, the next
section reports the resulting in-sample evidence.

\section{Results}
\label{sec:results}\label{sec:news-variance-rank}

\subsection{Main in-sample variance ranking}

The main result uses the prespecified, 4,096-coordinate
Qwen3-Embedding-8B representation.
Across the four reference laws, its variance percentile ranges from
\ppnum[2]{news-only-percentile-uniform-cap-15-pct}\% to
\ppnum[2]{news-only-percentile-concentrated-cap-15-pct}\%, compared with
\ppnum[1]{equal-percentile-concentrated-cap-15-pct}\% to
\ppnum[1]{equal-percentile-uniform-cap-12p5-pct}\% for equal risk weights.
The variance percentile is the share of feasible reference portfolios with
variance no greater than the candidate's.
Lower is therefore better: a smaller percentile means that fewer reference
portfolios attain equally low or lower variance.
\Cref{tab:news-percentile} reports this ranking under each prespecified
reference law.

\begin{table}[H]
  \centering
  \scriptsize
  \caption{In-sample variance percentile inside four prespecified feasible
    allocation populations. Entries are the share of
    \ppnum[0]{reference-draws} capped draws whose standardized variance is no
  greater than the candidate's. Lower is better.}
  \label{tab:news-percentile}
  \begin{tabular}{lrrr}
    \toprule
    Reference population & Mean effective $N$ & News-only & Equal risk \\
    \midrule
    Uniform, 12.5\% cap
    & \ppnum[2]{reference-effective-n-uniform-cap-12p5}
    & \ppnum[2]{news-only-percentile-uniform-cap-12p5-pct}\%
    & \ppnum[1]{equal-percentile-uniform-cap-12p5-pct}\% \\
    Uniform, 15\% cap
    & \ppnum[2]{reference-effective-n-uniform-cap-15}
    & \ppnum[2]{news-only-percentile-uniform-cap-15-pct}\%
    & \ppnum[1]{equal-percentile-uniform-cap-15-pct}\% \\
    Fixed $\alpha=0.8$, 15\% cap
    & \ppnum[2]{reference-effective-n-effective-n-matched}
    & \ppnum[2]{news-only-percentile-effective-n-matched-pct}\%
    & \ppnum[1]{equal-percentile-effective-n-matched-pct}\% \\
    Concentrated, 15\% cap
    & \ppnum[2]{reference-effective-n-concentrated-cap-15}
    & \ppnum[2]{news-only-percentile-concentrated-cap-15-pct}\%
    & \ppnum[1]{equal-percentile-concentrated-cap-15-pct}\% \\
    \bottomrule
  \end{tabular}
\end{table}

Under the two uniform laws, only
\ppnum[2]{news-only-percentile-uniform-cap-12p5-pct}\% and
\ppnum[2]{news-only-percentile-uniform-cap-15-pct}\% of draws have variance no
greater than the news-only allocation's.
The corresponding shares are
\ppnum[2]{news-only-percentile-effective-n-matched-pct}\% under the fixed
$\alpha=0.8$ law and
\ppnum[2]{news-only-percentile-concentrated-cap-15-pct}\% under the deliberately
more concentrated law.
Thus, a rule constructed without cross-asset return covariance lands near the
first percentile of conventional variance across all four declared feasible
populations.

The two concentrated populations make this ranking more informative than the
uniform comparisons alone, but they do not furnish an effective-$N$ match.
The fixed $\alpha=0.8$ law has mean effective $N$ of
\ppnum[2]{reference-effective-n-effective-n-matched}, compared with the
candidate's \ppnum[2]{news-only-effective-number-assets}, and only
\ppnum[2]{news-only-percentile-effective-n-matched-pct}\% of its draws attain
equally low or lower variance.
Under the more concentrated law, mean effective $N$ falls to
\ppnum[2]{reference-effective-n-concentrated-cap-15}, and the share is only
\ppnum[2]{news-only-percentile-concentrated-cap-15-pct}\%.
The low rank therefore survives comparisons that materially alter the
concentration profile within the declared feasible class.
Because both reference laws remain more diffuse than the candidate, this
sensitivity does not eliminate concentration as an explanation of the rank.

Equal risk weights provide a familiar point of comparison within the same laws.
Their percentiles are
\ppnum[1]{equal-percentile-uniform-cap-12p5-pct}\%,
\ppnum[1]{equal-percentile-uniform-cap-15-pct}\%,
\ppnum[1]{equal-percentile-effective-n-matched-pct}\%, and
\ppnum[1]{equal-percentile-concentrated-cap-15-pct}\%, respectively.
These ranks remain well above the news-only ranks under every reference law.
The comparison is a descriptive location within prespecified portfolio
populations, not a hypothesis test or evidence of prospective performance.

\subsection{Portfolio characteristics and conventional benchmarks}
\label{sec:news-variance-benchmark}

The percentile result says where the candidate lies within feasible reference
populations but not how far it sits from familiar portfolio strategies.
The equal-risk benchmark sets $q_i=1/N$.
Using the mapping $q_i(x)=x_i\sigma_i/A(x)$ gives
\[
  q_i=\frac1N
  \quad\Longleftrightarrow\quad
  x_i^{\mathrm{IV}}
  =\frac{\sigma_i^{-1}}{\sum_j\sigma_j^{-1}}.
\]
It is therefore the inverse-volatility capital portfolio expressed in the
normalized risk-weight coordinates of the empirical exercise, not an arbitrary
second equal-weight portfolio.
The sample GMV provides the opposite benchmark: it uses the complete in-sample
covariance matrix that the news-only construction excludes.
\Cref{tab:news-variance-benchmark} provides that calibration using the same
standardized return panel.

\begin{table}[H]
  \centering
  \scriptsize
  \caption{In-sample descriptive portfolio benchmarks. Standardized variance
    is computed on the complete 2018--2022 aligned return panel; relative
    variance indexes the long-only sample GMV to 100. Returns evaluate the
  news-only allocation after construction but do not determine its weights.}
  \label{tab:news-variance-benchmark}
  \begin{tabular}{lrrrr}
    \toprule
    Portfolio & Standardized variance & Rel. to GMV & Max. weight
    & Effective $N$ \\
    \midrule
    News-only
    & \ppnum[3]{news-only-variance}
    & \ppnum[1]{news-only-gmv-relative-variance-pct}
    & \ppnum[1]{news-only-maximum-weight-pct}\%
    & \ppnum[2]{news-only-effective-number-assets} \\
    Equal risk weights
    & \ppnum[3]{equal-standardized-variance}
    & \ppnum[1]{equal-gmv-relative-variance-pct}
    & -- & -- \\
    Sample GMV
    & \ppnum[3]{sample-gmv-variance}
    & 100.0 & -- & -- \\
    \bottomrule
  \end{tabular}
\end{table}

The news-only standardized variance is
\ppnum[1]{news-only-equal-variance-reduction-pct}\% lower than the variance of
the equal-risk benchmark.
It nevertheless remains
\ppnum[1]{news-only-gmv-excess-variance-pct}\% above the ex post long-only
sample GMV.
Together, these comparisons place the information-certified allocation on a
common scale with both the declared feasible-population laws and the
covariance-informed in-sample optimum.

The concentration diagnostics describe the decision object behind that rank.
Its largest holding is \ppnum[1]{news-only-maximum-weight-pct}\%, and its
effective number of names is
\ppnum[2]{news-only-effective-number-assets}.
The candidate therefore differs from equal risk weighting and is not reduced to
a one-name solution.
The low rank persists under reference laws that place more mass on concentrated
portfolios, but none matches the candidate's realized effective number of names.
These comparisons therefore assess sensitivity to concentration without
eliminating scalar concentration as a confound.

\subsection{Anatomy of the news-only allocation}
\label{sec:p3-portfolio-anatomy}

The scalar concentration diagnostics do not reveal which firms the rule
selects or which parts of the information geometry create its certificate.
\Cref{fig:p3-portfolio-anatomy} opens that decision object at three levels:
firm weights, additive firm-pair credits, and sector weights.

The pair layer is the informative attribution level for this optimum.
To see why, let $M=(W_{2,ij}^{2})_{i,j}$ and write the zero-slack objective as
$\mathcal C(q)=\tfrac12q^{\top}Mq$.
For every firm in the positive support of this solution, whose unit cap is
nonbinding, the simplex first-order condition gives
$(Mq^{\star})_i=\lambda$.
Multiplying by $q_i^{\star}$ and summing over the support yields
$\lambda=q^{\star\top}Mq^{\star}$.
Consequently, the natural firm-level share proposed by the quadratic
decomposition satisfies
\begin{equation}\label{eq:p3-firm-credit-identity}
  \frac{q_i^{\star}(Mq^{\star})_i}
  {q^{\star\top}Mq^{\star}}=q_i^{\star}
  \qquad\text{whenever }q_i^{\star}>0.
\end{equation}
Firm certificate shares therefore reproduce the portfolio weights rather than
provide a second diagnostic.
The unordered pair terms remain informative:
\begin{equation}\label{eq:p3-pair-credit-decomposition}
  c_{ij}:=q_i^{\star}q_j^{\star}W_{2,ij}^{2},
  \qquad
  \mathcal C(q^{\star})=\sum_{i<j}c_{ij}.
\end{equation}
They identify which weighted separations the optimizer actually uses.

\begin{figure}[H]
  \centering
  \import{images/}{portfolio_anatomy\ppfiguresuffix.pgf}
  \caption{Anatomy of the canonical news-only allocation. Panel (a)
    orders firms by sector and then by weight; horizontal lines mark the equal
    risk weight and the \ppnum[1]{reference-tight-cap-pct}\% cap used by the
    tightest reference population.
    Alternating ticker positions are labelled for legibility; all
    \ppnum[0]{n-tickers} firm weights remain plotted.
    Panel (b) reports the twelve largest unordered terms $c_{ij}$ in
    \eqref{eq:p3-pair-credit-decomposition} as shares of total certificate
    credit and distinguishes within- from cross-sector pairs. Panel (c)
    compares sector weights with the sector shares induced by equal risk
    weights. All panels describe the full-sample news-only allocation; they do
  not use returns or assign a causal role to sector membership.}
  \label{fig:p3-portfolio-anatomy}
\end{figure}

The solution assigns positive weight to
\ppnum[0]{anatomy-active-firms} of the \ppnum[0]{n-tickers} firms, so the
effective-$N$ statistic reflects a portfolio with many small positions and a
limited set of larger ones rather than diffuse weight on every name.
\Cref{tab:p3-top-holdings} makes the largest positions tangible.
For each row, the strongest partner is the holding that produces its largest
pair term in \eqref{eq:p3-pair-credit-decomposition}.
A large position need not have the largest pair credit: the latter also
depends on the counterpart's weight and their squared transport separation.

\input{generated/portfolio_holdings}

The sector panels supply a diagnostic rather than an explanation.
The news-only sector weights differ visibly from the sector shares implied by
equal risk weights.
Across the additive pair terms,
\ppnum[1]{anatomy-cross-sector-credit-pct}\% of certificate credit is
cross-sector and \ppnum[1]{anatomy-within-sector-credit-pct}\% is
within-sector.
Thus the reported certificate is not generated by within-sector separation
alone.
The split does not establish that transport geometry contains information
beyond sector labels: sector sizes, portfolio weights, and the number of
available cross-sector pairs also affect it.
It locates the certificate credit while leaving the economic source of those
separations open.

\subsection{Evolution across expanding information cutoffs}
\label{sec:p3-portfolio-evolution}

The canonical allocation pools articles through 2022.
To examine whether that decision is peculiar to the terminal corpus, the same
zero-slack problem is re-solved after expanding the article window through
each year-end from 2018 to 2022.
\Cref{fig:p3-portfolio-evolution} shows the resulting sequence.
These are expanding information cutoffs, not separate calendar-year article
samples.

\begin{figure}[H]
  \centering
  \import{images/}{portfolio_evolution\ppfiguresuffix.pgf}
  \caption{Evolution of the news-only allocation under expanding
    article cutoffs. Panel (a) reports firm weights for article windows ending
    in 2018, 2019, 2020, 2021, and 2022, with firms grouped by sector.
    Alternating ticker positions are labelled; every firm remains in each
    heatmap column.
    Panel (b) reports the effective number of names, the largest weight and its
    ticker, and one-way turnover
    $\tfrac12\lVert q_t-q_{t-1}\rVert_1$. The common frozen encoder is held
    fixed, and the terminal 2022 column reproduces the canonical allocation.
    The sequence measures decision sensitivity to accumulated articles; it is
  neither a prospective return test nor an event-study design.}
  \label{fig:p3-portfolio-evolution}
\end{figure}

The portfolio changes as the information set grows without becoming either
one-name concentrated or equal-risk weighted.
Across the five cutoffs, effective $N$ remains between
\ppnum[2]{anatomy-vintage-effective-n-min} and
\ppnum[2]{anatomy-vintage-effective-n-max}, while the largest weight ranges
from \ppnum[1]{anatomy-vintage-maximum-weight-min-pct}\% to
\ppnum[1]{anatomy-vintage-maximum-weight-max-pct}\%.
One-way turnover ranges from \ppnum[1]{anatomy-turnover-min-pct}\% to
\ppnum[1]{anatomy-turnover-max-pct}\%.
The largest reallocation occurs when the first one-year corpus is extended
through 2019:
\ppnum[1]{anatomy-turnover-2018-2019-pct}\%, compared with
\ppnum[1]{anatomy-turnover-2019-2020-pct}\% from 2019 to 2020.
The sequence therefore does not display a uniquely large break at the 2020
cutoff.
It cannot isolate a COVID effect in any case, because the design changes the
entire accumulated article distribution at once and holds a common encoder
fixed across cutoffs.

\subsection{Representation sensitivity}
\label{sec:p3-representation-results}

The expanding-cutoff sequence changes the information set while holding the
representation fixed, whereas the representation comparisons change the
measurement of that information.
\Cref{tab:p3-representation-main} shows the four measurement margins, with the
full ladder and paired bootstrap contrasts in
\Cref{sec:p3-representation-ablation}.
An allocation that violates the reference law's unchanged cap receives a dash
in the percentile column, while its relative-GMV index remains defined.

\input{generated/representation_ablation_main}

Moderate within-model compression leaves the outcome nearly unchanged:
Qwen3-8B at 1,024 coordinates is close to the full-width primary representation
in both reported metrics, and their paired interval includes zero.
At the same fixed width, Qwen3-4B ranks higher in conventional variance than
Qwen3-8B, but their paired difference does not survive Holm adjustment.
BGE-large at 1,024 coordinates is again close to Qwen3-8B, so the sample does
not support a simple model-family ordering.

The width ladder is non-monotone.
The 64-coordinate Qwen3-8B stress cell has the lowest conventional-variance
rank in the ladder, but its contrast with the 256-coordinate cell does not
survive Holm adjustment.
That outcome does not make 64 coordinates the preferred specification: the
full-width 8B representation was prespecified, and every arm is evaluated on
the same return sample.
Instead, its point ranking shows that extreme compression can materially alter
the geometry and selected allocation, in this case in a direction that
improves the in-sample variance benchmark without an adjusted rejection of
zero difference against the 256-coordinate cell.

The ChronoBERT annual ladder supplies a separate sensitivity.
The 2018-minus-2017 contrast raises the relative-GMV index, whereas the
2024-minus-2017 contrast lowers it; both remain distinguishable from zero
after vintage-family Holm adjustment.
Neither interval lies wholly inside the post-specified equivalence bound
(\Cref{tab:p3-representation-contrasts}).
Other annual pairs meet that bound, but the ladder does not establish
broad invariance or a monotone improvement with later training cutoffs.
These are differences in in-sample allocation risk, not prospective
performance gains or causal effects of later information alone.
Taken together, the ladder does not support a monotone ``larger model is
better'' account.
It shows instead that the primary result is empirically strong within the
declared benchmark but remains sensitive to the measurement layer that creates
the information geometry.

These comparisons are descriptive and in-sample: the same 2018--2022 return
panel evaluates the news-only allocations after their construction.
They describe how the rule ranks against the declared reference populations,
not how it will perform prospectively.
\Cref{sec:limitations} interprets what these results establish within the
maintained framework and where their boundaries lie.

%% file: generated/portfolio_holdings.tex
\begin{table}[H]
\centering
\scriptsize
\setlength{\tabcolsep}{3.0pt}
\begin{tabularx}{\linewidth}{lXlrrr}
\toprule
Ticker & Firm & Sector & Weight (\%) & Partner & Pair credit (\%) \\
\midrule
AZN & AstraZeneca PLC & Healthcare & 10.86 & GOOG & 1.34 \\
GOOG & Alphabet Inc. & Communication Services & 5.31 & AZN & 1.34 \\
NFLX & Netflix Inc. & Communication Services & 4.83 & AZN & 1.21 \\
XEL & Xcel Energy Inc. & Utilities & 4.72 & AZN & 1.20 \\
TSCO & Tractor Supply Company & Consumer Cyclical & 4.21 & AZN & 0.97 \\
UAL & United Airlines Holdings Inc. & Industrials & 4.18 & AZN & 0.99 \\
BMRN & BioMarin Pharmaceutical Inc. & Healthcare & 4.02 & AZN & 0.92 \\
TMUS & T-Mobile US Inc. & Communication Services & 3.46 & AZN & 0.86 \\
\bottomrule
\end{tabularx}
\caption{Largest positions in the canonical news-only allocation. Partner identifies the holding that forms the largest unordered additive certificate term with the row firm; pair credit reports that term as a percentage of total certificate credit.}
\label{tab:p3-top-holdings}
\end{table}

%% file: generated/representation_ablation_main.tex
\begin{table}[H]
\centering
\scriptsize
\caption{Selected representation sensitivity of the news-only allocation. The percentile column reports the percentage of allocations under the fixed $\alpha=0.8$ reference law whose standardized variance is no greater than the candidate's. Relative variance indexes the long-only sample GMV to 100. Lower is better in both columns. The complete ladder and paired bootstrap contrasts appear in \Cref{sec:p3-representation-ablation}.}
\label{tab:p3-representation-main}
\setlength{\tabcolsep}{7.0pt}
\begin{tabular}{lrrr}
\toprule
Encoder & Dim. & Fixed-$\alpha=0.8$ percentile & Rel. GMV \\
\midrule
Qwen3-8B full (primary) & 4096 & 0.19 & 147.5 \\
Qwen3-8B@1024 & 1024 & 0.08 & 146.7 \\
Qwen3-4B@1024 & 1024 & 0.29 & 148.5 \\
BGE-large-v1.5 full & 1024 & 0.03 & 145.3 \\
Qwen3-8B@64 & 64 & 0.01 & 144.6 \\
\midrule
ChronoBERT 2017 & 768 & -- & 164.4 \\
ChronoBERT 2018 & 768 & -- & 166.7 \\
ChronoBERT 2019 & 768 & -- & 160.7 \\
ChronoBERT 2020 & 768 & -- & 158.5 \\
ChronoBERT 2021 & 768 & -- & 160.2 \\
ChronoBERT 2022 & 768 & -- & 165.3 \\
ChronoBERT 2023 & 768 & -- & 158.7 \\
ChronoBERT 2024 & 768 & -- & 161.6 \\
\bottomrule
\end{tabular}
\end{table}

%% file: chapters/limitations.tex
\section{Discussion and Limitations}\label{sec:limitations}

Within its maintained transmission model and prespecified information geometry,
the paper establishes a bounded positive implication.
The certificate maps observed Wasserstein separation into a one-sided
restriction on portfolio variance, and the zero-slack allocation rule
minimizes the resulting certified upper bound without estimating cross-asset
return covariance.
In the \ppnum[0]{n-tickers}-firm panel, only
\ppnum[2]{news-only-percentile-uniform-cap-15-pct}\%--
\ppnum[2]{news-only-percentile-concentrated-cap-15-pct}\% of portfolios across
four prespecified reference populations have conventional variance no greater
than that allocation, while equal risk weights rank materially higher under
every comparison.
The evidence therefore places the allocation unusually low among the declared
feasible populations, and \Cref{tab:news-variance-benchmark} measures its distance
from the covariance-informed in-sample optimum.
Five boundaries qualify what this finding establishes and what it does not.

The first is identification.
The certificate is conditional on an antilipschitz carrier, firm-specific
slack, one coherent joint exposure law, and the return bridge.
Together, these maintained restrictions transmit observed information
separation into a statement about portfolio risk, but the text data do not
identify that transmission mechanism.
The cost is interpretive: the certificate is valid under the declared
restrictions rather than an unconditional implication of text geometry.
At the zero-slack boundary, however, the carrier constant drops out of the
normalized allocation through \Cref{cor:p3-carrier-invariance}, so the
selected risk weights depend only on the observed W$_2$ distances.
The maintained restrictions therefore determine what the low variance ranking
means economically---whether it reflects genuine risk reduction through
information separation---rather than whether the allocation exists or whether
its realized variance is low relative to the declared reference populations.
Isolating the transmission would require identified or calibrated counterparts
to the carrier, firm-specific slack, and return bridge.

The second boundary is representation dependence.
The article selection, embedding model, and output width determine the
measured distributions and may therefore alter the distances and portfolio
weights.
The representation ladder evaluates this sensitivity while holding the priced
firms and return evaluation fixed, but it does not make the canonical
Qwen3-Embedding-8B geometry invariant to measurement choices.
The width ladder is non-monotone: extreme compression to 64 coordinates
materially alters the point allocation and improves its in-sample variance
ranking, although the paired contrast with 256 coordinates does not survive
Holm adjustment, so the geometry is not simply degraded by truncation.
The ChronoBERT annual comparisons likewise change the in-sample allocation
nonmonotonically: several differences survive vintage-family Holm adjustment,
while only a subset meet the post-specified equivalence criterion.
A common architecture therefore does not establish broad representation
invariance, and cumulative continued training does not isolate a causal
effect of the information cutoff.
The primary representation was prespecified, and the ladder evaluates
sensitivity rather than selecting a winner from the same return data.
Prespecified alternatives using different ground metrics, unbalanced clouds,
or held-out evaluation windows would show which features of the geometry
survive those measurement choices.

The third boundary concerns comparison-set design.
The percentile ranking is relative to four prespecified Dirichlet-capped
reference populations, and the choice of concentration parameter and cap
determines what ``low'' means.
The two concentrated populations reduce the effective-$N$ gap relative to the
uniform laws, and the ranking survives, but neither matches the candidate's
realized effective $N$.
They therefore provide concentration sensitivity rather than a
concentration-adjusted comparison, and they do not eliminate concentration as
a confound.
These four laws nevertheless remain maintained design choices.
Alternative reference populations---market-capitalization-weighted draws,
factor-tilted draws, or draws from the empirical distribution of a historical
allocation universe---could rank the candidate differently.

The fourth boundary is portfolio implementation.
The empirical exercise implements the standardized-asset specialization, in
which unit marginal scales make capital weights coincide with normalized risk
weights.
The general raw-capital objective in \eqref{eq:p3-certified-objective}
retains the marginal volatility scales through $A(x)^2$ and the risk-weight
mapping $q(x)$.
Constructing a capital-weight allocation from the certificate therefore uses
the same marginal-scale inputs as an inverse-volatility portfolio.
The chronological benchmark applies this mapping using calibration-period
scales and evaluates the resulting constant-weight excess-return mixture on
the subsequent holdout period. It retains any capital-cap violation as an
explicit infeasibility status rather than changing the normalized certificate
rule by clipping. The benchmark places the certificate beside total-risk
covariance-based and spatial-covariance benchmarks while preserving its
information-geometry construction; residual-covariance GMV results are
diagnostics rather than matched total-risk competitors.

The fifth boundary is external validity.
The empirical universe is a coverage-screened survivor panel of
\ppnum[0]{n-tickers} firms with complete return histories.
The resulting evidence characterizes firms supported by the required source
coverage rather than every investable firm or a changing historical universe.
The information geometry and the return evaluation also draw on the same
survivor-conditioned roster, while the chronological benchmark uses a dated
return holdout and factor-complete excess-return panels.
The reported percentile is therefore descriptive, and the chronological
exercise is a retrospective return-holdout comparison conditional on the
frozen representation and selected universe. It is neither a full historical
point-in-time test nor an ex-ante guarantee of future realized variance.
Maintaining constant portfolio weights is a costless-rebalancing convention;
the calculation is not a buy-and-hold holdings simulation and includes no
turnover or transaction costs.
The expanding-cutoff sequence shows how the allocation evolves with
accumulated articles, but it does not constitute out-of-sample evidence
because the return evaluation period is unchanged.
A prospective performance claim requires a frozen point-in-time construction
in which the encoder, article corpus, and W$_2$ geometry are all fixed before
the evaluation returns are observed, together with genuinely out-of-sample
return evaluation.
The present evidence demonstrates the framework's implications for a specific
panel and representation; whether those implications survive prospective
evaluation is the immediate empirical priority.
The conclusion returns to the conditional portfolio-risk certificate and its
allocation rule within these boundaries.

%% file: chapters/conclusion.tex
\section{Conclusion}\label{sec:conclusion}

Mean--variance portfolio choice conventionally begins by estimating a return
covariance matrix, precisely the object that short and high-dimensional panels
make difficult to estimate reliably.
This paper takes a different route.
It asks what observed differences between firms' information distributions can
certify about portfolio risk before cross-asset covariance is estimated.
The resulting information certificate is a one-sided restriction on admissible
portfolio variance, not a point estimate of the covariance matrix.

Related work establishes neighbouring implications of the same distributional
geometry at other levels of aggregation.
At the pairwise level, \citet{gawronsky_continuous_2026} derive a covariance
envelope from Wasserstein separation.
At the cross-sectional level, \citet{gawronsky_spatial_2027} form a barycentric
interaction field through target-anchored Wasserstein barycentric
reconstruction.
The distinct step here is portfolio aggregation.
An antilipschitz carrier with bounded firm-specific slack transfers observed W2
separation into lower bounds on latent exposure separation; one coherent joint
exposure law then aggregates those restrictions into a portfolio-risk bound.
A maintained return bridge translates the exposure result into standardized and
raw-return variance statements.
The text determines the observed geometry, but it does not identify any of these
transmission restrictions.

Within this bridge, multi-firm transport dispersion supplies the sharp
certificate, while the weighted pairwise certificate provides the
computationally simpler decision rule used in the empirical analysis.
The investor minimizes the certified upper bound rather than an estimated
portfolio variance.
When firm-specific slack is zero, the common carrier scale changes how much
variance is certified away but not the selected normalized risk allocation,
which depends only on the observed W2 geometry.
Capital implementation still requires marginal volatility scales, but expected
returns and cross-asset return covariance do not enter the allocation's
construction.

In the \ppnum[0]{n-tickers}-firm 2018--2022 panel, the
Qwen3-Embedding-8B news-only allocation
falls between the
\ppnum[2]{news-only-percentile-uniform-cap-15-pct}\% and
\ppnum[2]{news-only-percentile-concentrated-cap-15-pct}\% variance percentiles
across four prespecified capped reference populations, while equal risk weights
rank higher under each corresponding comparison.
Its standardized variance is
\ppnum[1]{news-only-equal-variance-reduction-pct}\% below the variance of the
equal-risk benchmark but
\ppnum[1]{news-only-gmv-excess-variance-pct}\% above the ex post long-only
sample GMV.
This places the information-certified allocation at an unusually low location
among ordinary feasible allocations and quantifies its distance from the
covariance-informed in-sample optimum.
Portfolio anatomy further shows that
\ppnum[1]{anatomy-cross-sector-credit-pct}\% of the additive certificate credit
comes from cross-sector pairs.
This attribution locates the geometry used by the allocation; it does not
identify a sector mechanism.
Both the information geometry and the covariance used for evaluation draw on the
same 2018--2022 period.
The resulting ranking is therefore descriptive and in-sample: it illustrates
the allocation implied by the maintained framework rather than identifying the
information-to-risk transmission or predicting future portfolio variance.

As a theoretical restriction, the certificate can complement conventional
covariance estimation with information about admissible portfolio risk that
originates outside joint returns.
The certificate supplies the information-based allocation, while covariance
estimates provide a return-based evaluation and chronological benchmark.
Combining the two objects in a shrinkage or constrained covariance estimator is
a natural extension when return histories are short or unstable.

More broadly, the construction is not intrinsically tied to corporate news or
equities.
Where economically relevant objects admit distribution-valued representations
and a defensible transmission restriction links their observed geometry to
latent risk, the same logic can generate one-sided restrictions on aggregate
risk.
The immediate empirical priorities are point-in-time construction, genuinely
out-of-sample evaluation, and identified or calibrated counterparts to the
maintained carrier, firm-specific slack, and return bridge.
Those extensions would test the certificate's practical value; the present
contribution is the coherent portfolio-level object and the explicit bridge on
which it depends.

%% file: appendix.tex
\input{chapters/appendices}

%% file: chapters/appendices.tex
\section{Formal verification}\label{sec:lean-proofs}

The main text contains the human-readable proofs of the barycentre
representation, the aggregate transfer inequality, the information-certified
cap, the sharp certificate, envelope exactness, certificate convexity, and the
pairwise bridge.
Where the bound is sharp, attainment of the optimal multi-marginal plan is
supplied as a premise; the Lean record does not formalize the general existence
theorem.

The formal record checks the displayed identities and bounds under their
stated assumptions.
These results concern the algebraic structure of the coherent portfolio
certificate and do not validate the article-embedding proxy, the economic
content of the carrier-and-slack restrictions, the empirical distance
estimator, or the chronological evaluation design.

The exact machine-checked scope and assumptions are:
\ppLeanStatusList

\section{Barycentre representation of dispersion}\label{sec:barycentre-appendix}

\begin{proposition}[Barycentre representation]
  \label{prop:p3-barycenter-representation}
  For laws $P_1,\dots,P_n$ on $\mathcal H$ with finite second moments,
  \eqref{eq:p3-barycenter-representation} holds.
\end{proposition}

For weights $q$ on the simplex, set
$\bar x_q:=\sum_i q_i x_i$.
The pointwise weighted-variance identity
\[
  \sum_{i<j}q_i q_j\lVert x_i-x_j\rVert^2
  =\sum_i q_i\lVert x_i-\bar x_q\rVert^2
\]
is the bridge between the two representations.
Given any multimarginal coupling of $P_1,\dots,P_n$, the random barycentre
$\bar X_q=\sum_i q_i X_i$ induces a law $Q$.
Applying the identity and then minimizing over multimarginal couplings gives
the direction from the joint-coupling infimum to the barycentre infimum.

Conversely, take couplings of $(P_i,Q)$ whose costs approach
$W_2^2(P_i,Q)$ and glue their conditional laws given the common
$Q$-distributed variable.
The resulting multimarginal coupling has the required marginals.
Conditionally on that common variable, the pointwise weighted-variance
minimum gives
\[
  \sum_{i<j}q_i q_j\lVert X_i-X_j\rVert^2
  \le \sum_i q_i\lVert X_i-Q\rVert^2.
\]
Taking expectations, then infima over the couplings and over $Q$, gives the
reverse inequality.
Thus the infimum values agree without assuming that either infimum is
attained.

\section{Data and W2 artifact construction}\label{sec:data-appendix}

The empirical universe contains \ppnum[0]{n-tickers} firms with both article
embeddings and compatible local return histories. Return alignment leaves
\ppnum[0]{n-obs} common daily observations.
Each article carries a creation date, URL hash, firm identifier, and a
4,096-dimensional Qwen3-Embedding-8B embedding. The representation normalizes
every row to unit Euclidean norm, and the ground cost is Euclidean chord
distance.

The registry distinguishes transport order from ground-cost exponent. The W2
entries use balanced quadratic transport, rooted normalization, and ground
exponent one. The persisted artifact is rooted W2; Paper 3 squares it once at
the certificate boundary, with regression tests enforcing that transition.

Balanced assignment currently requires equal empirical cloud sizes. Rows are
first restricted by article date and then ordered by URL hash. At each annual
cutoff, the governed artifact registry fixes a common cloud size supported by
every firm. This rule retains all \ppnum[0]{n-tickers} priced firms rather than
deleting thin-coverage firms; the early vintages consequently use fewer
articles per firm and should be read as noisier geometries.

The canonical priced W$_2$ roster used by the certificate and its diagnostics
is listed in \Cref{tab:sample-universe}.

\input{generated/sample_universe}

\section{Supplementary empirical results}
\label{sec:p3-supplementary-results}

The main text reports the conventional benchmark values in
\Cref{tab:news-variance-benchmark}.
\Cref{fig:news-variance-benchmark} provides the corresponding visual comparison.

\begin{figure}[H]
  \centering
  \import{images/}{news_variance_benchmark\ppfiguresuffix.pgf}
  \caption{Standardized in-sample variance relative to the full-sample
    long-only GMV benchmark. The news-only allocation is a descriptive
  zero-slack maximizer of the news certificate.}
  \label{fig:news-variance-benchmark}
\end{figure}

\subsection{Complete embedding-representation sensitivity}
\label{sec:p3-representation-ablation}

The sensitivity tables hold the priced firms, standardized-return covariance,
reference laws, caps, and bootstrap scheme fixed while changing the
representation.
For representation $r$, let
$I_r=100\,V(q_r)/V(q_{\mathrm{GMV}})$ denote its relative-GMV index.
Each paired estimand is $\Delta_{c,r}=I_c-I_r$ in index points, so a negative
contrast favours the candidate allocation on conventional variance.
The common return bootstrap re-standardizes returns and recomputes the GMV
denominator in each draw.
Consequently, a dash means that the news-only allocation violates the fixed
law's cap; the cap is not relaxed to manufacture a percentile.

\input{generated/representation_ablation}

The full-width-to-1,024 Qwen3-8B contrast includes zero, so moderate truncation
does not reject zero difference in relative variance.
At fixed 1,024 width, the Qwen3-4B comparison and the 64-coordinate
Qwen3-8B stress-cell comparison with 256 coordinates do not survive Holm
adjustment.
Several ChronoBERT annual differences remain distinguishable from zero after
vintage-family Holm adjustment, including both the first annual step and the
2017-to-2024 endpoint.
Only a subset of annual pairs satisfy the benchmark-calibrated,
post-specified equivalence bound.
The annual point indices are nonmonotone, so neither a null vintage effect nor
broad invariance follows from using a common encoder architecture.

%% file: generated/sample_universe.tex

\begin{longtable}{lll}
\caption{Canonical empirical roster: ticker symbol, firm name, and sector label from Nasdaq summary metadata for the firms in the canonical priced $W_2$ geometry, sorted by sector then symbol.}
\label{tab:sample-universe} \\
\toprule
Symbol & Name & Sector \\
\midrule
\endfirsthead
\multicolumn{3}{c}{\tablename\ \thetable\ (continued)} \\
\toprule
Symbol & Name & Sector \\
\midrule
\endhead
\midrule
\endfoot
\bottomrule
\endlastfoot
BIDU & Baidu, Inc. & Communication Services \\
CHTR & Charter Communications, Inc. & Communication Services \\
CMCSA & Comcast Corporation & Communication Services \\
EA & Electronic Arts Inc. & Communication Services \\
GOOG & Alphabet Inc. & Communication Services \\
MTCH & Match Group Inc. & Communication Services \\
NFLX & Netflix Inc. & Communication Services \\
NTES & NetEase, Inc. & Communication Services \\
SIRI & Sirius XM Holdings Inc. & Communication Services \\
TMUS & T-Mobile US Inc. & Communication Services \\
TTWO & Take-Two Interactive Software, Inc. & Communication Services \\
BKNG & Booking Holdings Inc. & Consumer Cyclical \\
EBAY & eBay Inc. & Consumer Cyclical \\
EXPE & Expedia Group Inc. & Consumer Cyclical \\
HAS & Hasbro Inc. & Consumer Cyclical \\
JD & JD.com Inc. & Consumer Cyclical \\
LULU & Lululemon Athletica Inc. & Consumer Cyclical \\
MAR & Marriott International Inc. & Consumer Cyclical \\
MAT & Mattel Inc. & Consumer Cyclical \\
MELI & MercadoLibre Inc. & Consumer Cyclical \\
NCLH & Norwegian Cruise Line Holdings Ltd. & Consumer Cyclical \\
ORLY & O'Reilly Automotive Inc. & Consumer Cyclical \\
ROST & Ross Stores, Inc. & Consumer Cyclical \\
SBUX & Starbucks Corporation & Consumer Cyclical \\
TCOM & Trip.com Group Limited & Consumer Cyclical \\
TRIP & Tripadvisor, Inc. & Consumer Cyclical \\
TSCO & Tractor Supply Company & Consumer Cyclical \\
ULTA & Ulta Beauty Inc. & Consumer Cyclical \\
WYNN & Wynn Resorts Limited & Consumer Cyclical \\
COST & Costco Wholesale Corporation & Consumer Defensive \\
DLTR & Dollar Tree Inc. & Consumer Defensive \\
KDP & Keurig Dr Pepper Inc. & Consumer Defensive \\
KHC & The Kraft Heinz Company & Consumer Defensive \\
MDLZ & Mondelez International, Inc. & Consumer Defensive \\
MNST & Monster Beverage Corporation & Consumer Defensive \\
PEP & PepsiCo Inc. & Consumer Defensive \\
FANG & Diamondback Energy, Inc. & Energy \\
PYPL & PayPal Holdings Inc. & Financial Services \\
ALGN & Align Technology Inc. & Healthcare \\
AMGN & Amgen Inc. & Healthcare \\
AZN & AstraZeneca PLC & Healthcare \\
BIIB & Biogen Inc. & Healthcare \\
BMRN & BioMarin Pharmaceutical Inc. & Healthcare \\
DXCM & DexCom, Inc. & Healthcare \\
GILD & Gilead Sciences Inc. & Healthcare \\
HOLX & Hologic, Inc. & Healthcare \\
IDXX & IDEXX Laboratories, Inc. & Healthcare \\
ILMN & Illumina, Inc. & Healthcare \\
INCY & Incyte Corporation & Healthcare \\
ISRG & Intuitive Surgical Inc. & Healthcare \\
REGN & Regeneron Pharmaceuticals Inc. & Healthcare \\
VRTX & Vertex Pharmaceuticals Incorporated & Healthcare \\
AAL & American Airlines Group Inc. & Industrials \\
CSX & CSX Corporation & Industrials \\
FAST & Fastenal Company & Industrials \\
HON & Honeywell International Inc. & Industrials \\
JBHT & J.B. Hunt Transport Services, Inc. & Industrials \\
PCAR & PACCAR Inc & Industrials \\
UAL & United Airlines Holdings Inc. & Industrials \\
ADBE & Adobe Inc. & Technology \\
ADI & Analog Devices, Inc. & Technology \\
ADP & Automatic Data Processing Inc. & Technology \\
ADSK & Autodesk, Inc. & Technology \\
AKAM & Akamai Technologies, Inc. & Technology \\
AMAT & Applied Materials Inc. & Technology \\
AMD & Advanced Micro Devices Inc. & Technology \\
ASML & ASML Holding N.V. & Technology \\
AVGO & Broadcom Inc. & Technology \\
CDNS & Cadence Design Systems, Inc. & Technology \\
CHKP & Check Point Software Technologies Ltd. & Technology \\
CSCO & Cisco Systems Inc. & Technology \\
CTSH & Cognizant Technology Solutions Corporation & Technology \\
ENPH & Enphase Energy Inc. & Technology \\
FISV & Fiserv, Inc. & Technology \\
FTNT & Fortinet Inc. & Technology \\
INTC & Intel Corporation & Technology \\
INTU & Intuit Inc. & Technology \\
KLAC & KLA Corporation & Technology \\
LRCX & Lam Research Corporation & Technology \\
MCHP & Microchip Technology Incorporated & Technology \\
MRVL & Marvell Technology Inc. & Technology \\
MU & Micron Technology Inc. & Technology \\
NTAP & NetApp, Inc. & Technology \\
NVDA & NVIDIA Corporation & Technology \\
NXPI & NXP Semiconductors N.V. & Technology \\
OKTA & Okta, Inc. & Technology \\
PANW & Palo Alto Networks Inc. & Technology \\
PAYX & Paychex, Inc. & Technology \\
QCOM & QUALCOMM Incorporated & Technology \\
SNPS & Synopsys, Inc. & Technology \\
STX & Seagate Technology Holdings plc & Technology \\
SWKS & Skyworks Solutions, Inc. & Technology \\
TEAM & Atlassian Corporation & Technology \\
TXN & Texas Instruments Incorporated & Technology \\
WDAY & Workday Inc. & Technology \\
WDC & Western Digital Corporation & Technology \\
ZS & Zscaler Inc. & Technology \\
AEP & American Electric Power Company, Inc. & Utilities \\
EXC & Exelon Corporation & Utilities \\
XEL & Xcel Energy Inc. & Utilities \\
\end{longtable}

%% file: images/news_variance_benchmark.pgf
\begingroup%
\makeatletter%
\begin{pgfpicture}%
\pgfpathrectangle{\pgfpointorigin}{\pgfqpoint{4.251549in}{2.767046in}}%
\pgfusepath{use as bounding box, clip}%
\begin{pgfscope}%
\pgfsetbuttcap%
\pgfsetmiterjoin%
\definecolor{currentfill}{rgb}{1.000000,1.000000,1.000000}%
\pgfsetfillcolor{currentfill}%
\pgfsetlinewidth{0.000000pt}%
\definecolor{currentstroke}{rgb}{1.000000,1.000000,1.000000}%
\pgfsetstrokecolor{currentstroke}%
\pgfsetdash{}{0pt}%
\pgfpathmoveto{\pgfqpoint{0.000000in}{0.000000in}}%
\pgfpathlineto{\pgfqpoint{4.251549in}{0.000000in}}%
\pgfpathlineto{\pgfqpoint{4.251549in}{2.767046in}}%
\pgfpathlineto{\pgfqpoint{0.000000in}{2.767046in}}%
\pgfpathlineto{\pgfqpoint{0.000000in}{0.000000in}}%
\pgfpathclose%
\pgfusepath{fill}%
\end{pgfscope}%
\begin{pgfscope}%
\pgfsetbuttcap%
\pgfsetmiterjoin%
\definecolor{currentfill}{rgb}{1.000000,1.000000,1.000000}%
\pgfsetfillcolor{currentfill}%
\pgfsetlinewidth{0.000000pt}%
\definecolor{currentstroke}{rgb}{0.000000,0.000000,0.000000}%
\pgfsetstrokecolor{currentstroke}%
\pgfsetstrokeopacity{0.000000}%
\pgfsetdash{}{0pt}%
\pgfpathmoveto{\pgfqpoint{0.602522in}{0.179722in}}%
\pgfpathlineto{\pgfqpoint{4.231549in}{0.179722in}}%
\pgfpathlineto{\pgfqpoint{4.231549in}{2.747046in}}%
\pgfpathlineto{\pgfqpoint{0.602522in}{2.747046in}}%
\pgfpathlineto{\pgfqpoint{0.602522in}{0.179722in}}%
\pgfpathclose%
\pgfusepath{fill}%
\end{pgfscope}%
\begin{pgfscope}%
\pgfpathrectangle{\pgfqpoint{0.602522in}{0.179722in}}{\pgfqpoint{3.629027in}{2.567324in}}%
\pgfusepath{clip}%
\pgfsetbuttcap%
\pgfsetmiterjoin%
\definecolor{currentfill}{rgb}{0.000000,0.447059,0.698039}%
\pgfsetfillcolor{currentfill}%
\pgfsetlinewidth{0.351312pt}%
\definecolor{currentstroke}{rgb}{0.000000,0.000000,0.000000}%
\pgfsetstrokecolor{currentstroke}%
\pgfsetdash{}{0pt}%
\pgfpathmoveto{\pgfqpoint{0.767478in}{0.179722in}}%
\pgfpathlineto{\pgfqpoint{1.710082in}{0.179722in}}%
\pgfpathlineto{\pgfqpoint{1.710082in}{2.153397in}}%
\pgfpathlineto{\pgfqpoint{0.767478in}{2.153397in}}%
\pgfpathlineto{\pgfqpoint{0.767478in}{0.179722in}}%
\pgfpathclose%
\pgfusepath{stroke,fill}%
\end{pgfscope}%
\begin{pgfscope}%
\pgfpathrectangle{\pgfqpoint{0.602522in}{0.179722in}}{\pgfqpoint{3.629027in}{2.567324in}}%
\pgfusepath{clip}%
\pgfsetbuttcap%
\pgfsetmiterjoin%
\definecolor{currentfill}{rgb}{0.901961,0.623529,0.000000}%
\pgfsetfillcolor{currentfill}%
\pgfsetlinewidth{0.351312pt}%
\definecolor{currentstroke}{rgb}{0.000000,0.000000,0.000000}%
\pgfsetstrokecolor{currentstroke}%
\pgfsetdash{}{0pt}%
\pgfpathmoveto{\pgfqpoint{1.945733in}{0.179722in}}%
\pgfpathlineto{\pgfqpoint{2.888338in}{0.179722in}}%
\pgfpathlineto{\pgfqpoint{2.888338in}{2.355421in}}%
\pgfpathlineto{\pgfqpoint{1.945733in}{2.355421in}}%
\pgfpathlineto{\pgfqpoint{1.945733in}{0.179722in}}%
\pgfpathclose%
\pgfusepath{stroke,fill}%
\end{pgfscope}%
\begin{pgfscope}%
\pgfpathrectangle{\pgfqpoint{0.602522in}{0.179722in}}{\pgfqpoint{3.629027in}{2.567324in}}%
\pgfusepath{clip}%
\pgfsetbuttcap%
\pgfsetmiterjoin%
\definecolor{currentfill}{rgb}{0.000000,0.619608,0.450980}%
\pgfsetfillcolor{currentfill}%
\pgfsetlinewidth{0.351312pt}%
\definecolor{currentstroke}{rgb}{0.000000,0.000000,0.000000}%
\pgfsetstrokecolor{currentstroke}%
\pgfsetdash{}{0pt}%
\pgfpathmoveto{\pgfqpoint{3.123989in}{0.179722in}}%
\pgfpathlineto{\pgfqpoint{4.066593in}{0.179722in}}%
\pgfpathlineto{\pgfqpoint{4.066593in}{1.517733in}}%
\pgfpathlineto{\pgfqpoint{3.123989in}{1.517733in}}%
\pgfpathlineto{\pgfqpoint{3.123989in}{0.179722in}}%
\pgfpathclose%
\pgfusepath{stroke,fill}%
\end{pgfscope}%
\begin{pgfscope}%
\pgfsetbuttcap%
\pgfsetroundjoin%
\definecolor{currentfill}{rgb}{0.000000,0.000000,0.000000}%
\pgfsetfillcolor{currentfill}%
\pgfsetlinewidth{0.501875pt}%
\definecolor{currentstroke}{rgb}{0.000000,0.000000,0.000000}%
\pgfsetstrokecolor{currentstroke}%
\pgfsetdash{}{0pt}%
\pgfsys@defobject{currentmarker}{\pgfqpoint{0.000000in}{0.000000in}}{\pgfqpoint{0.000000in}{0.041667in}}{%
\pgfpathmoveto{\pgfqpoint{0.000000in}{0.000000in}}%
\pgfpathlineto{\pgfqpoint{0.000000in}{0.041667in}}%
\pgfusepath{stroke,fill}%
}%
\begin{pgfscope}%
\pgfsys@transformshift{1.238780in}{0.179722in}%
\pgfsys@useobject{currentmarker}{}%
\end{pgfscope}%
\end{pgfscope}%
\begin{pgfscope}%
\definecolor{textcolor}{rgb}{0.000000,0.000000,0.000000}%
\pgfsetstrokecolor{textcolor}%
\pgfsetfillcolor{textcolor}%
\pgftext[x=1.238780in,y=0.131111in,,top]{\color{textcolor}{\rmfamily\fontsize{8.000000}{9.600000}\selectfont\catcode`\^=\active\def^{\ifmmode\sp\else\^{}\fi}\catcode`\%=\active\def
\end{pgfscope}%
\begin{pgfscope}%
\pgfsetbuttcap%
\pgfsetroundjoin%
\definecolor{currentfill}{rgb}{0.000000,0.000000,0.000000}%
\pgfsetfillcolor{currentfill}%
\pgfsetlinewidth{0.501875pt}%
\definecolor{currentstroke}{rgb}{0.000000,0.000000,0.000000}%
\pgfsetstrokecolor{currentstroke}%
\pgfsetdash{}{0pt}%
\pgfsys@defobject{currentmarker}{\pgfqpoint{0.000000in}{0.000000in}}{\pgfqpoint{0.000000in}{0.041667in}}{%
\pgfpathmoveto{\pgfqpoint{0.000000in}{0.000000in}}%
\pgfpathlineto{\pgfqpoint{0.000000in}{0.041667in}}%
\pgfusepath{stroke,fill}%
}%
\begin{pgfscope}%
\pgfsys@transformshift{2.417036in}{0.179722in}%
\pgfsys@useobject{currentmarker}{}%
\end{pgfscope}%
\end{pgfscope}%
\begin{pgfscope}%
\definecolor{textcolor}{rgb}{0.000000,0.000000,0.000000}%
\pgfsetstrokecolor{textcolor}%
\pgfsetfillcolor{textcolor}%
\pgftext[x=2.417036in,y=0.131111in,,top]{\color{textcolor}{\rmfamily\fontsize{8.000000}{9.600000}\selectfont\catcode`\^=\active\def^{\ifmmode\sp\else\^{}\fi}\catcode`\%=\active\def
\end{pgfscope}%
\begin{pgfscope}%
\pgfsetbuttcap%
\pgfsetroundjoin%
\definecolor{currentfill}{rgb}{0.000000,0.000000,0.000000}%
\pgfsetfillcolor{currentfill}%
\pgfsetlinewidth{0.501875pt}%
\definecolor{currentstroke}{rgb}{0.000000,0.000000,0.000000}%
\pgfsetstrokecolor{currentstroke}%
\pgfsetdash{}{0pt}%
\pgfsys@defobject{currentmarker}{\pgfqpoint{0.000000in}{0.000000in}}{\pgfqpoint{0.000000in}{0.041667in}}{%
\pgfpathmoveto{\pgfqpoint{0.000000in}{0.000000in}}%
\pgfpathlineto{\pgfqpoint{0.000000in}{0.041667in}}%
\pgfusepath{stroke,fill}%
}%
\begin{pgfscope}%
\pgfsys@transformshift{3.595291in}{0.179722in}%
\pgfsys@useobject{currentmarker}{}%
\end{pgfscope}%
\end{pgfscope}%
\begin{pgfscope}%
\definecolor{textcolor}{rgb}{0.000000,0.000000,0.000000}%
\pgfsetstrokecolor{textcolor}%
\pgfsetfillcolor{textcolor}%
\pgftext[x=3.595291in,y=0.131111in,,top]{\color{textcolor}{\rmfamily\fontsize{8.000000}{9.600000}\selectfont\catcode`\^=\active\def^{\ifmmode\sp\else\^{}\fi}\catcode`\%=\active\def
\end{pgfscope}%
\begin{pgfscope}%
\pgfpathrectangle{\pgfqpoint{0.602522in}{0.179722in}}{\pgfqpoint{3.629027in}{2.567324in}}%
\pgfusepath{clip}%
\pgfsetbuttcap%
\pgfsetroundjoin%
\pgfsetlinewidth{0.803000pt}%
\definecolor{currentstroke}{rgb}{0.690196,0.690196,0.690196}%
\pgfsetstrokecolor{currentstroke}%
\pgfsetstrokeopacity{0.450000}%
\pgfsetdash{{0.800000pt}{1.320000pt}}{0.000000pt}%
\pgfpathmoveto{\pgfqpoint{0.602522in}{0.179722in}}%
\pgfpathlineto{\pgfqpoint{4.231549in}{0.179722in}}%
\pgfusepath{stroke}%
\end{pgfscope}%
\begin{pgfscope}%
\pgfsetbuttcap%
\pgfsetroundjoin%
\definecolor{currentfill}{rgb}{0.000000,0.000000,0.000000}%
\pgfsetfillcolor{currentfill}%
\pgfsetlinewidth{0.501875pt}%
\definecolor{currentstroke}{rgb}{0.000000,0.000000,0.000000}%
\pgfsetstrokecolor{currentstroke}%
\pgfsetdash{}{0pt}%
\pgfsys@defobject{currentmarker}{\pgfqpoint{0.000000in}{0.000000in}}{\pgfqpoint{0.041667in}{0.000000in}}{%
\pgfpathmoveto{\pgfqpoint{0.000000in}{0.000000in}}%
\pgfpathlineto{\pgfqpoint{0.041667in}{0.000000in}}%
\pgfusepath{stroke,fill}%
}%
\begin{pgfscope}%
\pgfsys@transformshift{0.602522in}{0.179722in}%
\pgfsys@useobject{currentmarker}{}%
\end{pgfscope}%
\end{pgfscope}%
\begin{pgfscope}%
\definecolor{textcolor}{rgb}{0.000000,0.000000,0.000000}%
\pgfsetstrokecolor{textcolor}%
\pgfsetfillcolor{textcolor}%
\pgftext[x=0.494882in, y=0.137513in, left, base]{\color{textcolor}{\rmfamily\fontsize{8.000000}{9.600000}\selectfont\catcode`\^=\active\def^{\ifmmode\sp\else\^{}\fi}\catcode`\%=\active\def
\end{pgfscope}%
\begin{pgfscope}%
\pgfpathrectangle{\pgfqpoint{0.602522in}{0.179722in}}{\pgfqpoint{3.629027in}{2.567324in}}%
\pgfusepath{clip}%
\pgfsetbuttcap%
\pgfsetroundjoin%
\pgfsetlinewidth{0.803000pt}%
\definecolor{currentstroke}{rgb}{0.690196,0.690196,0.690196}%
\pgfsetstrokecolor{currentstroke}%
\pgfsetstrokeopacity{0.450000}%
\pgfsetdash{{0.800000pt}{1.320000pt}}{0.000000pt}%
\pgfpathmoveto{\pgfqpoint{0.602522in}{0.514225in}}%
\pgfpathlineto{\pgfqpoint{4.231549in}{0.514225in}}%
\pgfusepath{stroke}%
\end{pgfscope}%
\begin{pgfscope}%
\pgfsetbuttcap%
\pgfsetroundjoin%
\definecolor{currentfill}{rgb}{0.000000,0.000000,0.000000}%
\pgfsetfillcolor{currentfill}%
\pgfsetlinewidth{0.501875pt}%
\definecolor{currentstroke}{rgb}{0.000000,0.000000,0.000000}%
\pgfsetstrokecolor{currentstroke}%
\pgfsetdash{}{0pt}%
\pgfsys@defobject{currentmarker}{\pgfqpoint{0.000000in}{0.000000in}}{\pgfqpoint{0.041667in}{0.000000in}}{%
\pgfpathmoveto{\pgfqpoint{0.000000in}{0.000000in}}%
\pgfpathlineto{\pgfqpoint{0.041667in}{0.000000in}}%
\pgfusepath{stroke,fill}%
}%
\begin{pgfscope}%
\pgfsys@transformshift{0.602522in}{0.514225in}%
\pgfsys@useobject{currentmarker}{}%
\end{pgfscope}%
\end{pgfscope}%
\begin{pgfscope}%
\definecolor{textcolor}{rgb}{0.000000,0.000000,0.000000}%
\pgfsetstrokecolor{textcolor}%
\pgfsetfillcolor{textcolor}%
\pgftext[x=0.435854in, y=0.472016in, left, base]{\color{textcolor}{\rmfamily\fontsize{8.000000}{9.600000}\selectfont\catcode`\^=\active\def^{\ifmmode\sp\else\^{}\fi}\catcode`\%=\active\def
\end{pgfscope}%
\begin{pgfscope}%
\pgfpathrectangle{\pgfqpoint{0.602522in}{0.179722in}}{\pgfqpoint{3.629027in}{2.567324in}}%
\pgfusepath{clip}%
\pgfsetbuttcap%
\pgfsetroundjoin%
\pgfsetlinewidth{0.803000pt}%
\definecolor{currentstroke}{rgb}{0.690196,0.690196,0.690196}%
\pgfsetstrokecolor{currentstroke}%
\pgfsetstrokeopacity{0.450000}%
\pgfsetdash{{0.800000pt}{1.320000pt}}{0.000000pt}%
\pgfpathmoveto{\pgfqpoint{0.602522in}{0.848727in}}%
\pgfpathlineto{\pgfqpoint{4.231549in}{0.848727in}}%
\pgfusepath{stroke}%
\end{pgfscope}%
\begin{pgfscope}%
\pgfsetbuttcap%
\pgfsetroundjoin%
\definecolor{currentfill}{rgb}{0.000000,0.000000,0.000000}%
\pgfsetfillcolor{currentfill}%
\pgfsetlinewidth{0.501875pt}%
\definecolor{currentstroke}{rgb}{0.000000,0.000000,0.000000}%
\pgfsetstrokecolor{currentstroke}%
\pgfsetdash{}{0pt}%
\pgfsys@defobject{currentmarker}{\pgfqpoint{0.000000in}{0.000000in}}{\pgfqpoint{0.041667in}{0.000000in}}{%
\pgfpathmoveto{\pgfqpoint{0.000000in}{0.000000in}}%
\pgfpathlineto{\pgfqpoint{0.041667in}{0.000000in}}%
\pgfusepath{stroke,fill}%
}%
\begin{pgfscope}%
\pgfsys@transformshift{0.602522in}{0.848727in}%
\pgfsys@useobject{currentmarker}{}%
\end{pgfscope}%
\end{pgfscope}%
\begin{pgfscope}%
\definecolor{textcolor}{rgb}{0.000000,0.000000,0.000000}%
\pgfsetstrokecolor{textcolor}%
\pgfsetfillcolor{textcolor}%
\pgftext[x=0.435854in, y=0.806518in, left, base]{\color{textcolor}{\rmfamily\fontsize{8.000000}{9.600000}\selectfont\catcode`\^=\active\def^{\ifmmode\sp\else\^{}\fi}\catcode`\%=\active\def
\end{pgfscope}%
\begin{pgfscope}%
\pgfpathrectangle{\pgfqpoint{0.602522in}{0.179722in}}{\pgfqpoint{3.629027in}{2.567324in}}%
\pgfusepath{clip}%
\pgfsetbuttcap%
\pgfsetroundjoin%
\pgfsetlinewidth{0.803000pt}%
\definecolor{currentstroke}{rgb}{0.690196,0.690196,0.690196}%
\pgfsetstrokecolor{currentstroke}%
\pgfsetstrokeopacity{0.450000}%
\pgfsetdash{{0.800000pt}{1.320000pt}}{0.000000pt}%
\pgfpathmoveto{\pgfqpoint{0.602522in}{1.183230in}}%
\pgfpathlineto{\pgfqpoint{4.231549in}{1.183230in}}%
\pgfusepath{stroke}%
\end{pgfscope}%
\begin{pgfscope}%
\pgfsetbuttcap%
\pgfsetroundjoin%
\definecolor{currentfill}{rgb}{0.000000,0.000000,0.000000}%
\pgfsetfillcolor{currentfill}%
\pgfsetlinewidth{0.501875pt}%
\definecolor{currentstroke}{rgb}{0.000000,0.000000,0.000000}%
\pgfsetstrokecolor{currentstroke}%
\pgfsetdash{}{0pt}%
\pgfsys@defobject{currentmarker}{\pgfqpoint{0.000000in}{0.000000in}}{\pgfqpoint{0.041667in}{0.000000in}}{%
\pgfpathmoveto{\pgfqpoint{0.000000in}{0.000000in}}%
\pgfpathlineto{\pgfqpoint{0.041667in}{0.000000in}}%
\pgfusepath{stroke,fill}%
}%
\begin{pgfscope}%
\pgfsys@transformshift{0.602522in}{1.183230in}%
\pgfsys@useobject{currentmarker}{}%
\end{pgfscope}%
\end{pgfscope}%
\begin{pgfscope}%
\definecolor{textcolor}{rgb}{0.000000,0.000000,0.000000}%
\pgfsetstrokecolor{textcolor}%
\pgfsetfillcolor{textcolor}%
\pgftext[x=0.435854in, y=1.141021in, left, base]{\color{textcolor}{\rmfamily\fontsize{8.000000}{9.600000}\selectfont\catcode`\^=\active\def^{\ifmmode\sp\else\^{}\fi}\catcode`\%=\active\def
\end{pgfscope}%
\begin{pgfscope}%
\pgfpathrectangle{\pgfqpoint{0.602522in}{0.179722in}}{\pgfqpoint{3.629027in}{2.567324in}}%
\pgfusepath{clip}%
\pgfsetbuttcap%
\pgfsetroundjoin%
\pgfsetlinewidth{0.803000pt}%
\definecolor{currentstroke}{rgb}{0.690196,0.690196,0.690196}%
\pgfsetstrokecolor{currentstroke}%
\pgfsetstrokeopacity{0.450000}%
\pgfsetdash{{0.800000pt}{1.320000pt}}{0.000000pt}%
\pgfpathmoveto{\pgfqpoint{0.602522in}{1.517733in}}%
\pgfpathlineto{\pgfqpoint{4.231549in}{1.517733in}}%
\pgfusepath{stroke}%
\end{pgfscope}%
\begin{pgfscope}%
\pgfsetbuttcap%
\pgfsetroundjoin%
\definecolor{currentfill}{rgb}{0.000000,0.000000,0.000000}%
\pgfsetfillcolor{currentfill}%
\pgfsetlinewidth{0.501875pt}%
\definecolor{currentstroke}{rgb}{0.000000,0.000000,0.000000}%
\pgfsetstrokecolor{currentstroke}%
\pgfsetdash{}{0pt}%
\pgfsys@defobject{currentmarker}{\pgfqpoint{0.000000in}{0.000000in}}{\pgfqpoint{0.041667in}{0.000000in}}{%
\pgfpathmoveto{\pgfqpoint{0.000000in}{0.000000in}}%
\pgfpathlineto{\pgfqpoint{0.041667in}{0.000000in}}%
\pgfusepath{stroke,fill}%
}%
\begin{pgfscope}%
\pgfsys@transformshift{0.602522in}{1.517733in}%
\pgfsys@useobject{currentmarker}{}%
\end{pgfscope}%
\end{pgfscope}%
\begin{pgfscope}%
\definecolor{textcolor}{rgb}{0.000000,0.000000,0.000000}%
\pgfsetstrokecolor{textcolor}%
\pgfsetfillcolor{textcolor}%
\pgftext[x=0.376825in, y=1.475524in, left, base]{\color{textcolor}{\rmfamily\fontsize{8.000000}{9.600000}\selectfont\catcode`\^=\active\def^{\ifmmode\sp\else\^{}\fi}\catcode`\%=\active\def
\end{pgfscope}%
\begin{pgfscope}%
\pgfpathrectangle{\pgfqpoint{0.602522in}{0.179722in}}{\pgfqpoint{3.629027in}{2.567324in}}%
\pgfusepath{clip}%
\pgfsetbuttcap%
\pgfsetroundjoin%
\pgfsetlinewidth{0.803000pt}%
\definecolor{currentstroke}{rgb}{0.690196,0.690196,0.690196}%
\pgfsetstrokecolor{currentstroke}%
\pgfsetstrokeopacity{0.450000}%
\pgfsetdash{{0.800000pt}{1.320000pt}}{0.000000pt}%
\pgfpathmoveto{\pgfqpoint{0.602522in}{1.852235in}}%
\pgfpathlineto{\pgfqpoint{4.231549in}{1.852235in}}%
\pgfusepath{stroke}%
\end{pgfscope}%
\begin{pgfscope}%
\pgfsetbuttcap%
\pgfsetroundjoin%
\definecolor{currentfill}{rgb}{0.000000,0.000000,0.000000}%
\pgfsetfillcolor{currentfill}%
\pgfsetlinewidth{0.501875pt}%
\definecolor{currentstroke}{rgb}{0.000000,0.000000,0.000000}%
\pgfsetstrokecolor{currentstroke}%
\pgfsetdash{}{0pt}%
\pgfsys@defobject{currentmarker}{\pgfqpoint{0.000000in}{0.000000in}}{\pgfqpoint{0.041667in}{0.000000in}}{%
\pgfpathmoveto{\pgfqpoint{0.000000in}{0.000000in}}%
\pgfpathlineto{\pgfqpoint{0.041667in}{0.000000in}}%
\pgfusepath{stroke,fill}%
}%
\begin{pgfscope}%
\pgfsys@transformshift{0.602522in}{1.852235in}%
\pgfsys@useobject{currentmarker}{}%
\end{pgfscope}%
\end{pgfscope}%
\begin{pgfscope}%
\definecolor{textcolor}{rgb}{0.000000,0.000000,0.000000}%
\pgfsetstrokecolor{textcolor}%
\pgfsetfillcolor{textcolor}%
\pgftext[x=0.376825in, y=1.810026in, left, base]{\color{textcolor}{\rmfamily\fontsize{8.000000}{9.600000}\selectfont\catcode`\^=\active\def^{\ifmmode\sp\else\^{}\fi}\catcode`\%=\active\def
\end{pgfscope}%
\begin{pgfscope}%
\pgfpathrectangle{\pgfqpoint{0.602522in}{0.179722in}}{\pgfqpoint{3.629027in}{2.567324in}}%
\pgfusepath{clip}%
\pgfsetbuttcap%
\pgfsetroundjoin%
\pgfsetlinewidth{0.803000pt}%
\definecolor{currentstroke}{rgb}{0.690196,0.690196,0.690196}%
\pgfsetstrokecolor{currentstroke}%
\pgfsetstrokeopacity{0.450000}%
\pgfsetdash{{0.800000pt}{1.320000pt}}{0.000000pt}%
\pgfpathmoveto{\pgfqpoint{0.602522in}{2.186738in}}%
\pgfpathlineto{\pgfqpoint{4.231549in}{2.186738in}}%
\pgfusepath{stroke}%
\end{pgfscope}%
\begin{pgfscope}%
\pgfsetbuttcap%
\pgfsetroundjoin%
\definecolor{currentfill}{rgb}{0.000000,0.000000,0.000000}%
\pgfsetfillcolor{currentfill}%
\pgfsetlinewidth{0.501875pt}%
\definecolor{currentstroke}{rgb}{0.000000,0.000000,0.000000}%
\pgfsetstrokecolor{currentstroke}%
\pgfsetdash{}{0pt}%
\pgfsys@defobject{currentmarker}{\pgfqpoint{0.000000in}{0.000000in}}{\pgfqpoint{0.041667in}{0.000000in}}{%
\pgfpathmoveto{\pgfqpoint{0.000000in}{0.000000in}}%
\pgfpathlineto{\pgfqpoint{0.041667in}{0.000000in}}%
\pgfusepath{stroke,fill}%
}%
\begin{pgfscope}%
\pgfsys@transformshift{0.602522in}{2.186738in}%
\pgfsys@useobject{currentmarker}{}%
\end{pgfscope}%
\end{pgfscope}%
\begin{pgfscope}%
\definecolor{textcolor}{rgb}{0.000000,0.000000,0.000000}%
\pgfsetstrokecolor{textcolor}%
\pgfsetfillcolor{textcolor}%
\pgftext[x=0.376825in, y=2.144529in, left, base]{\color{textcolor}{\rmfamily\fontsize{8.000000}{9.600000}\selectfont\catcode`\^=\active\def^{\ifmmode\sp\else\^{}\fi}\catcode`\%=\active\def
\end{pgfscope}%
\begin{pgfscope}%
\pgfpathrectangle{\pgfqpoint{0.602522in}{0.179722in}}{\pgfqpoint{3.629027in}{2.567324in}}%
\pgfusepath{clip}%
\pgfsetbuttcap%
\pgfsetroundjoin%
\pgfsetlinewidth{0.803000pt}%
\definecolor{currentstroke}{rgb}{0.690196,0.690196,0.690196}%
\pgfsetstrokecolor{currentstroke}%
\pgfsetstrokeopacity{0.450000}%
\pgfsetdash{{0.800000pt}{1.320000pt}}{0.000000pt}%
\pgfpathmoveto{\pgfqpoint{0.602522in}{2.521241in}}%
\pgfpathlineto{\pgfqpoint{4.231549in}{2.521241in}}%
\pgfusepath{stroke}%
\end{pgfscope}%
\begin{pgfscope}%
\pgfsetbuttcap%
\pgfsetroundjoin%
\definecolor{currentfill}{rgb}{0.000000,0.000000,0.000000}%
\pgfsetfillcolor{currentfill}%
\pgfsetlinewidth{0.501875pt}%
\definecolor{currentstroke}{rgb}{0.000000,0.000000,0.000000}%
\pgfsetstrokecolor{currentstroke}%
\pgfsetdash{}{0pt}%
\pgfsys@defobject{currentmarker}{\pgfqpoint{0.000000in}{0.000000in}}{\pgfqpoint{0.041667in}{0.000000in}}{%
\pgfpathmoveto{\pgfqpoint{0.000000in}{0.000000in}}%
\pgfpathlineto{\pgfqpoint{0.041667in}{0.000000in}}%
\pgfusepath{stroke,fill}%
}%
\begin{pgfscope}%
\pgfsys@transformshift{0.602522in}{2.521241in}%
\pgfsys@useobject{currentmarker}{}%
\end{pgfscope}%
\end{pgfscope}%
\begin{pgfscope}%
\definecolor{textcolor}{rgb}{0.000000,0.000000,0.000000}%
\pgfsetstrokecolor{textcolor}%
\pgfsetfillcolor{textcolor}%
\pgftext[x=0.376825in, y=2.479031in, left, base]{\color{textcolor}{\rmfamily\fontsize{8.000000}{9.600000}\selectfont\catcode`\^=\active\def^{\ifmmode\sp\else\^{}\fi}\catcode`\%=\active\def
\end{pgfscope}%
\begin{pgfscope}%
\pgfsetbuttcap%
\pgfsetroundjoin%
\definecolor{currentfill}{rgb}{0.000000,0.000000,0.000000}%
\pgfsetfillcolor{currentfill}%
\pgfsetlinewidth{0.501875pt}%
\definecolor{currentstroke}{rgb}{0.000000,0.000000,0.000000}%
\pgfsetstrokecolor{currentstroke}%
\pgfsetdash{}{0pt}%
\pgfsys@defobject{currentmarker}{\pgfqpoint{0.000000in}{0.000000in}}{\pgfqpoint{0.020833in}{0.000000in}}{%
\pgfpathmoveto{\pgfqpoint{0.000000in}{0.000000in}}%
\pgfpathlineto{\pgfqpoint{0.020833in}{0.000000in}}%
\pgfusepath{stroke,fill}%
}%
\begin{pgfscope}%
\pgfsys@transformshift{0.602522in}{0.246623in}%
\pgfsys@useobject{currentmarker}{}%
\end{pgfscope}%
\end{pgfscope}%
\begin{pgfscope}%
\pgfsetbuttcap%
\pgfsetroundjoin%
\definecolor{currentfill}{rgb}{0.000000,0.000000,0.000000}%
\pgfsetfillcolor{currentfill}%
\pgfsetlinewidth{0.501875pt}%
\definecolor{currentstroke}{rgb}{0.000000,0.000000,0.000000}%
\pgfsetstrokecolor{currentstroke}%
\pgfsetdash{}{0pt}%
\pgfsys@defobject{currentmarker}{\pgfqpoint{0.000000in}{0.000000in}}{\pgfqpoint{0.020833in}{0.000000in}}{%
\pgfpathmoveto{\pgfqpoint{0.000000in}{0.000000in}}%
\pgfpathlineto{\pgfqpoint{0.020833in}{0.000000in}}%
\pgfusepath{stroke,fill}%
}%
\begin{pgfscope}%
\pgfsys@transformshift{0.602522in}{0.313523in}%
\pgfsys@useobject{currentmarker}{}%
\end{pgfscope}%
\end{pgfscope}%
\begin{pgfscope}%
\pgfsetbuttcap%
\pgfsetroundjoin%
\definecolor{currentfill}{rgb}{0.000000,0.000000,0.000000}%
\pgfsetfillcolor{currentfill}%
\pgfsetlinewidth{0.501875pt}%
\definecolor{currentstroke}{rgb}{0.000000,0.000000,0.000000}%
\pgfsetstrokecolor{currentstroke}%
\pgfsetdash{}{0pt}%
\pgfsys@defobject{currentmarker}{\pgfqpoint{0.000000in}{0.000000in}}{\pgfqpoint{0.020833in}{0.000000in}}{%
\pgfpathmoveto{\pgfqpoint{0.000000in}{0.000000in}}%
\pgfpathlineto{\pgfqpoint{0.020833in}{0.000000in}}%
\pgfusepath{stroke,fill}%
}%
\begin{pgfscope}%
\pgfsys@transformshift{0.602522in}{0.380424in}%
\pgfsys@useobject{currentmarker}{}%
\end{pgfscope}%
\end{pgfscope}%
\begin{pgfscope}%
\pgfsetbuttcap%
\pgfsetroundjoin%
\definecolor{currentfill}{rgb}{0.000000,0.000000,0.000000}%
\pgfsetfillcolor{currentfill}%
\pgfsetlinewidth{0.501875pt}%
\definecolor{currentstroke}{rgb}{0.000000,0.000000,0.000000}%
\pgfsetstrokecolor{currentstroke}%
\pgfsetdash{}{0pt}%
\pgfsys@defobject{currentmarker}{\pgfqpoint{0.000000in}{0.000000in}}{\pgfqpoint{0.020833in}{0.000000in}}{%
\pgfpathmoveto{\pgfqpoint{0.000000in}{0.000000in}}%
\pgfpathlineto{\pgfqpoint{0.020833in}{0.000000in}}%
\pgfusepath{stroke,fill}%
}%
\begin{pgfscope}%
\pgfsys@transformshift{0.602522in}{0.447324in}%
\pgfsys@useobject{currentmarker}{}%
\end{pgfscope}%
\end{pgfscope}%
\begin{pgfscope}%
\pgfsetbuttcap%
\pgfsetroundjoin%
\definecolor{currentfill}{rgb}{0.000000,0.000000,0.000000}%
\pgfsetfillcolor{currentfill}%
\pgfsetlinewidth{0.501875pt}%
\definecolor{currentstroke}{rgb}{0.000000,0.000000,0.000000}%
\pgfsetstrokecolor{currentstroke}%
\pgfsetdash{}{0pt}%
\pgfsys@defobject{currentmarker}{\pgfqpoint{0.000000in}{0.000000in}}{\pgfqpoint{0.020833in}{0.000000in}}{%
\pgfpathmoveto{\pgfqpoint{0.000000in}{0.000000in}}%
\pgfpathlineto{\pgfqpoint{0.020833in}{0.000000in}}%
\pgfusepath{stroke,fill}%
}%
\begin{pgfscope}%
\pgfsys@transformshift{0.602522in}{0.581125in}%
\pgfsys@useobject{currentmarker}{}%
\end{pgfscope}%
\end{pgfscope}%
\begin{pgfscope}%
\pgfsetbuttcap%
\pgfsetroundjoin%
\definecolor{currentfill}{rgb}{0.000000,0.000000,0.000000}%
\pgfsetfillcolor{currentfill}%
\pgfsetlinewidth{0.501875pt}%
\definecolor{currentstroke}{rgb}{0.000000,0.000000,0.000000}%
\pgfsetstrokecolor{currentstroke}%
\pgfsetdash{}{0pt}%
\pgfsys@defobject{currentmarker}{\pgfqpoint{0.000000in}{0.000000in}}{\pgfqpoint{0.020833in}{0.000000in}}{%
\pgfpathmoveto{\pgfqpoint{0.000000in}{0.000000in}}%
\pgfpathlineto{\pgfqpoint{0.020833in}{0.000000in}}%
\pgfusepath{stroke,fill}%
}%
\begin{pgfscope}%
\pgfsys@transformshift{0.602522in}{0.648026in}%
\pgfsys@useobject{currentmarker}{}%
\end{pgfscope}%
\end{pgfscope}%
\begin{pgfscope}%
\pgfsetbuttcap%
\pgfsetroundjoin%
\definecolor{currentfill}{rgb}{0.000000,0.000000,0.000000}%
\pgfsetfillcolor{currentfill}%
\pgfsetlinewidth{0.501875pt}%
\definecolor{currentstroke}{rgb}{0.000000,0.000000,0.000000}%
\pgfsetstrokecolor{currentstroke}%
\pgfsetdash{}{0pt}%
\pgfsys@defobject{currentmarker}{\pgfqpoint{0.000000in}{0.000000in}}{\pgfqpoint{0.020833in}{0.000000in}}{%
\pgfpathmoveto{\pgfqpoint{0.000000in}{0.000000in}}%
\pgfpathlineto{\pgfqpoint{0.020833in}{0.000000in}}%
\pgfusepath{stroke,fill}%
}%
\begin{pgfscope}%
\pgfsys@transformshift{0.602522in}{0.714926in}%
\pgfsys@useobject{currentmarker}{}%
\end{pgfscope}%
\end{pgfscope}%
\begin{pgfscope}%
\pgfsetbuttcap%
\pgfsetroundjoin%
\definecolor{currentfill}{rgb}{0.000000,0.000000,0.000000}%
\pgfsetfillcolor{currentfill}%
\pgfsetlinewidth{0.501875pt}%
\definecolor{currentstroke}{rgb}{0.000000,0.000000,0.000000}%
\pgfsetstrokecolor{currentstroke}%
\pgfsetdash{}{0pt}%
\pgfsys@defobject{currentmarker}{\pgfqpoint{0.000000in}{0.000000in}}{\pgfqpoint{0.020833in}{0.000000in}}{%
\pgfpathmoveto{\pgfqpoint{0.000000in}{0.000000in}}%
\pgfpathlineto{\pgfqpoint{0.020833in}{0.000000in}}%
\pgfusepath{stroke,fill}%
}%
\begin{pgfscope}%
\pgfsys@transformshift{0.602522in}{0.781827in}%
\pgfsys@useobject{currentmarker}{}%
\end{pgfscope}%
\end{pgfscope}%
\begin{pgfscope}%
\pgfsetbuttcap%
\pgfsetroundjoin%
\definecolor{currentfill}{rgb}{0.000000,0.000000,0.000000}%
\pgfsetfillcolor{currentfill}%
\pgfsetlinewidth{0.501875pt}%
\definecolor{currentstroke}{rgb}{0.000000,0.000000,0.000000}%
\pgfsetstrokecolor{currentstroke}%
\pgfsetdash{}{0pt}%
\pgfsys@defobject{currentmarker}{\pgfqpoint{0.000000in}{0.000000in}}{\pgfqpoint{0.020833in}{0.000000in}}{%
\pgfpathmoveto{\pgfqpoint{0.000000in}{0.000000in}}%
\pgfpathlineto{\pgfqpoint{0.020833in}{0.000000in}}%
\pgfusepath{stroke,fill}%
}%
\begin{pgfscope}%
\pgfsys@transformshift{0.602522in}{0.915628in}%
\pgfsys@useobject{currentmarker}{}%
\end{pgfscope}%
\end{pgfscope}%
\begin{pgfscope}%
\pgfsetbuttcap%
\pgfsetroundjoin%
\definecolor{currentfill}{rgb}{0.000000,0.000000,0.000000}%
\pgfsetfillcolor{currentfill}%
\pgfsetlinewidth{0.501875pt}%
\definecolor{currentstroke}{rgb}{0.000000,0.000000,0.000000}%
\pgfsetstrokecolor{currentstroke}%
\pgfsetdash{}{0pt}%
\pgfsys@defobject{currentmarker}{\pgfqpoint{0.000000in}{0.000000in}}{\pgfqpoint{0.020833in}{0.000000in}}{%
\pgfpathmoveto{\pgfqpoint{0.000000in}{0.000000in}}%
\pgfpathlineto{\pgfqpoint{0.020833in}{0.000000in}}%
\pgfusepath{stroke,fill}%
}%
\begin{pgfscope}%
\pgfsys@transformshift{0.602522in}{0.982529in}%
\pgfsys@useobject{currentmarker}{}%
\end{pgfscope}%
\end{pgfscope}%
\begin{pgfscope}%
\pgfsetbuttcap%
\pgfsetroundjoin%
\definecolor{currentfill}{rgb}{0.000000,0.000000,0.000000}%
\pgfsetfillcolor{currentfill}%
\pgfsetlinewidth{0.501875pt}%
\definecolor{currentstroke}{rgb}{0.000000,0.000000,0.000000}%
\pgfsetstrokecolor{currentstroke}%
\pgfsetdash{}{0pt}%
\pgfsys@defobject{currentmarker}{\pgfqpoint{0.000000in}{0.000000in}}{\pgfqpoint{0.020833in}{0.000000in}}{%
\pgfpathmoveto{\pgfqpoint{0.000000in}{0.000000in}}%
\pgfpathlineto{\pgfqpoint{0.020833in}{0.000000in}}%
\pgfusepath{stroke,fill}%
}%
\begin{pgfscope}%
\pgfsys@transformshift{0.602522in}{1.049429in}%
\pgfsys@useobject{currentmarker}{}%
\end{pgfscope}%
\end{pgfscope}%
\begin{pgfscope}%
\pgfsetbuttcap%
\pgfsetroundjoin%
\definecolor{currentfill}{rgb}{0.000000,0.000000,0.000000}%
\pgfsetfillcolor{currentfill}%
\pgfsetlinewidth{0.501875pt}%
\definecolor{currentstroke}{rgb}{0.000000,0.000000,0.000000}%
\pgfsetstrokecolor{currentstroke}%
\pgfsetdash{}{0pt}%
\pgfsys@defobject{currentmarker}{\pgfqpoint{0.000000in}{0.000000in}}{\pgfqpoint{0.020833in}{0.000000in}}{%
\pgfpathmoveto{\pgfqpoint{0.000000in}{0.000000in}}%
\pgfpathlineto{\pgfqpoint{0.020833in}{0.000000in}}%
\pgfusepath{stroke,fill}%
}%
\begin{pgfscope}%
\pgfsys@transformshift{0.602522in}{1.116330in}%
\pgfsys@useobject{currentmarker}{}%
\end{pgfscope}%
\end{pgfscope}%
\begin{pgfscope}%
\pgfsetbuttcap%
\pgfsetroundjoin%
\definecolor{currentfill}{rgb}{0.000000,0.000000,0.000000}%
\pgfsetfillcolor{currentfill}%
\pgfsetlinewidth{0.501875pt}%
\definecolor{currentstroke}{rgb}{0.000000,0.000000,0.000000}%
\pgfsetstrokecolor{currentstroke}%
\pgfsetdash{}{0pt}%
\pgfsys@defobject{currentmarker}{\pgfqpoint{0.000000in}{0.000000in}}{\pgfqpoint{0.020833in}{0.000000in}}{%
\pgfpathmoveto{\pgfqpoint{0.000000in}{0.000000in}}%
\pgfpathlineto{\pgfqpoint{0.020833in}{0.000000in}}%
\pgfusepath{stroke,fill}%
}%
\begin{pgfscope}%
\pgfsys@transformshift{0.602522in}{1.250131in}%
\pgfsys@useobject{currentmarker}{}%
\end{pgfscope}%
\end{pgfscope}%
\begin{pgfscope}%
\pgfsetbuttcap%
\pgfsetroundjoin%
\definecolor{currentfill}{rgb}{0.000000,0.000000,0.000000}%
\pgfsetfillcolor{currentfill}%
\pgfsetlinewidth{0.501875pt}%
\definecolor{currentstroke}{rgb}{0.000000,0.000000,0.000000}%
\pgfsetstrokecolor{currentstroke}%
\pgfsetdash{}{0pt}%
\pgfsys@defobject{currentmarker}{\pgfqpoint{0.000000in}{0.000000in}}{\pgfqpoint{0.020833in}{0.000000in}}{%
\pgfpathmoveto{\pgfqpoint{0.000000in}{0.000000in}}%
\pgfpathlineto{\pgfqpoint{0.020833in}{0.000000in}}%
\pgfusepath{stroke,fill}%
}%
\begin{pgfscope}%
\pgfsys@transformshift{0.602522in}{1.317031in}%
\pgfsys@useobject{currentmarker}{}%
\end{pgfscope}%
\end{pgfscope}%
\begin{pgfscope}%
\pgfsetbuttcap%
\pgfsetroundjoin%
\definecolor{currentfill}{rgb}{0.000000,0.000000,0.000000}%
\pgfsetfillcolor{currentfill}%
\pgfsetlinewidth{0.501875pt}%
\definecolor{currentstroke}{rgb}{0.000000,0.000000,0.000000}%
\pgfsetstrokecolor{currentstroke}%
\pgfsetdash{}{0pt}%
\pgfsys@defobject{currentmarker}{\pgfqpoint{0.000000in}{0.000000in}}{\pgfqpoint{0.020833in}{0.000000in}}{%
\pgfpathmoveto{\pgfqpoint{0.000000in}{0.000000in}}%
\pgfpathlineto{\pgfqpoint{0.020833in}{0.000000in}}%
\pgfusepath{stroke,fill}%
}%
\begin{pgfscope}%
\pgfsys@transformshift{0.602522in}{1.383932in}%
\pgfsys@useobject{currentmarker}{}%
\end{pgfscope}%
\end{pgfscope}%
\begin{pgfscope}%
\pgfsetbuttcap%
\pgfsetroundjoin%
\definecolor{currentfill}{rgb}{0.000000,0.000000,0.000000}%
\pgfsetfillcolor{currentfill}%
\pgfsetlinewidth{0.501875pt}%
\definecolor{currentstroke}{rgb}{0.000000,0.000000,0.000000}%
\pgfsetstrokecolor{currentstroke}%
\pgfsetdash{}{0pt}%
\pgfsys@defobject{currentmarker}{\pgfqpoint{0.000000in}{0.000000in}}{\pgfqpoint{0.020833in}{0.000000in}}{%
\pgfpathmoveto{\pgfqpoint{0.000000in}{0.000000in}}%
\pgfpathlineto{\pgfqpoint{0.020833in}{0.000000in}}%
\pgfusepath{stroke,fill}%
}%
\begin{pgfscope}%
\pgfsys@transformshift{0.602522in}{1.450832in}%
\pgfsys@useobject{currentmarker}{}%
\end{pgfscope}%
\end{pgfscope}%
\begin{pgfscope}%
\pgfsetbuttcap%
\pgfsetroundjoin%
\definecolor{currentfill}{rgb}{0.000000,0.000000,0.000000}%
\pgfsetfillcolor{currentfill}%
\pgfsetlinewidth{0.501875pt}%
\definecolor{currentstroke}{rgb}{0.000000,0.000000,0.000000}%
\pgfsetstrokecolor{currentstroke}%
\pgfsetdash{}{0pt}%
\pgfsys@defobject{currentmarker}{\pgfqpoint{0.000000in}{0.000000in}}{\pgfqpoint{0.020833in}{0.000000in}}{%
\pgfpathmoveto{\pgfqpoint{0.000000in}{0.000000in}}%
\pgfpathlineto{\pgfqpoint{0.020833in}{0.000000in}}%
\pgfusepath{stroke,fill}%
}%
\begin{pgfscope}%
\pgfsys@transformshift{0.602522in}{1.584633in}%
\pgfsys@useobject{currentmarker}{}%
\end{pgfscope}%
\end{pgfscope}%
\begin{pgfscope}%
\pgfsetbuttcap%
\pgfsetroundjoin%
\definecolor{currentfill}{rgb}{0.000000,0.000000,0.000000}%
\pgfsetfillcolor{currentfill}%
\pgfsetlinewidth{0.501875pt}%
\definecolor{currentstroke}{rgb}{0.000000,0.000000,0.000000}%
\pgfsetstrokecolor{currentstroke}%
\pgfsetdash{}{0pt}%
\pgfsys@defobject{currentmarker}{\pgfqpoint{0.000000in}{0.000000in}}{\pgfqpoint{0.020833in}{0.000000in}}{%
\pgfpathmoveto{\pgfqpoint{0.000000in}{0.000000in}}%
\pgfpathlineto{\pgfqpoint{0.020833in}{0.000000in}}%
\pgfusepath{stroke,fill}%
}%
\begin{pgfscope}%
\pgfsys@transformshift{0.602522in}{1.651534in}%
\pgfsys@useobject{currentmarker}{}%
\end{pgfscope}%
\end{pgfscope}%
\begin{pgfscope}%
\pgfsetbuttcap%
\pgfsetroundjoin%
\definecolor{currentfill}{rgb}{0.000000,0.000000,0.000000}%
\pgfsetfillcolor{currentfill}%
\pgfsetlinewidth{0.501875pt}%
\definecolor{currentstroke}{rgb}{0.000000,0.000000,0.000000}%
\pgfsetstrokecolor{currentstroke}%
\pgfsetdash{}{0pt}%
\pgfsys@defobject{currentmarker}{\pgfqpoint{0.000000in}{0.000000in}}{\pgfqpoint{0.020833in}{0.000000in}}{%
\pgfpathmoveto{\pgfqpoint{0.000000in}{0.000000in}}%
\pgfpathlineto{\pgfqpoint{0.020833in}{0.000000in}}%
\pgfusepath{stroke,fill}%
}%
\begin{pgfscope}%
\pgfsys@transformshift{0.602522in}{1.718434in}%
\pgfsys@useobject{currentmarker}{}%
\end{pgfscope}%
\end{pgfscope}%
\begin{pgfscope}%
\pgfsetbuttcap%
\pgfsetroundjoin%
\definecolor{currentfill}{rgb}{0.000000,0.000000,0.000000}%
\pgfsetfillcolor{currentfill}%
\pgfsetlinewidth{0.501875pt}%
\definecolor{currentstroke}{rgb}{0.000000,0.000000,0.000000}%
\pgfsetstrokecolor{currentstroke}%
\pgfsetdash{}{0pt}%
\pgfsys@defobject{currentmarker}{\pgfqpoint{0.000000in}{0.000000in}}{\pgfqpoint{0.020833in}{0.000000in}}{%
\pgfpathmoveto{\pgfqpoint{0.000000in}{0.000000in}}%
\pgfpathlineto{\pgfqpoint{0.020833in}{0.000000in}}%
\pgfusepath{stroke,fill}%
}%
\begin{pgfscope}%
\pgfsys@transformshift{0.602522in}{1.785335in}%
\pgfsys@useobject{currentmarker}{}%
\end{pgfscope}%
\end{pgfscope}%
\begin{pgfscope}%
\pgfsetbuttcap%
\pgfsetroundjoin%
\definecolor{currentfill}{rgb}{0.000000,0.000000,0.000000}%
\pgfsetfillcolor{currentfill}%
\pgfsetlinewidth{0.501875pt}%
\definecolor{currentstroke}{rgb}{0.000000,0.000000,0.000000}%
\pgfsetstrokecolor{currentstroke}%
\pgfsetdash{}{0pt}%
\pgfsys@defobject{currentmarker}{\pgfqpoint{0.000000in}{0.000000in}}{\pgfqpoint{0.020833in}{0.000000in}}{%
\pgfpathmoveto{\pgfqpoint{0.000000in}{0.000000in}}%
\pgfpathlineto{\pgfqpoint{0.020833in}{0.000000in}}%
\pgfusepath{stroke,fill}%
}%
\begin{pgfscope}%
\pgfsys@transformshift{0.602522in}{1.919136in}%
\pgfsys@useobject{currentmarker}{}%
\end{pgfscope}%
\end{pgfscope}%
\begin{pgfscope}%
\pgfsetbuttcap%
\pgfsetroundjoin%
\definecolor{currentfill}{rgb}{0.000000,0.000000,0.000000}%
\pgfsetfillcolor{currentfill}%
\pgfsetlinewidth{0.501875pt}%
\definecolor{currentstroke}{rgb}{0.000000,0.000000,0.000000}%
\pgfsetstrokecolor{currentstroke}%
\pgfsetdash{}{0pt}%
\pgfsys@defobject{currentmarker}{\pgfqpoint{0.000000in}{0.000000in}}{\pgfqpoint{0.020833in}{0.000000in}}{%
\pgfpathmoveto{\pgfqpoint{0.000000in}{0.000000in}}%
\pgfpathlineto{\pgfqpoint{0.020833in}{0.000000in}}%
\pgfusepath{stroke,fill}%
}%
\begin{pgfscope}%
\pgfsys@transformshift{0.602522in}{1.986036in}%
\pgfsys@useobject{currentmarker}{}%
\end{pgfscope}%
\end{pgfscope}%
\begin{pgfscope}%
\pgfsetbuttcap%
\pgfsetroundjoin%
\definecolor{currentfill}{rgb}{0.000000,0.000000,0.000000}%
\pgfsetfillcolor{currentfill}%
\pgfsetlinewidth{0.501875pt}%
\definecolor{currentstroke}{rgb}{0.000000,0.000000,0.000000}%
\pgfsetstrokecolor{currentstroke}%
\pgfsetdash{}{0pt}%
\pgfsys@defobject{currentmarker}{\pgfqpoint{0.000000in}{0.000000in}}{\pgfqpoint{0.020833in}{0.000000in}}{%
\pgfpathmoveto{\pgfqpoint{0.000000in}{0.000000in}}%
\pgfpathlineto{\pgfqpoint{0.020833in}{0.000000in}}%
\pgfusepath{stroke,fill}%
}%
\begin{pgfscope}%
\pgfsys@transformshift{0.602522in}{2.052937in}%
\pgfsys@useobject{currentmarker}{}%
\end{pgfscope}%
\end{pgfscope}%
\begin{pgfscope}%
\pgfsetbuttcap%
\pgfsetroundjoin%
\definecolor{currentfill}{rgb}{0.000000,0.000000,0.000000}%
\pgfsetfillcolor{currentfill}%
\pgfsetlinewidth{0.501875pt}%
\definecolor{currentstroke}{rgb}{0.000000,0.000000,0.000000}%
\pgfsetstrokecolor{currentstroke}%
\pgfsetdash{}{0pt}%
\pgfsys@defobject{currentmarker}{\pgfqpoint{0.000000in}{0.000000in}}{\pgfqpoint{0.020833in}{0.000000in}}{%
\pgfpathmoveto{\pgfqpoint{0.000000in}{0.000000in}}%
\pgfpathlineto{\pgfqpoint{0.020833in}{0.000000in}}%
\pgfusepath{stroke,fill}%
}%
\begin{pgfscope}%
\pgfsys@transformshift{0.602522in}{2.119837in}%
\pgfsys@useobject{currentmarker}{}%
\end{pgfscope}%
\end{pgfscope}%
\begin{pgfscope}%
\pgfsetbuttcap%
\pgfsetroundjoin%
\definecolor{currentfill}{rgb}{0.000000,0.000000,0.000000}%
\pgfsetfillcolor{currentfill}%
\pgfsetlinewidth{0.501875pt}%
\definecolor{currentstroke}{rgb}{0.000000,0.000000,0.000000}%
\pgfsetstrokecolor{currentstroke}%
\pgfsetdash{}{0pt}%
\pgfsys@defobject{currentmarker}{\pgfqpoint{0.000000in}{0.000000in}}{\pgfqpoint{0.020833in}{0.000000in}}{%
\pgfpathmoveto{\pgfqpoint{0.000000in}{0.000000in}}%
\pgfpathlineto{\pgfqpoint{0.020833in}{0.000000in}}%
\pgfusepath{stroke,fill}%
}%
\begin{pgfscope}%
\pgfsys@transformshift{0.602522in}{2.253639in}%
\pgfsys@useobject{currentmarker}{}%
\end{pgfscope}%
\end{pgfscope}%
\begin{pgfscope}%
\pgfsetbuttcap%
\pgfsetroundjoin%
\definecolor{currentfill}{rgb}{0.000000,0.000000,0.000000}%
\pgfsetfillcolor{currentfill}%
\pgfsetlinewidth{0.501875pt}%
\definecolor{currentstroke}{rgb}{0.000000,0.000000,0.000000}%
\pgfsetstrokecolor{currentstroke}%
\pgfsetdash{}{0pt}%
\pgfsys@defobject{currentmarker}{\pgfqpoint{0.000000in}{0.000000in}}{\pgfqpoint{0.020833in}{0.000000in}}{%
\pgfpathmoveto{\pgfqpoint{0.000000in}{0.000000in}}%
\pgfpathlineto{\pgfqpoint{0.020833in}{0.000000in}}%
\pgfusepath{stroke,fill}%
}%
\begin{pgfscope}%
\pgfsys@transformshift{0.602522in}{2.320539in}%
\pgfsys@useobject{currentmarker}{}%
\end{pgfscope}%
\end{pgfscope}%
\begin{pgfscope}%
\pgfsetbuttcap%
\pgfsetroundjoin%
\definecolor{currentfill}{rgb}{0.000000,0.000000,0.000000}%
\pgfsetfillcolor{currentfill}%
\pgfsetlinewidth{0.501875pt}%
\definecolor{currentstroke}{rgb}{0.000000,0.000000,0.000000}%
\pgfsetstrokecolor{currentstroke}%
\pgfsetdash{}{0pt}%
\pgfsys@defobject{currentmarker}{\pgfqpoint{0.000000in}{0.000000in}}{\pgfqpoint{0.020833in}{0.000000in}}{%
\pgfpathmoveto{\pgfqpoint{0.000000in}{0.000000in}}%
\pgfpathlineto{\pgfqpoint{0.020833in}{0.000000in}}%
\pgfusepath{stroke,fill}%
}%
\begin{pgfscope}%
\pgfsys@transformshift{0.602522in}{2.387440in}%
\pgfsys@useobject{currentmarker}{}%
\end{pgfscope}%
\end{pgfscope}%
\begin{pgfscope}%
\pgfsetbuttcap%
\pgfsetroundjoin%
\definecolor{currentfill}{rgb}{0.000000,0.000000,0.000000}%
\pgfsetfillcolor{currentfill}%
\pgfsetlinewidth{0.501875pt}%
\definecolor{currentstroke}{rgb}{0.000000,0.000000,0.000000}%
\pgfsetstrokecolor{currentstroke}%
\pgfsetdash{}{0pt}%
\pgfsys@defobject{currentmarker}{\pgfqpoint{0.000000in}{0.000000in}}{\pgfqpoint{0.020833in}{0.000000in}}{%
\pgfpathmoveto{\pgfqpoint{0.000000in}{0.000000in}}%
\pgfpathlineto{\pgfqpoint{0.020833in}{0.000000in}}%
\pgfusepath{stroke,fill}%
}%
\begin{pgfscope}%
\pgfsys@transformshift{0.602522in}{2.454340in}%
\pgfsys@useobject{currentmarker}{}%
\end{pgfscope}%
\end{pgfscope}%
\begin{pgfscope}%
\pgfsetbuttcap%
\pgfsetroundjoin%
\definecolor{currentfill}{rgb}{0.000000,0.000000,0.000000}%
\pgfsetfillcolor{currentfill}%
\pgfsetlinewidth{0.501875pt}%
\definecolor{currentstroke}{rgb}{0.000000,0.000000,0.000000}%
\pgfsetstrokecolor{currentstroke}%
\pgfsetdash{}{0pt}%
\pgfsys@defobject{currentmarker}{\pgfqpoint{0.000000in}{0.000000in}}{\pgfqpoint{0.020833in}{0.000000in}}{%
\pgfpathmoveto{\pgfqpoint{0.000000in}{0.000000in}}%
\pgfpathlineto{\pgfqpoint{0.020833in}{0.000000in}}%
\pgfusepath{stroke,fill}%
}%
\begin{pgfscope}%
\pgfsys@transformshift{0.602522in}{2.588141in}%
\pgfsys@useobject{currentmarker}{}%
\end{pgfscope}%
\end{pgfscope}%
\begin{pgfscope}%
\pgfsetbuttcap%
\pgfsetroundjoin%
\definecolor{currentfill}{rgb}{0.000000,0.000000,0.000000}%
\pgfsetfillcolor{currentfill}%
\pgfsetlinewidth{0.501875pt}%
\definecolor{currentstroke}{rgb}{0.000000,0.000000,0.000000}%
\pgfsetstrokecolor{currentstroke}%
\pgfsetdash{}{0pt}%
\pgfsys@defobject{currentmarker}{\pgfqpoint{0.000000in}{0.000000in}}{\pgfqpoint{0.020833in}{0.000000in}}{%
\pgfpathmoveto{\pgfqpoint{0.000000in}{0.000000in}}%
\pgfpathlineto{\pgfqpoint{0.020833in}{0.000000in}}%
\pgfusepath{stroke,fill}%
}%
\begin{pgfscope}%
\pgfsys@transformshift{0.602522in}{2.655042in}%
\pgfsys@useobject{currentmarker}{}%
\end{pgfscope}%
\end{pgfscope}%
\begin{pgfscope}%
\pgfsetbuttcap%
\pgfsetroundjoin%
\definecolor{currentfill}{rgb}{0.000000,0.000000,0.000000}%
\pgfsetfillcolor{currentfill}%
\pgfsetlinewidth{0.501875pt}%
\definecolor{currentstroke}{rgb}{0.000000,0.000000,0.000000}%
\pgfsetstrokecolor{currentstroke}%
\pgfsetdash{}{0pt}%
\pgfsys@defobject{currentmarker}{\pgfqpoint{0.000000in}{0.000000in}}{\pgfqpoint{0.020833in}{0.000000in}}{%
\pgfpathmoveto{\pgfqpoint{0.000000in}{0.000000in}}%
\pgfpathlineto{\pgfqpoint{0.020833in}{0.000000in}}%
\pgfusepath{stroke,fill}%
}%
\begin{pgfscope}%
\pgfsys@transformshift{0.602522in}{2.721942in}%
\pgfsys@useobject{currentmarker}{}%
\end{pgfscope}%
\end{pgfscope}%
\begin{pgfscope}%
\definecolor{textcolor}{rgb}{0.000000,0.000000,0.000000}%
\pgfsetstrokecolor{textcolor}%
\pgfsetfillcolor{textcolor}%
\pgftext[x=0.127483in, y=0.843886in, left, base,rotate=90.000000]{\color{textcolor}{\rmfamily\fontsize{9.000000}{10.800000}\selectfont\catcode`\^=\active\def^{\ifmmode\sp\else\^{}\fi}\catcode`\%=\active\def
\end{pgfscope}%
\begin{pgfscope}%
\definecolor{textcolor}{rgb}{0.000000,0.000000,0.000000}%
\pgfsetstrokecolor{textcolor}%
\pgfsetfillcolor{textcolor}%
\pgftext[x=0.277507in, y=0.734382in, left, base,rotate=90.000000]{\color{textcolor}{\rmfamily\fontsize{9.000000}{10.800000}\selectfont\catcode`\^=\active\def^{\ifmmode\sp\else\^{}\fi}\catcode`\%=\active\def
\end{pgfscope}%
\begin{pgfscope}%
\pgfpathrectangle{\pgfqpoint{0.602522in}{0.179722in}}{\pgfqpoint{3.629027in}{2.567324in}}%
\pgfusepath{clip}%
\pgfsetbuttcap%
\pgfsetroundjoin%
\pgfsetlinewidth{0.803000pt}%
\definecolor{currentstroke}{rgb}{0.400000,0.400000,0.400000}%
\pgfsetstrokecolor{currentstroke}%
\pgfsetdash{{2.960000pt}{1.280000pt}}{0.000000pt}%
\pgfpathmoveto{\pgfqpoint{0.602522in}{1.517733in}}%
\pgfpathlineto{\pgfqpoint{4.231549in}{1.517733in}}%
\pgfusepath{stroke}%
\end{pgfscope}%
\begin{pgfscope}%
\pgfsetrectcap%
\pgfsetmiterjoin%
\pgfsetlinewidth{0.501875pt}%
\definecolor{currentstroke}{rgb}{0.000000,0.000000,0.000000}%
\pgfsetstrokecolor{currentstroke}%
\pgfsetdash{}{0pt}%
\pgfpathmoveto{\pgfqpoint{0.602522in}{0.179722in}}%
\pgfpathlineto{\pgfqpoint{0.602522in}{2.747046in}}%
\pgfusepath{stroke}%
\end{pgfscope}%
\begin{pgfscope}%
\pgfsetrectcap%
\pgfsetmiterjoin%
\pgfsetlinewidth{0.501875pt}%
\definecolor{currentstroke}{rgb}{0.000000,0.000000,0.000000}%
\pgfsetstrokecolor{currentstroke}%
\pgfsetdash{}{0pt}%
\pgfpathmoveto{\pgfqpoint{0.602522in}{0.179722in}}%
\pgfpathlineto{\pgfqpoint{4.231549in}{0.179722in}}%
\pgfusepath{stroke}%
\end{pgfscope}%
\begin{pgfscope}%
\definecolor{textcolor}{rgb}{0.000000,0.000000,0.000000}%
\pgfsetstrokecolor{textcolor}%
\pgfsetfillcolor{textcolor}%
\pgftext[x=1.238780in,y=2.153397in,,bottom]{\color{textcolor}{\rmfamily\fontsize{7.000000}{8.400000}\selectfont\catcode`\^=\active\def^{\ifmmode\sp\else\^{}\fi}\catcode`\%=\active\def
\end{pgfscope}%
\begin{pgfscope}%
\definecolor{textcolor}{rgb}{0.000000,0.000000,0.000000}%
\pgfsetstrokecolor{textcolor}%
\pgfsetfillcolor{textcolor}%
\pgftext[x=2.417036in,y=2.355421in,,bottom]{\color{textcolor}{\rmfamily\fontsize{7.000000}{8.400000}\selectfont\catcode`\^=\active\def^{\ifmmode\sp\else\^{}\fi}\catcode`\%=\active\def
\end{pgfscope}%
\begin{pgfscope}%
\definecolor{textcolor}{rgb}{0.000000,0.000000,0.000000}%
\pgfsetstrokecolor{textcolor}%
\pgfsetfillcolor{textcolor}%
\pgftext[x=3.595291in,y=1.517733in,,bottom]{\color{textcolor}{\rmfamily\fontsize{7.000000}{8.400000}\selectfont\catcode`\^=\active\def^{\ifmmode\sp\else\^{}\fi}\catcode`\%=\active\def
\end{pgfscope}%
\end{pgfpicture}%
\makeatother%
\endgroup%

%% file: generated/representation_ablation.tex
\begin{table}[H]
\centering
\scriptsize
\setlength{\tabcolsep}{3.2pt}
\begin{tabular}{lrrrrrr}
\toprule
Encoder & Dim. & Uniform 12.5\% & Uniform 15\% & Fixed $\alpha=0.8$ & Concentrated 15\% & Rel. GMV \\
\midrule
Qwen3-8B full (primary) & 4096 & 0.10 & 0.08 & 0.19 & 0.46 & 147.5 \\
Qwen3-4B full & 2560 & 0.07 & 0.06 & 0.08 & 0.33 & 146.7 \\
Qwen3-8B@1024 & 1024 & 0.07 & 0.06 & 0.08 & 0.33 & 146.7 \\
Qwen3-4B@1024 & 1024 & 0.17 & 0.17 & 0.29 & 0.73 & 148.5 \\
BGE-large-v1.5 full & 1024 & -- & 0.01 & 0.03 & 0.18 & 145.3 \\
Qwen3-8B@256 & 256 & 0.27 & 0.29 & 0.46 & 0.98 & 149.4 \\
Qwen3-8B@64 & 64 & 0.01 & 0.01 & 0.01 & 0.13 & 144.6 \\
ChronoBERT 2017 & 768 & -- & -- & -- & -- & 164.4 \\
ChronoBERT 2018 & 768 & -- & -- & -- & -- & 166.7 \\
ChronoBERT 2019 & 768 & -- & -- & -- & -- & 160.7 \\
ChronoBERT 2020 & 768 & -- & -- & -- & -- & 158.5 \\
ChronoBERT 2021 & 768 & -- & -- & -- & -- & 160.2 \\
ChronoBERT 2022 & 768 & -- & -- & -- & -- & 165.3 \\
ChronoBERT 2023 & 768 & -- & -- & -- & -- & 158.7 \\
ChronoBERT 2024 & 768 & -- & -- & -- & -- & 161.6 \\
\midrule
Equal risk weights & -- & 34.73 & 34.02 & 32.84 & 28.33 & 162.6 \\
\bottomrule
\end{tabular}
\caption{Representation sensitivity of the news-only allocation across seven base representations and eight published annual ChronoBERT encoder vintages on the common 100-firm, 1,207-date standardized-return panel. The four middle columns report the percentage of fixed-law allocations with standardized variance no greater than the candidate's; lower is better. Relative variance indexes the long-only sample GMV to 100, so values above 100 measure percentage excess variance and lower is better. A dash marks a candidate that violates the law's unchanged cap. The $\alpha=0.8$ law is a fixed moderately concentrated comparison; its realized mean effective $N$ is reported rather than assumed to match the primary allocation. Returns evaluate every fixed news-only allocation in sample; they do not construct it.}
\label{tab:p3-representation-ablation}
\end{table}
\medskip
\begingroup
\scriptsize
\setlength{\tabcolsep}{3.0pt}
\begin{longtable}{@{}p{0.35\linewidth}rrrrrr@{}}
\caption{Paired stationary-return-bootstrap contrasts for the 15-cell representation ladder. The seven base contrasts are descriptive candidate-minus-reference relative-GMV index-point differences, so a negative estimate favours the candidate. $p$ denotes two-sided add-one sign-tail probabilities; $p_H$ denotes Holm adjustment separately within the seven base and 28 vintage contrasts. The ChronoBERT rows use the BGE-large-minus-Qwen3-8B@1024 interval as a benchmark-calibrated, post-specified robustness threshold. Eq. requires strict containment of the full marginal interval; intervals are not simultaneous. Published annual cumulative training is not a matched-compute causal or architecture-only comparison.}
\label{tab:p3-representation-contrasts} \\
\toprule
Contrast ($c-r$) & Estimate & 95\% CI & $p$ & $p_H$ & Bound & Eq. \\
\midrule
\endfirsthead
\multicolumn{7}{c}{\tablename\ \thetable\ (continued)} \\
\toprule
Contrast ($c-r$) & Estimate & 95\% CI & $p$ & $p_H$ & Bound & Eq. \\
\midrule
\endhead
\midrule
\endfoot
\bottomrule
\endlastfoot
Qwen3-4B full $-$ Qwen3-8B full & -0.842 & [-2.664, 1.649] & 0.454 & 0.873 & -- & -- \\
Qwen3-4B@1024 $-$ Qwen3-8B@1024 & 1.811 & [-0.863, 3.486] & 0.192 & 0.768 & -- & -- \\
Qwen3-8B@1024 $-$ Qwen3-8B full & -0.794 & [-1.934, 0.762] & 0.291 & 0.873 & -- & -- \\
Qwen3-8B@256 $-$ Qwen3-8B@1024 & 2.641 & [0.076, 4.830] & 0.046 & 0.322 & -- & -- \\
Qwen3-8B@64 $-$ Qwen3-8B@256 & -4.774 & [-8.324, 0.365] & 0.065 & 0.390 & -- & -- \\
BGE-large-v1.5 full $-$ Qwen3-8B@1024 & -1.446 & [-3.807, 1.674] & 0.349 & 0.873 & -- & -- \\
BGE-large-v1.5 full $-$ Qwen3-4B@1024 & -3.257 & [-5.526, 0.733] & 0.115 & 0.575 & -- & -- \\
\addlinespace[1.5pt]
ChronoBERT 2018 $-$ ChronoBERT 2017 & 2.252 & [1.170, 4.326] & 0.001 & 0.028 & 3.807 & No \\
ChronoBERT 2019 $-$ ChronoBERT 2017 & -3.795 & [-6.322, -2.501] & 0.001 & 0.028 & 3.807 & No \\
ChronoBERT 2019 $-$ ChronoBERT 2018 & -6.047 & [-10.132, -4.313] & 0.001 & 0.028 & 3.807 & No \\
ChronoBERT 2020 $-$ ChronoBERT 2017 & -5.960 & [-11.077, -3.918] & 0.001 & 0.028 & 3.807 & No \\
ChronoBERT 2020 $-$ ChronoBERT 2018 & -8.212 & [-14.985, -5.533] & 0.001 & 0.028 & 3.807 & No \\
ChronoBERT 2020 $-$ ChronoBERT 2019 & -2.165 & [-5.258, -0.789] & 0.001 & 0.028 & 3.807 & No \\
ChronoBERT 2021 $-$ ChronoBERT 2017 & -4.209 & [-9.080, -2.089] & 0.001 & 0.028 & 3.807 & No \\
ChronoBERT 2021 $-$ ChronoBERT 2018 & -6.461 & [-13.015, -3.710] & 0.001 & 0.028 & 3.807 & No \\
ChronoBERT 2021 $-$ ChronoBERT 2019 & -0.414 & [-3.500, 1.169] & 0.576 & 1.000 & 3.807 & Yes \\
ChronoBERT 2021 $-$ ChronoBERT 2020 & 1.751 & [0.486, 3.072] & 0.013 & 0.117 & 3.807 & Yes \\
ChronoBERT 2022 $-$ ChronoBERT 2017 & 0.893 & [-2.228, 3.110] & 0.535 & 1.000 & 3.807 & Yes \\
ChronoBERT 2022 $-$ ChronoBERT 2018 & -1.359 & [-5.718, 1.175] & 0.305 & 1.000 & 3.807 & No \\
ChronoBERT 2022 $-$ ChronoBERT 2019 & 4.688 & [3.284, 6.615] & 0.001 & 0.028 & 3.807 & No \\
ChronoBERT 2022 $-$ ChronoBERT 2020 & 6.852 & [5.381, 10.737] & 0.001 & 0.028 & 3.807 & No \\
ChronoBERT 2022 $-$ ChronoBERT 2021 & 5.102 & [3.913, 8.582] & 0.001 & 0.028 & 3.807 & No \\
ChronoBERT 2023 $-$ ChronoBERT 2017 & -5.703 & [-10.542, -3.385] & 0.001 & 0.028 & 3.807 & No \\
ChronoBERT 2023 $-$ ChronoBERT 2018 & -7.955 & [-14.412, -5.048] & 0.001 & 0.028 & 3.807 & No \\
ChronoBERT 2023 $-$ ChronoBERT 2019 & -1.908 & [-4.976, -0.125] & 0.035 & 0.280 & 3.807 & No \\
ChronoBERT 2023 $-$ ChronoBERT 2020 & 0.257 & [-0.967, 1.629] & 0.643 & 1.000 & 3.807 & Yes \\
ChronoBERT 2023 $-$ ChronoBERT 2021 & -1.494 & [-3.093, 0.113] & 0.068 & 0.408 & 3.807 & Yes \\
ChronoBERT 2023 $-$ ChronoBERT 2022 & -6.595 & [-10.218, -4.922] & 0.001 & 0.028 & 3.807 & No \\
ChronoBERT 2024 $-$ ChronoBERT 2017 & -2.821 & [-5.549, -1.127] & 0.001 & 0.028 & 3.807 & No \\
ChronoBERT 2024 $-$ ChronoBERT 2018 & -5.073 & [-9.026, -2.862] & 0.001 & 0.028 & 3.807 & No \\
ChronoBERT 2024 $-$ ChronoBERT 2019 & 0.974 & [-0.358, 2.294] & 0.146 & 0.730 & 3.807 & Yes \\
ChronoBERT 2024 $-$ ChronoBERT 2020 & 3.139 & [1.980, 6.280] & 0.001 & 0.028 & 3.807 & No \\
ChronoBERT 2024 $-$ ChronoBERT 2021 & 1.388 & [0.041, 4.493] & 0.048 & 0.336 & 3.807 & No \\
ChronoBERT 2024 $-$ ChronoBERT 2022 & -3.713 & [-5.678, -2.351] & 0.001 & 0.028 & 3.807 & No \\
ChronoBERT 2024 $-$ ChronoBERT 2023 & 2.882 & [1.991, 5.585] & 0.001 & 0.028 & 3.807 & No \\
\end{longtable}
\endgroup